\documentclass[useams,useamsmath,usenatbib]{mnras}
\usepackage{amsmath}
\usepackage{setspace}
\usepackage{pdfpages}
\graphicspath{{./images/}}

\def\ie{\,{\rm i.e.}\,}
\def\eg{\,{\rm e.g.}\,}
\def\um{{\rm $\mu$m}\,}
\def\sdss{{\sc sdss}}
\def\viking{{\sc viking}}
\def\hatlas{{\sc h-atlas}}
\def\wise{{\em WISE}}

\def\galex{{\em GALEX}}
\def\lambdar{{\sc lambdar}}

\def\bbf{}
\begin{document}

\twocolumn

\title[GAMA \lambdar]{Galaxy And Mass Assembly (GAMA): Accurate Panchromatic
Photometry from Optical Priors using {\sc lambdar}}

%Authors {{{
\author[Wright et al.]
{A.H.~Wright$^{1}$\thanks{e-mail:Angus.Wright@icrar.org},
A.S.G.~Robotham$^{1}$,
N.~Bourne$^{2}$,
S.P.~Driver$^{1,3}$,
L.~Dunne$^{2,4}$,
\newauthor
S.J.~Maddox$^{2,4}$,
M.~Alpaslan$^{5}$,
S.K.~Andrews$^{1}$,
A.E.~Bauer$^{6}$,
J.~Bland-Hawthorn$^{7}$,
\newauthor
S.~Brough$^{6}$,
M.J.I.~Brown$^{8}$,
M.~Cluver$^{9}$,
L.J.M.~Davies$^{1}$,
B.W.~Holwerda$^{10}$,
\newauthor
A.M.~Hopkins$^{6}$,
T.H.~Jarrett$^{11}$,
P.R.~Kafle$^{1}$,
R.~Lange$^{1}$,
J.~Liske$^{12}$,
J.~Loveday$^{13}$,
\newauthor
A.J.~Moffett$^{1}$,
P.~Norberg$^{14}$,
C.C.~Popescu$^{15,16}$,
M. Smith$^{4}$,
E.N.~Taylor$^{17}$,
\newauthor
R.J.~Tuffs$^{18}$,
L.~Wang$^{19}$,
S.M.~Wilkins$^{13}$ \\
$^{1}$ICRAR, The
University of Western Australia, 35 Stirling Highway, Crawley, WA 6009, Australia\\
$^{2}$SUPA, Institute for Astronomy, University of Edinburgh, Royal Observatory, Blackford Hill, Edinburgh, EH9 3HJ, UK \\
$^{3}$SUPA, School of Physics
\& Astronomy, University of St Andrews, North Haugh, St Andrews, KY16 9SS, UK \\
$^{4}$School of Physics and Astronomy, Cardiff University, The Parade, Cardiff, CF24 3AA, UK \\
$^{5}$NASA Ames Research Center, N232, Moffett Field, Mountain View, CA 94035, United States \\
$^{6}$Australian Astronomical Observatory, PO Box 915, North Ryde, NSW 1670, Australia \\
$^{7}$Sydney Institute for Astronomy, School of Physics A28, University of Sydney, NSW 2006, Australia \\
$^{8}$School of Physics and Astronomy, Monash University, Clayton, Victoria 3800, Australia \\
$^{9}$Department of Physics and Astronomy, University of the Western Cape, Robert Sobukwe Road, Bellville, 7535, South Africa \\
$^{10}$University of Leiden, Sterrenwacht Leiden, Niels Bohrweg 2, NL-2333 CA Leiden, The Netherlands \\
$^{11}$Astronomy Dept, University of Cape Town, Rondebosch, 7701, RSA \\
$^{12}$Hamburger Sternwarte, Universit{\"at} Hamburg, Gojenbergsweg 112, 21029 Hamburg, Germany \\
$^{13}$Astronomy Centre, University of Sussex, Falmer, Brighton BN1 9QH, UK \\
$^{14}$ICC \& CEA, Department of Physics, Durham University, South Road, Durham DH1 3LE, UK \\
$^{15}$Jeremiah Horrocks Institute, University of Central Lancashire, PR1 2HE Preston, UK \\
$^{16}$The Astronomical Institute of the Romanian Academy, Str. Cutitul de Argint 5, Bucharest, Romania \\
$^{17}$School of Physics, the University of Melbourne, VIC 3010, Australia \\
$^{18}$Max Planck Institut fuer Kernphysik, Saupfercheckweg 1, 69117 Heidelberg, Germany \\
$^{19}$SRON Netherlands Institute for Space Research, Landleven 12, 9747 AD, Groningen, The Netherlands \vspace{-10pt}}
\date{Received XXXX; Accepted XXXX}
\pubyear{2016} \volume{000}
\pagerange{\pageref{firstpage}--\pageref{lastpage}}
%}}}
\maketitle
\label{firstpage}
\begin{abstract}%{{{
{
We present the Lambda Adaptive Multi-Band Deblending Algorithm in R (\lambdar), a novel code for calculating matched aperture photometry across
images that are neither pixel- nor PSF-matched, using prior aperture definitions derived from high resolution optical imaging.
The development of this program is motivated by the desire for consistent photometry and uncertainties across large ranges of photometric imaging,
for use in calculating spectral energy distributions.
We describe the program, specifically key features required for robust determination
of panchromatic photometry: propagation of apertures to images with arbitrary resolution, local background estimation, aperture normalisation,
uncertainty determination and propagation, and object deblending. Using simulated images, we demonstrate that the program
is able to recover accurate photometric measurements in both high-resolution, low-confusion, and low-resolution, high-confusion, regimes.
We apply the program to the 21-band photometric dataset from the Galaxy And Mass
Assembly (GAMA) Panchromatic Data Release (PDR; \citealt{Driver2016}), which contains imaging spanning the far-UV to the far-IR.
We compare photometry derived from \lambdar\ with that presented in \cite{Driver2016},
finding broad agreement between the datasets.
Nonetheless, we demonstrate that the photometry from \lambdar\ is superior to that from the GAMA PDR, as determined by a
reduction in the outlier rate and intrinsic scatter of colours in the \lambdar\ dataset.
We similarly find a decrease in the outlier rate of stellar masses and star formation rates using \lambdar\ photometry.
Finally, we note an exceptional increase in the number of UV and mid-IR sources able to be constrained, which is accompanied by a significant
increase in the mid-IR colour-colour parameter-space able to be explored.
}
\end{abstract}
%}}}

%}}}

\section{Introduction}\label{sec: intro} %{{{
Over the past decade, the existence of large multi-wavelength collaborations such as
the Galaxy and Mass Assembly (GAMA; \citealt{Driver2011,Driver2016,Liske2015}) survey,
{\em Herschel} Astrophysical Terahertz Large Area Survey (H-ATLAS; \cite{Eales2010}),
{\em Herschel} Extragalactic Legacy Project (HELP; \citealt{Vaccari2015}),
the Cosmological Evolution Survey (COSMOS; \citealt{Scoville2007}), the Cosmic Assembly Near-infrared
Deep Extragalactic Legacy Survey (CANDLES; \citealt{Grogin2011}), and the Great Observatories Origins
Deep Survey (GOODS; \citealt{Elbaz2011}), has enabled scientists to probe an
increasing array of extra-galactic environments, and eras in an increasingly
comprehensive and systematic manner.

One area of interest in multi-wavelength extra-galactic studies is the determination of self-consistent galactic parameters
such as stellar mass \citep{Taylor2011}, dust mass \citep{Dunne2011}, and star formation rate measures \citep{Davies2015}.
Using statistically robust samples of these parameters, we can populate global distributions of
interest, such as the galaxy stellar mass function (GSMF; \citealt{Baldry2012}) and evolution of the cosmic
star formation rate \citep{Madau2014}. By combining self-consistent measures of these distributions with
HI mass estimates, we can examine the galactic baryonic mass function (BMF; \citealt{Papastergis2012}).
While these individual parameters are able to be calculated to high accuracy without the fitting of complex models
(indeed, adding more information than is explicitly necessary can act to detriment the measurement of individual parameters;
see \citealt{Taylor2011}), in order to calculate these parameters self-consistently measurement of individual
galactic spectral energy distributions (SEDs) is nominally best-practice. This is because modelling the SED
allows all galactic parameters to be optimised simultaneously with consideration of how they impact
one-another and co-evolve \citep{Walcher2011,Conroy2013}.

Measurement of these parameters requires quantification of the flux emitted by an object in one or more photometric images, and in
particular the management of data with very different sensitivity limits and spatial resolutions.
To measure total object fluxes robustly, it is important to determine a sensible metric of measurement, and then to quantify any flux systematically
missed because of this chosen method. The simplest approach to measuring total object photometry involves using circular apertures to capture a known fraction of an object's flux,
which can then be corrected to a total flux \citep{Kron1980,Petrosian1976}, or by extending these methods to elliptical apertures \citep{Bertin1996,Jarrett2000}.
Measurement can be refined by fitting observed structure when calculating photometry, either by assuming a fixed profile shape, \eg an exponential profile
\citep{Patterson1940,Freeman1970}, \cite{deVaucouleurs1948} profile, or by fitting for the profile shape using a generalised Sersic
profile \citep{Sersic1963,Graham2005b,Kelvin2014,Jarrett2013}. These methods, however, can cause systematic under-estimation of total fluxes as a
function of morphology \citep{Graham2005b}.

Unfortunately, there is no `standard' photometric method that is used, or even necessarily able, to extract photometry from a wide range of
photometric images \citep{Hill2010,Driver2016}. As a result, compilation of large samples of multi-wavelength
photometry is typically achieved in one of three ways: by using a cross-matching scheme that combines photometric measurements
(often from different methods) at the catalogue level (`table matching', see \eg \citealt{Bundy2012});
by degrading the resolution of all images to that of the lowest resolution image, and performing matched aperture photometry on these degraded
images (`forced aperture photometry', see \eg \citealt{Bertin1996,Capak2007,Hildebrandt2012,Hill2010,Driver2016});
or by using information in a high-resolution band to inform the extraction of photometry at lower resolutions,
either by matching flux ratios (`flux fitting', see \eg \citealt{Laidler2007,deSantis2007,Merlin2015,Mancone2013})
or by matching structure (`profile fitting', see \eg \citealt{Erwin2014,Vika2013,Strauss2002,Kuijken2008,Kelvin2012}).

These methods of analysis each have benefits and detriments. `Forced aperture photometry' is implemented widely
but has limited use when the quality of images needing to be analysed varies significantly \citep{Hill2011},
as the method discards spatial information in the image degradation. `Flux fitting' and `profile fitting'
are both very sophisticated, and are useful in cases where there exists a large disparity between photometric images
and the highest resolution image is able to reliably determine object structure \citep{Kelvin2012}. In cases where it
is not possible to reliably determine object structure in all bands, however, one must propagate an
observed profile in one band to lower resolution, often longer wavelength, images. As physical processes vary greatly as a
function of wavelength it is not clear how the profiles might be linked across such large wavelength regions. Accounting for this
change across wavelength likely involves assuming complex models, which may not hold for arbitrary galaxy populations.
Finally, `table matching' is quick, easy, and requires no further analysis of photometric imaging \citep{Bundy2012,Driver2016},
however it does not guarantee that individual measurements will be consistent across multiple facilities and/or wavelengths
(see Section \ref{sec: PDR}).

The point of consistency is an important one and is the reason why so much effort has been invested in developing
programs for matched aperture, forced aperture, flux fitting, and profile fitting photometry. In order to
model the SED of any object, photometric data are compared to physically motivated models of panchromatic emission that
are either pre-constructed (as is the case in energy-balance programs, see \eg \citealt{daCunha2008,Boquien2013}) or developed
dynamically (as in radiative transfer programs, see \eg \citealt{Camps2015,Popescu2011}). In any case, it is
assumed that the data have measurements and uncertainties that are consistent, so that no measurement is unfairly
weighted with respect to any other during least-squares optimisation. For the specific goals of GAMA, in particular
the careful measurement of the SEDs from the UV to the FIR, such consistency is vital.
For this reason we are required to conduct an analysis that is more sophisticated than simple table matching.

For this purpose, we have developed a bespoke program for calculating consistent photometry for objects
across imaging with arbitrary resolutions, using prior information derived from a highest resolution band; the Lambda Adaptive
Multi-Band Deblending Algorithm in R (\lambdar).

In Section \ref{sec: PDR} we discuss the GAMA photometric dataset. In Section \ref{sec: lambdar} we discuss the program
and its many features, detailing the function of the more important or complex routines. Sections
\ref{sec: rband sims} and \ref{sec: fir sims} detail our testing of the program on simulated optical
and far-IR imaging respectively. Section \ref{sec: comparison to PDR} details the photometry that we measure for all GAMA
objects, and how our measurements compare to those presented in the GAMA Panchromatic Data Release (PDR; \citealt{Driver2016}).
In Section \ref{sec: comparison to PDR II} we examine how the new photometry compares to the PDR with regard to derived galactic properties
such as stellar mass and star formation rate. In Section \ref{sec: Data Release} we detail the data release to accompany this publication.
Finally, we present a summary and concluding remarks in Section \ref{sec: conclusions}.
%}}}

\section{The GAMA Panchromatic Data Release}\label{sec: PDR} %{{{
Photometry in GAMA spans 5 different observatories, 21 different broad-band filters, and has pixel resolutions ranging from 0.4 to
12 arcseconds. Each filter has its own characteristic point spread function (PSF), which in GAMA natively range in Full Width
at Half Maximum (FWHM) from $0.85^{\prime\prime}$ to $36^{\prime\prime}$. Finally, each observatory typically implements a
different image calibration scheme, specifically regarding estimation and removal of local sky-backgrounds.

With the exception of imaging in the two {\em Herschel} {\sc pacs} bands, the imaging used for measurement of photometry here
is the same as that used in the GAMA PDR \citep{Driver2016}. Here we give a brief review of the photometry used in this analysis, and direct
the interested reader to publications cited for detailed descriptions of the data and their genesis. {\bbf A
summary of the imaging properties in the GAMA PDR is given in Table \ref{tab: bands}.}

Imaging in the UV domain is from The GALaxy Evolution eXplorer (\galex, \citealt{Martin2010}) satellite,
a medium-class explorer mission operated by NASA and launched in April 2003.
Data collected by \galex\ in the GAMA equatorial fields was observed throughout both the medium imaging survey (MIS) and an additional dedicated survey, lead by
R.J. Tuffs, to MIS depth. \galex\ imaging has a pixel resolution of $1.5^{\prime\prime}$, and has a PSF FWHM of $4.2^{\prime\prime}$ and $5.3^{\prime\prime}$ in the FUV
($153$nm) and NUV ($230$nm) channels respectively \citep{Morrissey2007}. \galex\ imagery has approximately $92\%$ and $95\%$ coverage in the equatorial fields.
A detailed description of the GAMA \galex\ dataset is presented in \cite{Andrae2014}, and is summarised in \cite{Liske2015, Driver2016}.

The Sloan Digital Sky Survey (SDSS, \citealt{York2000}) provides uniform optical imaging in the GAMA equatorial fields in {\em ugriz} bands, at a
pixel resolution of $0.4^{\prime\prime}$ and a typical PSF FWHM of $1.4^{\prime\prime}$. Imaging used here is from SDSS DR7 data \citep{Abazajian2009},
and is described originally in \cite{Hill2011}, updated in \cite{Liske2015}. Importantly, the imaging used here has been Gaussianised to a PSF FWHM
of $2^{\prime\prime}$.

Near-IR imaging is from the Visible and Infrared Telescope for Astronomy (VISTA, \citealt{Sutherland2015}), forming part of the VIsta Kilo-degree
INfrared Galaxy survey (VIKING). VISTA has a pixel resolution of $0.4^{\prime\prime}$, and a typical PSF FWHM of $0.85^{\prime\prime}$. These data have
also undergone Gaussianisation to a common $2^{\prime\prime}$ PSF FWHM. While there is $100\%$ observational coverage from VISTA as part of the VIKING
survey, quality control required that $\sim 2.2\%$ of the imaging frames be removed prior to mosaicing. As a result the final coverage varies slightly,
but is typically better than $99\%$ in each of {\sc zyjhk}. Details of the VIKING quality control are given in \cite{Driver2016}.

Mid-IR imaging is from the Wide-Field Infrared Survey Explorer (WISE, \citealt{Wright2010}) satellite, a medium-class explorer mission operated
by NASA and launched in December 2009. Imaging used by GAMA has been `drizzled' (see \citealt{Jarrett2012,Cluver2014}), reaching a final
PSF FWHM of $5.9^{\prime\prime}$, $6.5^{\prime\prime}$, $7.0^{\prime\prime}$, and $12.4^{\prime\prime}$ in the
W1 (3.4$\mu$m), W2 (4.6$\mu$m), W3 (12$\mu$m) and W4 (22$\mu$m) bands respectively.

The {\em Herschel} space observatory \citep{Pilbratt2010} is operated by the European Space Agency and was launched in May 2009. Imaging used by
GAMA from {\em Herschel} was observed as part of The Herschel Astrophysical Terahertz Large Area Survey ({\sc hatlas}, \citealt{Eales2010}).
{\sc hatlas} imaging in the GAMA equatorial fields utilises co-ordinated observations using both the {\sc pacs} \citep{Poglitsch2010} and {\sc spire}
\citep{Griffin2010} instruments to obtain scans at $100\mu$m, $160\mu$m, $250\mu$m, $350\mu$m, $500\mu$m. Details of the imaging used are given in
\cite{Valiante2016}. Note that, due to ongoing investigation into the impact of the nebuliser scale on the final imaging properties, we opt to
use the pre-nebulised maps for analysis here. Small scale variations in the sky, which are removed by the nebuliser, are instead removed as part of the
sky estimate routine; Appendix \ref{sec: Nebuliser} shows an example of the small variations measured by the nebuliser compared to those measured
by \lambdar.

Details of the methods for measuring photometry across all 21-bands in the PDR are given in \cite{Driver2016}. Briefly,
per-object photometry was collated in a number of ways. UV photometry from \galex\ was calculated using a combination
of aperture photometry and measurement using a curve of growth (CoG). Optical and near-IR photometry from {\sc sdss} and {\sc
vista} were calculated by forced aperture photometry \citep{Hill2011,Driver2016}, using SExtractor \citep{Bertin1996}.
Mid-IR photometry from \wise\ were calculated using a combination of aperture photometry and PSF
modelling \citep{Cluver2014}. Far-IR photometry from the {\em Herschel} spacecraft were calculated using deblended, PSF-weighted aperture photometry
\citep{Bourne2012}. Each of these datasets is subsequently table-matched to create the final PDR photometric dataset.

\begin{figure}
\centering
\includegraphics[scale=0.25]{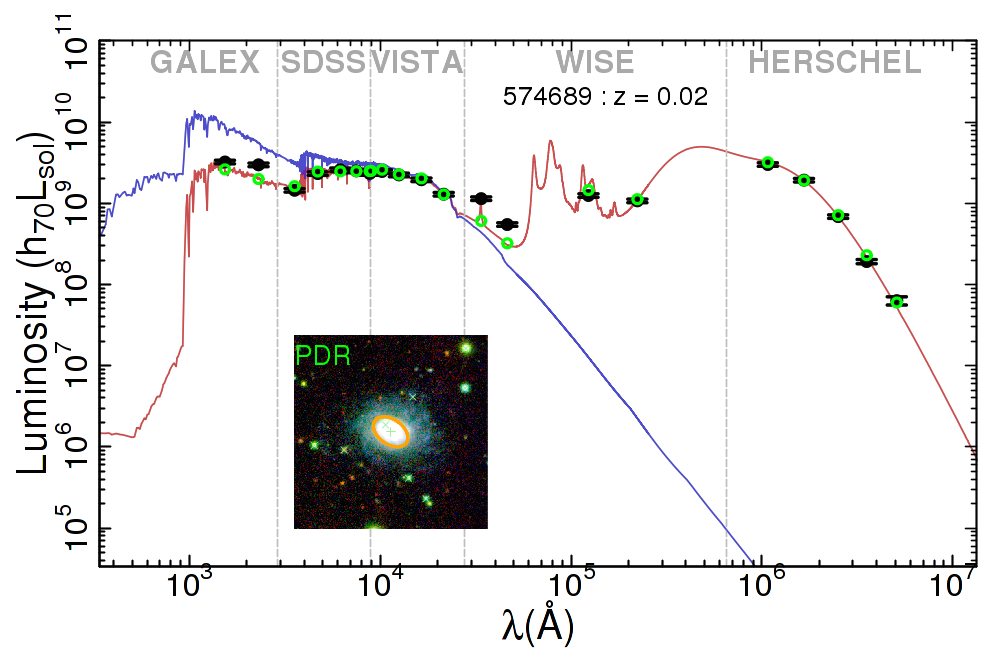}
\caption{A simple example of how inconsistent photometry can result in incorrect measurement of the
SED. Input photometry to the SED fit is shown in black,
the model photometry in green, the obscured SED in red, and the unobscured SED in blue. For this object, the UV
data were measured using an aperture which encompasses the entire galaxy, while the optical and Far-IR data
have been measured in the shrunken aperture due to aperture shredding in source extraction. This aperture is shown in
the inset 3-colour image (made using {\bbf the {\sc viking} H - {\sc sdss} i - {\sc sdss} g
bands for red, green, and blue respectively}). The MIR
data have been measured within a standard aperture with $8.25^{\prime\prime}$ radius.
The SED has then been fit to the inconsistent photometry, giving the SED shown above. }\label{fig: bad aperture}
\end{figure}

To demonstrate how multi-wavelength table-matched photometry can produce incorrect measurements
of the galactic SED, Figure \ref{fig: bad aperture} shows a fit to inconsistent photometry as present in
the GAMA PDR. This example shows inconsistency across instrument/facility
boundaries (for example, the {\sc galex - sdss} boundary) but roughly consistent photometry within an instrument or
facility's bandpass. While this has been chosen because it is a particularly dramatic case, we note that similar effects will be present at a
lower level in all photometric measurements that are not made in a consistent manner across the entire frequency bandpass.
We are therefore required to develop a method for measuring consistent photometry across the highly diverse GAMA PDR dataset.

\begin{table}
\centering
\begin{tabular}{ccccc}
Band & Survey/           & Central         & Pixel Scale          & Native (conv.)     \\
     & Facility                 &     Wavelength  &             (${}^{\prime\prime}$) & PSF FWHM (${}^{\prime\prime}$)\\
\hline
FUV  & \galex            & 1550\AA   &  1.5    &  4.1 \\
NUV  & \galex            & 2275\AA   &  1.5    &  5.2 \\
u    & \sdss             & 3540\AA   &  0.339  &  1.4 (2.0)   \\
g    & \sdss             & 4770\AA   &  0.339  &  1.4 (2.0)   \\
r    & \sdss             & 6230\AA   &  0.339  &  1.4 (2.0)   \\
i    & \sdss             & 7630\AA   &  0.339  &  1.4 (2.0)   \\
z    & \sdss             & 9134\AA   &  0.339  &  1.4 (2.0)   \\
Z    & \viking           & 8770\AA   &  0.339  &  0.9 (2.0)  \\
Y    & \viking           & 1.020\um  &  0.339  &  0.9 (2.0)  \\
J    & \viking           & 1.252\um  &  0.339  &  0.9 (2.0)  \\
H    & \viking           & 1.645\um  &  0.339  &  0.9 (2.0)  \\
K    & \viking           & 2.147\um  &  0.339  &  0.9 (2.0)  \\
W1   & \wise             & 3.4\um    &  1      &  5.9   \\
W2   & \wise             & 4.6\um    &  1      &  6.5   \\
W3   & \wise             & 12\um     &  1      &  7.0   \\
W4   & \wise             & 22\um     &  1      &  12.4   \\
100  & \hatlas          & 100\um    &  3      &  9.6   \\
160  & \hatlas          & 160\um    &  4      &  12.5   \\
250  & \hatlas          & 150\um    &  6      &  18   \\
350  & \hatlas          & 350\um    &  8      &  25   \\
500  & \hatlas          & 500\um    &  12     &  36   \\
\hline
\end{tabular}
\caption{Details of the 21 bands included in the GAMA database, and that are used for the
creation of galactic SEDs. In the SDSS optical and VIKING Near-IR, the PSF FWHM values are shown for both the
native imaging and the post-Gaussianised (i.e. convolved) imaging.}\label{tab: bands}
\end{table}

%%}}}

\section[\lambdar]{{\em \lambdar}: Lambda Adaptive Multi-Band Deblending Algorithm in R}\label{sec: lambdar} %Program %{{{
%Details {{{
The \lambdar\ program is a development of a package detailed in \cite{Bourne2012}.
We have modified and evolved much of the internal
mechanics, introduced scalability, and ported the program from IDL to
an open source platform, R \citep{R}.

The program has been designed for flexibility, scalability, and accuracy.
\lambdar\ is available on the collaborative
build network GitHub (\url{https://github.com/AngusWright/LAMBDAR}), to facilitate rapid updates.
It is the hope of the authors that, by releasing the program publicly to the astronomical community,
it will be tested, scrutinised, and hopefully improved, in a transparent and thorough fashion.

The program is essentially a tool for
performing aperture photometry. The user supplies a FITS image and a
catalogue (containing object locations and aperture parameters), which the program uses
to compute and output individual object fluxes.
The program is designed to include functionality that incorporates behaviour similar to other matched aperture programs,
such as the matched-aperture function within SExtractor,
while allowing increased levels of sophistication and flexibility if desired.
This is done for two reasons; firstly, it allows checks for
consistency with other matched aperture codes; and secondly,
to allow flexibility for the user to perform precisely the type of matched aperture photometry
they require. Note that the \lambdar\ package does not perform a source
detection, but rather requires an input catalogue of apertures (\ie the `priors', see Section \ref{sec: inputs}).

In the following Sections (\ref{sec: inputs} to \ref{sec: errors}), we outline the technical details of the program. The program follows the
following broad process:
\begin{enumerate}\label{list: program}
\item read the required inputs, such as aperture priors and images (\S \ref{sec: inputs});
\item place input aperture priors onto the same pixel-grid as the image being analysed (\S \ref{sec: aperture creation});
\item convolve these aperture priors with the image PSF (\S \ref{sec: psf conv});
\item perform object deblending using convolved aperture priors (\S \ref{sec: deblending});
\item perform estimation of local sky-backgrounds (\S \ref{sec: Sky Estimate});
\item perform estimation of noise correlation using random/blank apertures (\S \ref{sec: randoms/blanks});
\item calculate object fluxes using deblended convolved aperture priors, accounting for local backgrounds (\S \ref{sec: measurement type});
\item calculate and apply required normalisation of fluxes to account for aperture weighting and/or missed flux (\S \ref{sec: minimum aperture correction});
\item calculate final flux uncertainties, incorporating errors from each of the above steps (\S \ref{sec: errors}).
\end{enumerate}

Additionally, individual routine
descriptions (and instructions on how to run the program) are available in the package documentation.
We direct the interested reader to the download page listed previously, where this and other documentation
can be found. Alternatively, the reader can install the program directly into R using the following simple commands
within the R environment:
\lstset{language=R}
\begin{lstlisting}
install.packages(devtools)
library(devtools)
install_github(`AngusWright/LAMBDAR')
library(LAMBDAR)
\end{lstlisting}

%}}}

\subsection{Inputs} \label{sec: inputs} %{{{
The program does not perform an object detection, but rather requires an input catalogue from a source detection on the
user's chosen `prior' image. This list of prior targets remains static while analysing all images of interest; only a single source
detection is required for the definition of prior targets. As such, for any successful flux measurement the user must specify (within the parameter file)
at least:
\begin{enumerate}
\item A catalogue of object Right Ascensions, Declinations, and Aperture Parameters (semi-major axis, semi-minor axis, position-angle);
\item A FITS image with an unrotated tan gnomonic or orthographic WCS Astrometry.
\end{enumerate}

While the input catalogue need only contain the list of prior-based targets, it is often the case that we also want to mask and deblend contaminating
sources which do not form part of the prior-list. As such, the input catalogue can contain an additional parameter for identifying sources in the
catalogue that are contaminants. However, as contaminating sources vary over a broad frequency range (\eg stars in the optical, and high-redshift
galaxies in the far-IR), these additional sources often need to be tailored to specific images, separate to the static list of prior-based targets.
Details of how these full catalogues are determined for GAMA are supplied in Section \ref{sec: Catalogues}.

In addition to the required parameters, the user can specify any of a large number of optional parameters in order to
perform various functions designed to improve the flux determination and/or allow for flexibility. Many of these parameters
are discussed in the Sections below, and all have descriptions within the program's documentation and default parameter file.

{\bbf For reference, Table \ref{tab: lambdar params} outlines the parameter settings used in the GAMA run of \lambdar, as well as a
short description of each parameters' purpose. We include a brief justification of these chosen settings in Section \ref{sec: input params}.}

\begin{table*}
\centering
\begin{tabular}{c|c|l|l}
Parameter & Setting & Caveats & Description \\
\hline
ResampleAper & TRUE & FALSE in {\sc sdss/viking} & Perform recursive descent aperture placement \\
ResamplingRes & 3  & & Resolution of each recursive descent step \\
ResamplingIters & 4 &  & Number of recursive descent iterations \\
PSFConvolve   & TRUE & & Perform a convolution of apertures with the PSF\\
DoSkyEst   & TRUE & FALSE in FUV only & Perform a local sky estimate for each source\\
SkyEstProbCut & 3 & & Sigma-value used in clipping of sky pixels \\
SkyEstIters   & 5 & & Number of sigma-clipping iterations in sky estimate\\
BlankCor  & TRUE & & Estimate correlation in noise using blank apertures\\
nBlanks  & 50 & & Number of blank apertures to measure for every source\\
PSFWeighted & TRUE  & & Use `weighted' apertures for flux measurements \\
PixelFluxWgt & TRUE  & FALSE from $12\mu$m redward & Use pixel-flux to weight apertures at the $0^{\rm th}$ Iteration \\
IterateFluxes  & TRUE  & & Iteratively measure fluxes, weighting by mean surface brightness \\
nIterations & 15 & & Number of iterations to perform\\
\hline
\end{tabular}
\caption{\bbf Settings used in the GAMA \lambdar\ run. While this is not every setting in the program, these are all the settings
that are of importance to the flux/error determination, discussed below, and/or set to a value that is not default.}\label{tab: lambdar params}
\end{table*}

To create unrotated imaging we choose to use the SWarp software \citep{Bertin2002}, and specify a {\sc manual} astrometric output.
%}}}

\subsection{Aperture Placement}\label{sec: aperture creation}%{{{
When provided with the parameters required to define an elliptical aperture (as described above),
how one goes about placing that aperture on a finite grid of pixels can be non-trivial.
To allow for varying levels of complexity, \lambdar\ implements three different methods of
placing elliptical apertures: binary, quaternary, and recursive descent aperture placement.

\begin{figure}
\centering
\includegraphics[scale=0.23]{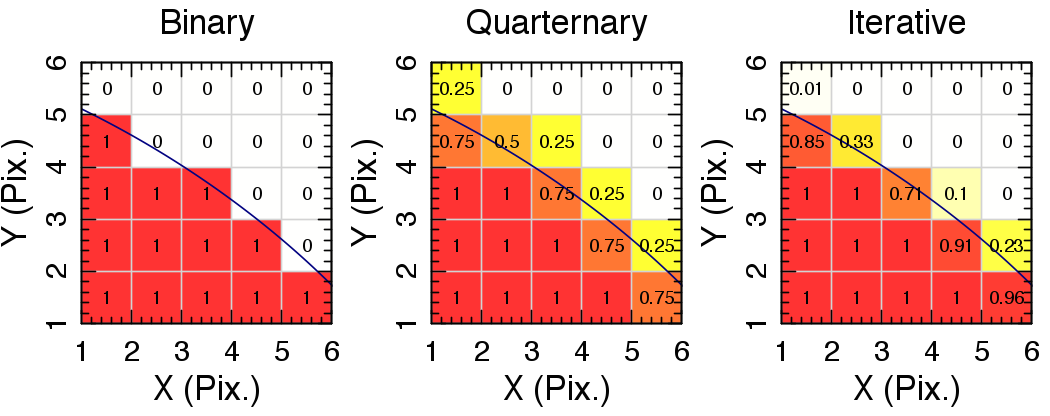}
\caption{A demonstration of the three types of aperture placement that can be employed in
\lambdar. The left-hand panel shows how aperture pixels are assigned using the binary aperture placement method;
the centre panel shows pixel assignments using the quaternary aperture placement method; and
the right-hand panel shows pixel assignments using the recursive descent method, implemented (by default) in \lambdar.}\label{fig: Aperture toymod}
\end{figure}

Given a 0-filled matrix/grid of pixels, binary aperture placement involves the allocation of 1s to all
matrix elements (pixels) whose centres lie within the boundary of the elliptical aperture.
For quaternary placement, pixels
are valued as either $\{0,\frac{1}{4},\frac{1}{2},\frac{3}{4},1\}$, depending on how many corners of the pixel lie
within the aperture boundary; $\{0,1,2,3,4\}$ respectively. Finally, the quaternary method can be implemented recursively, such that
pixels that are neither entirely
inside nor outside the aperture are sub-divided into smaller pixels, and
are re-evaluated. The resultant subpixels are then summed together using their value multiplied by
how many subdivisions down the tree they lie; \ie
\begin{equation}
\int\limits_0^r\int\limits_0^{2\pi}A(r,\theta)drd\theta\approx\sum_{i}\sum_{j}A(i,j)\times\frac{1}{(n\times d)},
\end{equation}
where $n$ is the number of orders in the recursive descent, used in calculating the coverage of the $(i,j)^{th}$ pixel, and $d$ is the degree of
sub-division of the pixels, per step. These three methods of aperture placement are shown in Figure \ref{fig: Aperture toymod}.

Binary aperture placement is a very efficient and effective method of defining apertures provided
that the size of the aperture, compared to the resolution of the grid, is large. As this is often not the case, using quaternary or iterative
placement is recommended. In practice, however, systematic effects induced by the choice of aperture placement are small, and can be mitigated
entirely by implementing aperture corrections (discussed at length in Section \ref{sec: minimum aperture correction}). \lambdar\ allows the user to
choose which placement method is best suited to their imaging. For GAMA imaging, we use quaternary aperture generation, with recursive descent
implemented in all but the highest resolution bands (see Table \ref{tab: lambdar params}).
%}}}

\subsection{PSF Convolution}\label{sec: psf conv} %{{{
After aperture placement, the program performs a convolution of the aperture with the PSF of the
image being analysed. Convolution of apertures and point sources occurs after both the aperture and PSF have been placed on
the same pixel grid as the image being analysed. Conversely, in real observations the convolution of an object's emission with the PSF happens prior
to pixelisation. This introduces a fundamental difference in how we treat objects approaching the point source
limit, and how they behave under observation. As such, we identify the impact of this treatment, and how it affects
the program's flux measurements.

The problem with performing pixelisation before convolution is that it is possible to lose positional information during pixelisation.
As soon as an aperture has any axis that fails to cover multiple pixels, its effective centre will artificially shift to the pixel
centre, and information will be lost. This is particularly problematic in images where pixels are large (compared to the aperture definitions).
As such, we define the set of sources that can be adversely affected by performing the pixelisation before convolution as those
with aperture minor-axis smaller than half the image pixel diagonal:
\begin{equation}\label{eq: apbound}
r_{\rm m} \le \Delta p \frac{\sqrt{2}}{2}.
\end{equation}

Below this limit, aperture positional information may be lost under pixelisation. To account for this loss of information, we do
not actively convolve apertures below this limit with the PSF.
Instead we simply duplicate the PSF and interpolate it onto the same sub-pixel centroid as the source in question.

Above this limit, the aperture is Nyquist sampled under pixelisation, and
subsequently positional information cannot be lost. As such, for these sources we are able to create the
normalised PSF convolved aperture model, $M_i \left(x,y\right)$, from the PSF function,
$f_{\rm PSF} \left(x,y\right)$, and the prior aperture function, $f_{\rm ap,i} \left(x,y\right)$, as;
\begin{equation}
M_i = {\rm Re}\left[\mathcal{F}^{-1}\left(S\right)
/n_{S}\right]
\end{equation}
where
\begin{equation}
S = {\rm Mod}\left[\mathcal{F}\left(f_{\rm PSF}
\right)\right]\times\mathcal{F}\left(f_{\rm ap,i}\right),
\end{equation}
$\mathcal{F}\left(f\right)$ is the Fourier transform of $f$,
$\mathcal{F}^{-1}\left(f\right)$ is the Inverse Fourier transform of $f$,
Mod$\left[f\right]$ is the complex modulus of $f$, Re$\left[f\right]$
is the real-part extraction of $f$, and $n_S$ is the number of pixels in the
image $S$.

The complex modulus in this equation serves the purpose of removing the spatial information of the
PSF after convolution, thus ensuring all positional information of the convolved aperture originates from the aperture itself,
and is not impacted by whether the supplied PSF is centred on a pixel centre, pixel corner, or anywhere in-between. This application of the
complex modulus can adversely affect the structure of the PSF, particularly in cases where the PSF contains discrete steps in flux or multiple frequency
components with different spatial centres. However as this is not typically the case with observational PSFs, we opt to perform the complex modulus (and
therefore correct for possible PSF centroid issues) while acknowledging the limitations of this implementation. Furthermore, we test all the PSFs that are
empirically determined in GAMA for adverse effects caused by the above. We find that there is typically a small residual (of a few percent or less in the
brightest pixels) between the pre- and post-convolution PSF, but that this residual is dominated by the centroid shift that the modulus is designed
to introduce.
%}}}

\subsection{Object Deblending}\label{sec: deblending}%{{{
\begin{figure}
\centering
\includegraphics[scale=0.45]{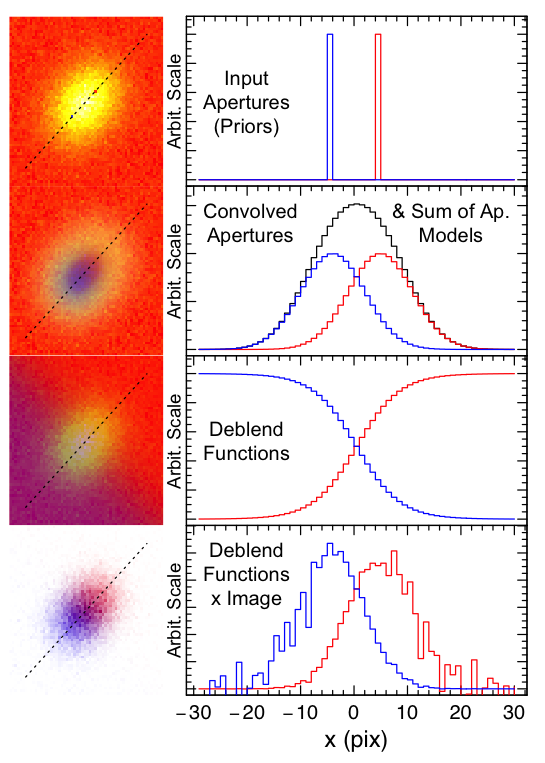}
\caption{Demonstration of how deblending of apertures is performed. Two point sources (i.e. objects
perfectly modelled by the PSF) were simulated using an example PSF and noise profile (image top left). Model parameters
for two point sources are provided to the program, at the known locations of the two objects (thereby simulating
use of a known optical prior; top right). Using these priors, and the known PSF, models for the two sources are generated
(second row). The deblend function for each object is determined by the ratio between each model and the sum of all models (third row).
The image is then deblended for each object, through multiplication by the deblend function (fourth row). This final image is then used for flux
measurement, using the user-desired measurement method (see Section \ref{sec: measurement type}). In each row, the right column shows the
slice through the left-hand image along the dotted black line.
 }\label{fig: deblend1}
\end{figure}

After convolution of the apertures with the PSF, the program performs a complex deblending of sources.
\lambdar\ implements a method of deblending whereby flux in any given pixel is fractionally split between all sources with
aperture models within that pixel. In order to accurately determine how much flux belongs to a given object, in any pixel,
we make a few simple assumptions. Firstly, the PSF-convolved aperture models, $M_i\left(\rm x,y\right)$, are assumed to be a tracer
of the emission profile of each source (for the purposes of deblending only).
Secondly, we can define the total modelled flux of any given pixel, $T\left(x,y\right)$, as the sum of all
$n$ object models, evaluated at that pixel:
\begin{equation}
T\left({\rm x,y}\right) = \displaystyle\sum\limits_i M_i\left({\rm x,y}\right).
\end{equation}
Using this total modelled flux, we can define the fractional contribution
of the $i^{\rm th}$ model, at pixel $\rm\left(x,y\right)$, as:
\begin{equation}
W_i\left({\rm x,y}\right) = \frac{M_i\left({\rm x,y}\right)}{T\left({\rm x,y}\right)}.
\end{equation}
We call $W\left({\rm x,y}\right)$ the deblending weight function.
Combining these two formulae, we define the $i^{\rm th}$ `deblended' model as:
\begin{equation}
D_i\left({\rm x,y}\right) = M_i\left({\rm x,y}\right)W_i\left({\rm x,y}\right).
\end{equation}
Using this model, we are able to calculate the flux of individual objects in the blended
regime:
\begin{equation}\label{eqn: deblended flux}
F^D_i = \displaystyle\sum\limits_{x,y} \Big( D_i\left(x,y\right) \times I\left(x,y\right)\Big),
\end{equation}
where $I\left(x,y\right)$ is the data image.
Note that this prescription is identical to using the deblend weight function to create a `deblended image':
\begin{equation}
I^D_i\left({\rm x,y}\right) = W_i\left({\rm x,y}\right)I\left({\rm x,y}\right),
\end{equation}
and then simply applying the original model $M_i\left({\rm x,y}\right)$ to this image. In terms of description, the
former is more useful for calculating uncertainties and corrections on aperture fluxes, and is used in
Section \ref{sec: minimum aperture correction}. Conversely, the latter makes more sense intuitively, and as a result
we often choose to show it in visualisations. For example, Figure \ref{fig: deblend1} demonstrates the deblending process using
this latter description of the deblending procedure. In the figure, we simulate two point sources (with equal flux) in a low resolution image that
are separated by less than the PSF FWHM (and which are therefore unresolved). Using the high resolution priors ($f_{\rm ap,i}$, first panel),
which we then convolve with the low resolution PSF to create the aperture models ($M_i$, second panel), we can then calculate
the deblend weights ($W_i$, third panel) for each object. This is done by dividing the
aperture model ($M_i$,the red and blue lines in the second panel, respectively) by the sum of all models ($T$, the black line in the second panel).
Finally, we multiply the simulated image, $I$, by the deblend weights to generate the deblended image ($I^D_i$, bottom panel).

\subsubsection{Flux Weighting \& Iterative Deblending} %{{{
The process of deblending objects can be improved when an additional object weighting mechanism is
applied to objects, such as weighting based on relative surface brightness. The program allows this additional
weighting in three ways. Firstly, it allows initially unweighted models to be refined (using information from the image being analysed) through iteration,
where the previous iteration's measured mean surface brightness per pixel is used as a weight for the subsequent iteration. Secondly, it
allows users to use the central-pixel-flux of each object as a weight. Finally,
it allows users to specify their own input weights, allowing, for example, information from other bands to influence flux measurements.
Each of these methods has benefits and detriments, and
it is often useful for the user to explore multiple options when attempting to extract the best photometry from
their data. The program allows users to combine the latter two weighting options with the iterative improvement
mechanism, and outputs fluxes measured at each stage of the iteration. An important caveat to the iterative flux determination
procedure is the behaviour when an object is measured to have a flux less than or equal to zero. As these objects are deemed to have
no contribution to the flux in the image, their weights are set to zero and the object is effectively discarded. It is not possible for
the objects to return to the measurement space after being assigned a weight of 0, as no further measurements take place.
These objects are assigned the flux as measured at the last iteration (prior to being discarded), and a photometry warning in the
catalogue accompanies the measurement. Examples of the iterative deblending process are provided in Appendix \ref{sec: Iterative Deblending},
for a range of blended-object flux ratios. As this is a simple example, we also note that a real, complex deblend is shown (in 2D) in panel `d' of
Figure \ref{fig: CoG} (this figure is discussed at length in Section \ref{sec: CoGs}).

\begin{figure*}
\includegraphics[scale=0.40]{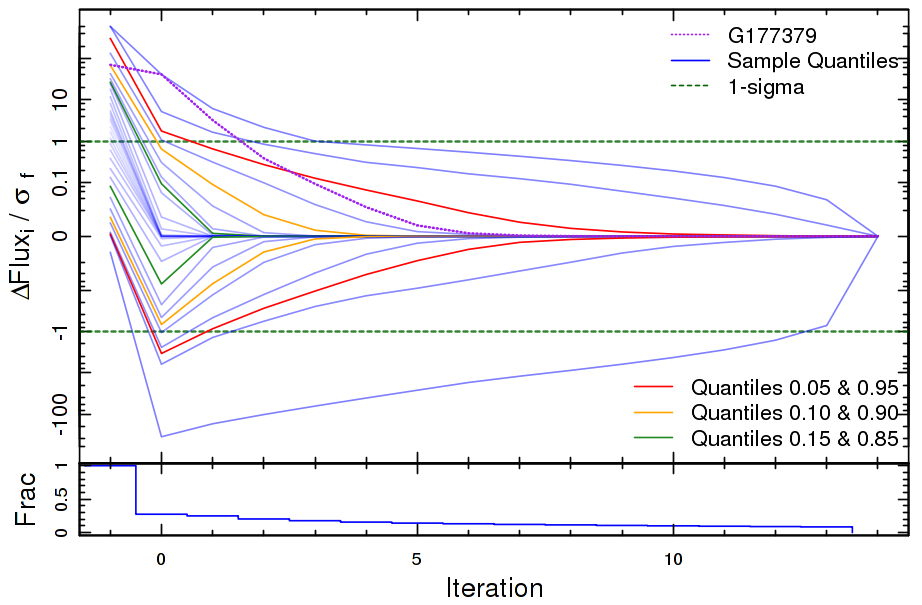}
\caption{Demonstration of the convergence of flux iteration in the program for a sample of 953 blended galaxies located in 1 square degree
centred on G177379 (shown in purple). At each iteration, we calculate a residual between every object's flux and the final measured flux. We then normalise
these residuals by the final flux uncertainty, $\sigma_f$. We draw lines showing the distribution of 30 evenly distributed quantiles
(from 99\% to 1\%), as a function of iteration. {\bbf The outermost $10$, $20$, and $30$ percent are highlighted with red, orange, and green lines
respectively.} Here the $-1^{\rm th}$ iteration is the flux measured in a blended aperture, the $0^{\rm th}$ iteration is that
measured in an aperture whose deblend is based solely on the object apertures and their on-sky positions (i.e. it does not incorporate flux
information), and subsequent iterations are deblended according to iterative average surface brightness. The histogram beneath the main figure
shows the fraction of sources that have yet to converge at each iteration, as determined by whether their flux at the $i^{\rm th}$ iteration is
not equal to the final estimate. We see that the majority (\ie $\geq 95\%$) of fluxes have converged to within $1\sigma$ of their final estimate
within 5 iterations.}\label{fig: Iteration}
\end{figure*}

As described in the Section above, the program optionally uses an iterative deblending of object apertures, based on the
measured object average surface brightnesses. Figure \ref{fig: Iteration} shows the impact of this procedure for 953 galaxies
in the GAMA {\sc sdss} {\em r}-band imaging. These galaxies are all located within 1 square degree, centred on our example galaxy G177379.
In this figure, we demonstrate the impact of iterated deblending on the convergence (as a population) of object fluxes as a function of
iteration. We calculate the residual between every object's flux at the $i^{\rm th}$ iteration and its final flux (measured at the $15^{\rm th}$
iteration), normalised by the object's final uncertainty. We then calculate 60 evenly distributed quantiles (from 99\% to 1\%) for the
population of all objects, and draw contours along these quantiles. From this figure, we can see that by iteration ~5 all but the most extreme
few percent of objects are converged to within the uncertainty of their final flux.

%}}}

\subsubsection{Quantifying Deblend Solutions using Curve of Growth analysis} \label{sec: CoGs} %{{{
In order to demonstrate the importance and effectiveness of our deblend method, the program has the ability to output
a CoG for each catalogued object. A CoG is a description of enclosed flux as a function of radius. In the
program, CoGs are output as a diagnostic that can be used to investigate deblend solutions or galaxies that
appear to have anomalous photometry. Currently CoGs are not used to assist with flux determination, however this addition
is likely to occur in the near future.

\begin{figure*}
\includegraphics[scale=0.40]{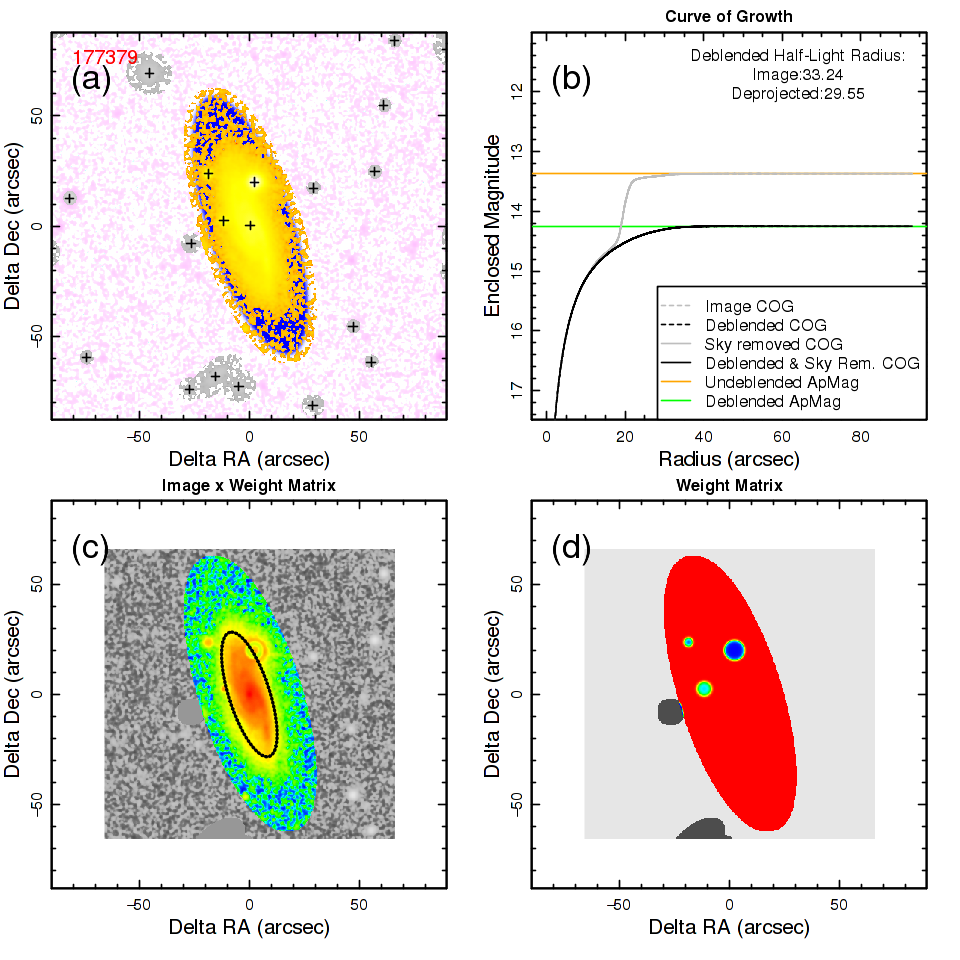}
\caption{A demonstration of the impact of object deblending on the CoG flux of GAMA object G177379. Panel (a) shows the
input image (greyscale), with object aperture beneath in blue. Positive flux within the aperture is shown in yellow. Pixels deemed to be
part of the `sky' are shown in pink. Panel (b) shows the object  Curve of Growth (CoG). The grey lines show the object CoG without deblending,
and the black lines show the CoG with deblending (here the dotted lines are not visible as they are immediately behind the solid lines).
Horizontal orange and green lines mark the measured aperture magnitude for the object before and
after deblending respectively. The text in the panel describes the circular and deprojected half-light radii, in arcseconds, with the deprojection being
based on the input aperture (prior to convolution). Panel (c) shows the image stamp after deblending. Coloured pixels mark those within the object aperture, and greyscale
pixels mark those beyond the aperture. The black dotted line marks the measured deblended and deprojected half-light radius, as described in panel (b). Panel (d)
shows the deblend weights for this object. Again, coloured pixels mark those within the aperture, and
greyscale pixels mark those beyond. Essentially, the grey and black CoGs in panel (b) are the radial integrals of panels (a) and (c) respectively.
This 4-panelled figure is a data-product optionally output by the program.}\label{fig: CoG}
\end{figure*}

An example of CoG output is shown in Figure \ref{fig: CoG}, where we show the GAMA object G177379, which is contaminated
by a nearby bright star. In the figure, we show the image for our sample object (panel `a') with the location of sources within the image, and
colouring to show the object's model aperture and which pixels were used in measuring the sky-estimate for this source. In panel `b' we show the
CoG for this source, both with and without deblending of nearby sources. In panel `c' we show the deblended image $I^D_i$ for this source, and
include an estimate of the object's deprojected, deblended, half-light radius. Finally, panel `d' shows the 2D deblend weights for this source,
and is coloured by what is within the object's aperture. The impact of contamination on the CoG prior to deblending is evident, with large
steps in the flux integral as a function of radius clearly apparent.
After deblending, however, the CoG is much more well behaved and plateaus to a final flux without large
steps.

%}}}

\subsubsection{Quantifying Deblend Uncertainty} %{{{
Finally, the program incorporates an uncertainty term to quantify the confidence in a deblend solution, $\Delta W_i$.
This deblend uncertainty term is of the form:
\begin{equation}\label{eqn: deblend uncertainty}
\Delta W_i = \left[1-\frac{\displaystyle\sum\limits_{x,y} D_i\left({\rm x,y}\right)}{\displaystyle\sum\limits_{x,y} M_i\left({\rm x,y}\right)}\right]\times{\mathcal D},
\end{equation}
where $\mathcal D$ is the `deblend uncertainty factor'. We chose to use
\begin{equation}\label{eqn: deblend factor}
{\mathcal D} = \frac{1}{\sqrt{12}}\times |F^M_i|
\end{equation}
where $F^M_i$ is the flux measured within the $i^{\rm th}$ source aperture prior to deblending, defined as
$\displaystyle\sum\limits_{x,y} \Big( M_i\left(x,y\right) \times I\left(x,y\right)\Big)$. This is the $-1^{\rm th}$ iteration shown in Figure \ref{fig: Iteration}.
Here $I\left(x,y\right)$ is the data image. The definition of the deblend uncertainty is such that an object that is determined to contribute 0 flux to the image (and
which therefore has a $\displaystyle\sum\limits_{x,y} D_i\left({\rm x,y}\right)=0$), will be given an uncertainty of $1/\sqrt{12}$ times the blended flux in the aperture.
The factor $1/\sqrt{12}$ is the standard deviation of the uniform distribution over $U \in [0,1]$, which is used to incorporate the (conservative) assumption
that the distribution of deblend fractions is uniform over $[0,1]$. This will not be the case (in fact, the distribution likely follows a beta distribution; see
\citealt{Cameron2011}),
but we nonetheless choose to use this uniform approximation to be conservative. The result is that, for highly deblended sources, our deblend
uncertainty is likely slightly over-estimated.
%}}}

%}}}

\subsection{Sky Estimate}\label{sec: Sky Estimate} %{{{
An important step in any aperture photometry measurement is a reliable determination of the
local sky-background around each aperture. As such, \lambdar\ has an internal routine
for determining the local sky-background around every aperture provided to the program, and
returns relevant information such as the mean and median sky values, the associated median absolute deviation (MAD) root-mean-squares (RMSs),
and the Pearson chi-square normality test p-value. In this way, the function provides an indication of
the local sky value, its uncertainty, and a quantification of the sky's Gaussianity.

\begin{figure*}
\centering
\includegraphics[scale=0.22]{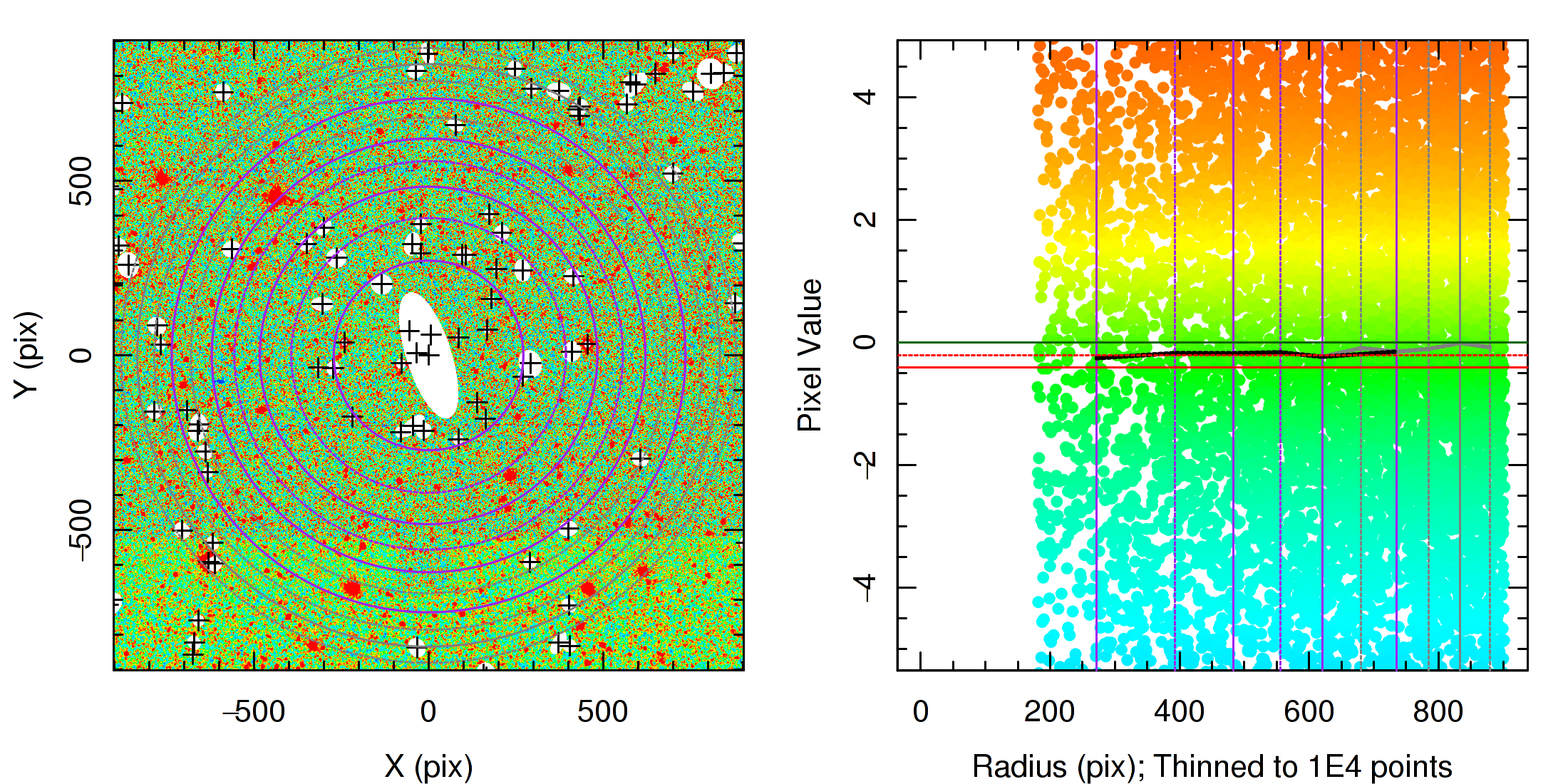}
\caption{Demonstration of the Sky Estimate measured around GAMA object G177379 in the SDSS r-band. The left panel shows the image, masked pixels
are shaded in black. In the right panel, pixels values are shown as a function of radius, with the range on the y-axis set to
be twice the measured Sky RMS (which is $4.76 ADU$ for this object and image).
The black lines show the binned running median (solid) and
the uncertainty on the median (dashed). {\bbf Here the
uncertainty is small, so the dashed line is hard to distinguish from the solid.
Horizontal red lines indicate the sky estimate using mean (solid) and median (dashed) statistics. The horizontal dark green
line indicates the 0-line, for reference.
Both panels are coloured by pixel values, on the same scale, and bin centres used in the estimate are shown as alternating
solid and dashed purple/grey lines. Purple bins correspond to those whose means are within 1-sigma of the
final sky estimate, and grey are those outside 1-sigma (and so were discarded when calculating the final estimate; See
Section \ref{sec: Sky Estimate}). }
This 2-panelled figure is a data-product optionally output by the program. }\label{fig: skyest}
\end{figure*}

In order to ensure that the function returns an accurate measure of the sky and is not contaminated by
object flux, the program performs both a masking of all catalogued objects and (by default) an aggressive sigma-clipping of
sky-pixels. After masking and sigma-clipping, the program bins pixels into 10 radial bins (such that each bin contains
an equal number of unmasked pixels). The radii are arranged with minimum bin edge at a radius equal to the object semimajor axis length,
and the largest bin edge at $10\times$ this radius.
In addition, the bins have hard minima and maxima, such that the
innermost bin-edge is at least $3$ PSF FWHM from the
object centre and the outermost bin edge is at least $10$ PSF FWHM from the object centre.
{\bbf If an aperture occupies a large fraction of the image, such that the largest bin radius
would extend beyond the image edge, the function will generate the 10 equal-N bins using the pixels between the lower bin radius and the image edge. }
After binning using both a mean and median, the program then calculates the weighted mean
of each to determine the sky estimate. When performing the weighted mean, the program uses weighting in both the confidence on the
bin's individual mean/median, and in distance from the aperture centre:
\begin{equation}\label{eqn: sky est}
w_i = \left[r_{\rm i,cen}\times\sigma_i\right]^{-1}
\end{equation}
with $r_{\rm i,cen}$ is the central radius of the $i^{\rm th}$ bin, and $\sigma_i$ is the uncertainty on the bin's mean/median.
As such, the estimate is weighted to be more representative of bins with better estimates and at lower radii.
The uncertainty on the estimate is the standard deviation of the binned values, without weighting (and thus, is the largest possible uncertainty).
If there exist bins whose values are beyond the measured $1\sigma$ limit of the sky, these bins are discarded and the sky-estimate
recalculated. Finally, the program determines the number of bins that
are within $1\sigma$ of the final sky estimate, and returns this diagnostic for reference of the user. Figure
\ref{fig: skyest} shows an example of the sky estimate and diagnostic images output by the program. The figure shows
GAMA object G177379 imaged in the SDSS r-band, the binned values for this galaxy, and the estimate for this object. Note the masked pixels
in the image and the grey bins that have mean/median beyond the $1\sigma$ of the final estimate (and were therefore discarded). {\bbf In this example, we can see that bins which have been excluded from the sky estimate are those
which have been contaminated by pixels with different noise properties, from an adjacent stripe. }
In Section \ref{sec: testing skyest} we demonstrate that the sky estimate routine is robust to strong gradients in the sky,
and variations in the uniformity of the sky RMS.

%}}}

\subsection{Randoms/Blanks Estimation}\label{sec: randoms/blanks} %{{{
A measurement of the local sky, as described in Section \ref{sec: Sky Estimate}, fails to account for correlations in the sky (which can systematically
impact the actual sky RMS as a function of aperture geometry). As such, the program has two mechanisms for accounting for correlations in
sky pixels around objects of interest: users can simply specify a multiplicative sky-correlation factor in the parameter file, or the program
can perform a per-object randoms/blanks estimation. The multiplicative factor is used to increase the measured sky-error from the previous Section
to reflect the impact of correlations, whereas the randoms/blanks estimations uses each object's aperture to empirically measure the correlated sky noise
around the object.

The randoms/blanks estimation is calculated for every aperture by taking the masked image stamp $I^m_i(x,y)$ and transposing it in x and y as determined by
quasi-random draws from a uniform distribution with boundaries $[0,N_{\rm x,y}]$, where $N_{\rm x,y}$ is the width of the image stamp in pixels in x and y.
Using this transposed image stamp $I^m_i(x^*,y^*)$, the program measures the post-masking aperture-weighted flux at that point;
\begin{equation}\label{eqn: randoms calculation}
f_{i}=\displaystyle\sum\limits_{x,y} I^m_i(x^*,y^*)\times M^m_i(x,y)
\end{equation}
where $M^m_i(x,y)$ denotes the object aperture after removal of masked pixels. To calculate the final mean
the program performs this measurement $N_{\rm rand}$ times and then calculates the weighted mean and unbiased weighted standard deviation,
using the following equations respectively:
\begin{eqnarray}\label{eqn: randoms calculation2}
F_b&=\frac{\displaystyle\sum\limits_i f_i \times T^m_i}{ \displaystyle\sum\limits_i T^m_i }, \\
\sigma_b&=\sqrt{\frac{\left(\displaystyle\sum\limits_i T^m_i\times(f_i - F_b)^2\right)\times\left(\displaystyle\sum\limits_i T^m_i\right)}{
\left(\displaystyle\sum\limits_i T^m_i\right)^2 - \displaystyle\sum\limits_i (T^m_i)^2}},
\end{eqnarray}
where $T^m_i = \displaystyle\sum\limits_{\rm x,y} M^m_i(x,y)$. In addition to these, the program also returns {\bbf an independently calculated }
weighted median absolute deviation (MAD), $\sigma_{b,mad}$.
The reason for the inclusion of a MAD based $\sigma_{b,mad}$ is that the standard deviation determined can
sometimes be unreasonably over-estimated. {\bbf Standard deviations calculated via the MAD provide a more conservative measurement that is less impacted
by outliers. In the case of gaussian noise, the MAD is related to the standard deviation: }
${\rm SD}={\rm MAD RMS}/\Phi^{-1}(\frac{3}{4})\approx1.4826\times {\rm MAD RMS}$,
where $\Phi^{-1} (P)$ is the inverse of the cumulative distribution function of the normal function.
{\bbf This conversion is performed internally. By providing both the weighted MAD derived SD and the unbiased weighted standard deviation, the
program provides a check for the validity of the SDs. }
In the case of blanks, it also returns the number of blank apertures for which a post-masking
aperture-weighted flux was successfully measured (because of heavy masking in crowded areas, entire apertures can be masked and hence provide no information).
The randoms estimation and blanks estimation differ only
in that the blanks estimation masks all catalogued sources in the image stamp before calculation, while the randoms function masks out only the
object for which the correction is being calculated. This is done because the program uses image cutouts which are, by definition, centred on
a source. As a result, randoms can be biased from being true reflections of random apertures because of this systematic image cropping.

As a result of the masking of all catalogued objects, the blanks estimation provides fundamentally different information to the randoms estimation.
The blanks estimation details the flux contained within this aperture when placed over a
part of the image that contains no sources brighter than the catalogue limit (and is therefore believed to be sky),
whereas the latter details the flux contained in this aperture when randomly
placed on the image, agnostic of all sources (catalogued or otherwise). The distinction between randoms and blanks is a useful one, as a comparison of randoms and blanks
can indicate the influence of source masking on your correlated-noise estimate. If the randoms and blanks return equivalent standard deviations, then this can
indicate that the input catalogue is too shallow for reliable sky-estimation, or that you are masking the wrong pixels (\ie your catalogue has been improperly defined for
the image being analysed).

\begin{figure}
\centering
\includegraphics[scale=0.40]{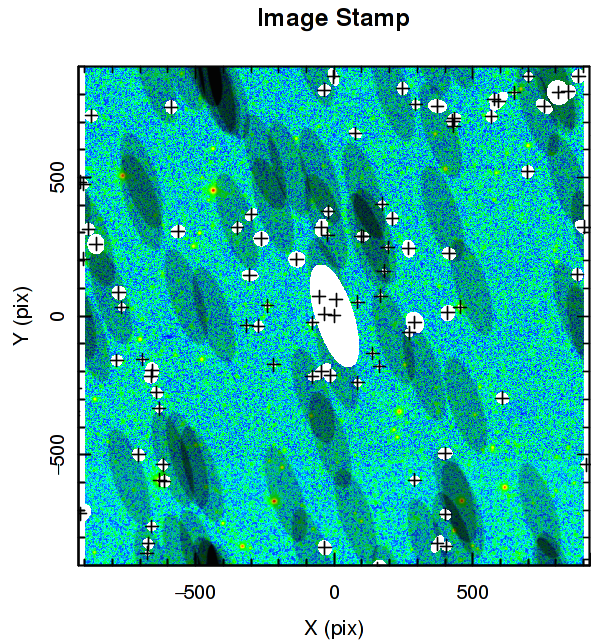}
\caption{Demonstration of how blanks are measured within the program, using GAMA object G177379 in the SDSS r-band as an example.
Blank apertures are shaded in black (darker shades highlight pixels that went into multiple randoms). Masked pixels are white.
The total flux in each blank aperture is measured, corrected for masking, and is used to calculate the weighted mean and standard deviation
blank flux for this source, which is returned in the final catalogue. This figure is a data-product optionally output by the program,
and is useful for diagnostic checks. }\label{fig: randomsCorr}
\end{figure}

Additionally, measurement of the aperture flux values means that the randoms/blanks routine can
also provide a rudimentary check for the measured sky estimate. An example of the blanks estimation is shown in Figure \ref{fig: randomsCorr}, performed on a
convolved SDSS r-band image.
Comparing this to the sky estimate for this same object (and band) shown in Figure \ref{fig: skyest}, the annular sky estimate returns a mean sky value
of $-0.20\pm0.07$ ADU per pixel, with a pixel-to-pixel RMS of $4.76$ ADU. Conversely, the blanks estimation returns an effective mean pixel
value of $0.73$, with an effective pixel-to-pixel MAD RMS of $72.39$ ADU (using 50 blanks). This suggests
that, at this aperture scale, pixel-to-pixel correlations reduce the number of effective samples of the noise measured within the aperture
by a factor of $15.21$. We expect correlations in the SDSS background to be present because of our process of Gaussianisation, and we can estimate that there
should be correlations on the same order as the area of the Gaussianisation kernel. A reduction in effective samples on the order of $15\times$ requires a
Gaussian convolution kernel with FWHM $\le 1.5^{\prime\prime}$, which is the domain of the convolution kernel which was used. As such, we believe this to be a
successful verification of the procedure.

%}}}

\subsection{Flux Calculation}\label{sec: measurement type} %{{{
Once the deblended model has been determined, the next step is to convert the model aperture shown in
Figure \ref{fig: deblend1} to the form desired for calculation of flux. The program is able to perform
two types of flux measurement: simple aperture photometry and profile-weighted photometry.

For performing simple aperture photometry, the program uses the model aperture generated after convolution, $M_i\left(x,y\right)$,
and converts it back to standard boxcar form. To achieve this, a user defined
aperture fraction, $f \in \left(0,1\right]$, is used. The aperture model is integrated outward until the point where $f$ of the aperture
is contained, and at this point a binary cut is imposed; all pixels with value greater than or equal to the pixel value at the cut
point are given value $1$,
and all pixels with values lower are given value $0$. This converts the model aperture from being a
constantly varying aperture with domain $M_i\left(x,y\right) \in \left[0,1\right]$, to being a boxcar-like
aperture with domain $M^*_i\left(x,y\right)\in \{0,1\}$. This binary aperture is then multiplied by the deblending weighting
function, $W_i\left(x,y\right)$, giving the final deblended aperture $D_i\left(x,y\right) \in \left[0,1\right]$.
The image is then simply multiplied by final aperture and summed to return the deblended
object flux, $F^D_{i}$:
\begin{equation}
F^D_i = \displaystyle\sum\limits_{x,y} \Big( D_i\left(x,y\right) \times I\left(x,y\right)\Big).
\end{equation}
In the case of isolated objects, \ie where $W_i\left(x,y\right)=1\,\, \forall\,\, \left(x,y\right)$,
$D_i\left(x,y\right)=M^*_i\left(x,y\right)$ and $F^D_i$ is simply the sum of the aperture multiplied by the image.

\begin{figure}
\includegraphics[scale=0.45]{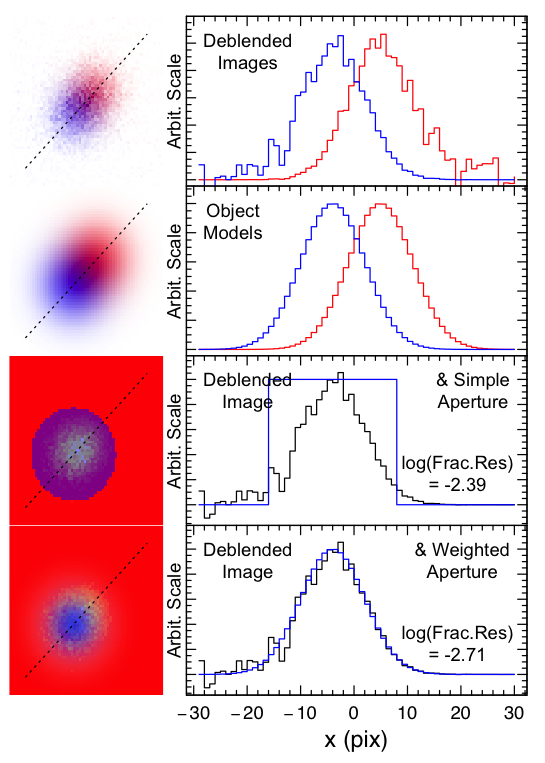}
\caption{Demonstration of the two different measurement methods, being applied to the simulated objects in Figure \ref{fig: deblend1}. The top panel
is the same as the final panel in Figure \ref{fig: deblend1}. The second panel shows models for the two sources. The third panel shows the `simple'
measurement aperture for the blue source (after passing the model through the binary filter detailed in Section \ref{sec: measurement type}),
overlaid on the ``deblended image'' ($I\left(x,y\right)\times W\left(x,y\right)$) in black.
The bottom panel shows the `weighted' measurement aperture, which is identical to the model aperture $M\left(x,y\right)$. Again, this aperture is overlaid on the
``deblended image''. In the bottom two panels, the text inset shows the fractional residual between the input flux and the flux measured by
the aperture after accounting for aperture normalisation (Section \ref{sec: minimum aperture correction}).
As is discussed in Section \ref{sec: deblending}, in practice the program constructs individual ``deblended apertures'' rather than the
``deblended images'', shown here, as they are equivalent. We demonstrate deblended images here simply for clarity,
to better explain the process. }\label{fig: measurement methods}
\end{figure}

For weighted photometry, the program skips the step of converting the aperture back to its standard boxcar form;
\ie $M^*_i\left(x,y\right)=M_i\left(x,y\right)$. Instead, the program
uses the aperture model as a weighting function to extract a measurement. This allows for more reliable detections in cases where flux
may otherwise be swamped by noise (particularly in the point-source limit). A demonstration of the different measurement
methods can be seen in Figure \ref{fig: measurement methods}. The use of the weighting function is then corrected for using an aperture normalisation
detailed in Section \ref{sec: minimum aperture correction}).

After the flux measurement, the program subtracts the sky estimate measured in Section \ref{sec: Sky Estimate}. This is simply the deblended flux $F^D_i$ minus
the sky-flux within the aperture: $F^s_i = f_{\rm s}\times \displaystyle\sum\limits_{x,y} M^*_i\left(x,y\right)$. The uncertainty on the flux is discussed
in Section \ref{sec: errors}.

%}}}

\subsection{Aperture Normalisation} \label{sec: minimum aperture correction}%{{{
When performing aperture photometry, it is important to consider the impact of the choice of
aperture weighting and size on the final photometric measurement. In the zero-noise regime,
we want a measurement such that the choice of aperture weighting, and any aperture truncation, has no impact on the
final object flux. In order to achieve this, {\bbf the program normalises aperture fluxes }
to account for any use of weighting or truncation. This normalisation is akin to a traditional aperture correction for missed flux when
performing simple aperture photometry, and to a weighting normalisation when performing weighted aperture photometry.
In practice, calculating the required correction/normalisation can be done using a single method, regardless of measurement type.

The program calculates two different factors that can be used to normalise the measured fluxes.
To calculate the factors, the
program makes two limiting assumptions about the distribution of source flux. The
first factor, denoted the ``maximum correction'', assumes that the distribution of source flux follows exactly the shape of the
object model (\ie a PSF in the point source limit, and an aperture convolved PSF in the aperture limit). For the maximum correction,
the program then measures how much of this flux is missed when measured using the model aperture;
\begin{equation}
C_{\rm max}= \frac{\sum\limits_{x,y} M_i(x,y)}{\sum\limits_{x,y} \Big( M^*_i(x,y)\times M_i(x,y)\Big)}.
\end{equation}
Here $M_i\left(x,y\right)$ is the PSF-convolved aperture model, and $M^*_i\left(x,y\right)$ is the aperture after possibly
going through process of box-car conversion detailed in Section \ref{sec: measurement type}.

In addition to this maximum correction, the program returns a second factor, the ``minimum correction''. This factor
instead assumes that the distribution of object flux follows the smallest possible distribution, a PSF:
\begin{equation}
C_{\rm min}= \frac{\sum\limits_{x,y} P_i(x,y)}{\sum\limits_{x,y} \Big( M^*_i(x,y)\times P_i(x,y)\Big)}
\end{equation}
where $P_i(x,y)$ is the PSF function, re-interpolated onto the same pixel grid and centroid as the aperture $M^*_i(x,y)$.
This correction factor can be expressed as follows: for every aperture (resolved or otherwise), the minimum correction $C_{\rm min}$
recovers all flux missed because of aperture weighting or truncation in the limit where the true source is a point source.
In this way, the minimum correction can only help the flux determination, by doing the most conservative correction possible. { \bbf This
correction is incorporated automatically into the fluxes output by the program, and both the minimum and maximum corrections are
included in the output catalogue. }

We note that, when performing PSF-weighted photometry of point sources, because the Aperture function $M_i(x,y)$ is equal to the
PSF function $P_i(x,y)$, both the minimum and maximum corrections reduce to:
\begin{equation}
C_{\rm min} = \frac{\sum\limits_{x,y} M_i(x,y)}{\sum\limits_{x,y} \left(M_i(x,y)\right)^2}.
\end{equation}

These factors are calculated whenever an empirical PSF or analytic Gaussian FWHM is supplied.
It does not require PSF convolution of the aperture to have taken place, which is useful when investigating apertures of standard sizes,
such as the $8.25^{\prime\prime}$ radius `standard apertures' used in \wise\ \citep{Cluver2014}.
Note that these factors are defined such that they are multiplicative; that is the final flux is defined as
\begin{equation}
F_{\rm final} = F_{\rm meas}\times C_{\rm min/max}.
\end{equation}

\begin{figure*}
\includegraphics[scale=0.50]{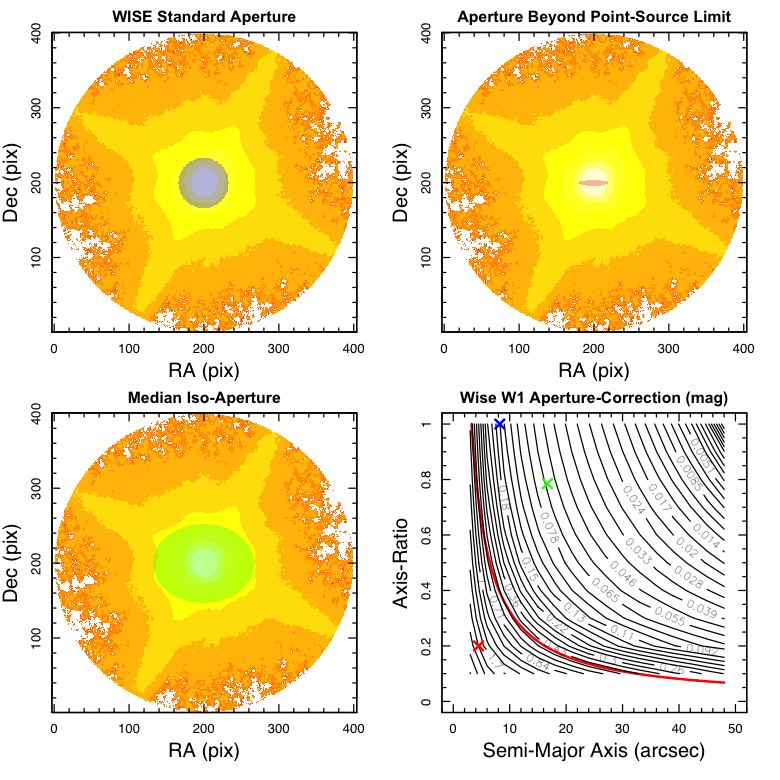}
\caption{Demonstration of the minimum aperture correction implemented by the program. Here we use the \wise\ W1 G12 PSF, and generate the
aperture correction for a range of aperture sizes and ellipticities. Sample apertures were generated at 8000 uniformly distributed points in
radius:axis-ratio:position-angle(PA) space. PA was found to have the least impact on variation in the aperture correction, and as such here we
show the correction in the radius:axis-ratio space. Three examples of the apertures generated are shown here: Panel (a) shows the \wise\ $8.25^{\prime\prime}$
radius `standard aperture'; Panel (b) shows an aperture that is unreasonably small, given the size of the image PSF;
and Panel (c) shows an aperture with the median semi-major axis length and ellipticity in {\sc gama}.
Panel (d) shows the aperture correction value as a function of semi-major axis and axis-ratio, as PA was found to have the smallest impact on the
aperture correction. Coloured crosses show value of the corrections for each of the three sample aperture (colours are matched). The solid red line shows the
limit where the aperture semi-minor axis is equal to half the PSF FWHM; this is an indicator of the minimum sensible aperture that someone might use for
measuring fluxes in \wise.}\label{fig: ApCorr}
\end{figure*}

To demonstrate the minimum correction, and its importance, we calculate the factor empirically for a range of
simple apertures using the \wise\ W1 G12 PSF,
which was derived from observations of Neptune throughout the \wise\ campaign observing the GAMA 12hr field.
Figure \ref{fig: ApCorr} shows how the factor (which is an aperture correction, because we have simple apertures)
varies for a range of aperture sizes and ellipticities. This figure shows that aperture corrections can be substantial ($> 0.1$ mag) when apertures are
small ($r_{\rm min} \le 12^{\prime\prime}$)  and/or highly elliptical ($b/a \le 0.2$).
We note, however, that generating an aperture so small
and/or highly elliptical is unlikely (except when intentionally using fixed-size apertures), because the PSF begins to dictate the aperture shape as
radius and axis-ratio approach 0.

%}}}

\subsection{Error Propagation}\label{sec: errors} %{{{
Measurements of the various types of uncertainty associated with each flux measurement are generated by the
program, such that they can be combined by the user (depending on what they feel is appropriate).
While we have detailed the various uncertainties incorporated into each of the various measurements in each Section, here
we provide an example of two cases (optical + resolved, and FIR + confused) and how we would derive the uncertainties for measurements in each
of these cases.

Internally, the program combines errors as are outlined in this Section. That is, the final ``deblended flux error'' output by the program will
contain, at most, the terms specified here. Other possible error terms are output in the catalogue, but are not combined internally. If the user
does not request measurement of one or more of the terms required for this calculation then that/those terms are neglected from the error calculation.

In the optical regime, we typically have high SNR, high resolution, and little contamination from blended sources. As a result, the principle
sources of uncertainty are typically from pixel-to-pixel noise variations, correlated noise variations, uncertainty in estimation of the sky, and
photonic shot noise. As a result, for any typical flux measurement in the optical regime, we would derive an associated uncertainty as follows:
\begin{multline}
   \Delta F_i = {\Bigg [}\displaystyle\sum\limits_{x,y} \left(D_i\times E_i\right)^2  + \left(\sigma_{b}\right)^2 \\
           + \left(\left[1-\frac{\displaystyle\sum\limits_{x,y} D_i}{\displaystyle\sum\limits_{x,y} M_i}\right]\times{\mathcal D}\right)^2
           + \left(\sigma_{\rm bg}\displaystyle\sum\limits_{x,y} M_i\right)^2 {\Bigg]}^{\frac{1}{2}}
\end{multline}\label{eqn: opticalfluxerr}
where $E_i$ is the sigma map associated with the image at the location of the aperture, $\sigma_{b}$ is the standard deviation of the
sky blanks derived from the randoms/blanks routine, $\sigma_{\rm bg}$ is the uncertainty of the sky estimate returned from the
sky estimate routine (per pixel), and ${\mathcal D}$ is the deblend uncertainty factor given in Equation \ref{eqn: deblend factor}.
In cases where blanks are not run, but the sky estimate has been calculated, $\sigma_{\rm rms}\times\sqrt{\displaystyle\sum\limits_{x,y} M_i}$
(i.e. the sky RMS per aperture derived from the sky estimation routine) is used as a proxy for $\sigma_b$.
The sigma map $E_i$ constrains the uncertainty caused by shot noise, and is defined using some variant of the following equation:
\begin{equation}
E_i\left(x,y\right)=\sqrt{\frac{\big| I_i\left(x,y\right)\big|}{ G\left(x,y\right)}},
\end{equation}
where $G\left(x,y\right)$ is the gain per pixel.
In cases where a single gain value is supplied, or found in the FITS header, the program uses this value for $G\,\forall \left(x,y\right)$.
If a weight-map is provided, and the program is also provided a single maximum gain value, or one is found in the FITS header, the program will assume that the
weight-map is inversely proportional to the pixel variance, and will use the gain value in tandem with the variance map to derive a sigma map with varying gain $G(x,y)$.
If the program is provided with a weight-map and not a maximum gain value, it will assume that the weight map is identically equal to the inverse pixel variance, and
will use this to generate the sigma map. If the program is not provided a weight map or a gain value, then the gain is assumed to be $1$ and purely Poissonian
uncertainties are derived. Finally, the user may bypass the generation of a sigma map entirely by providing their own as a separate FITS image.

In the FIR regime, where we have confusion of sources, complex and extensive blending of sources, smooth backgrounds, and correlated noise, the
major sources of uncertainty are typically pixel-to-pixel noise variations, correlated noise variations, boosting from confusion, and
uncertainty in estimation of the sky. However, calculation of the standard deviation of blanks measurements is effective for determination of
the RMS uncertainty, the pixel-pixel correlation, instrumental noise, and the contamination from confusing sources. As such, we are able to calculate the FIR uncertainties
from the blanks uncertainty, sky estimate uncertainty, and deblend uncertainty alone. These three components nonetheless account for all the relevant sources
of uncertainty in the FIR, identical to those in the optical (with the exception of the shot-noise, as this concept ceases being applicable at low photon energies).
This consistency of uncertainty determination is particularly useful when performing $\chi^2$ fits to the panchromatic data, as we do
not unfairly weight any point over another. This is one of the primary benefits of measuring photometry in this consistent manner.

If the user has an appropriate analytic value for the level of uncertainty in the image introduced by confusion, then this can be specified
directly in the parameter file ($C$, in units ADU/pixel), instead of having the program perform randoms/blanks.
This additional uncertainty term, $\mathcal{C}$, is then calculated using the
aperture model, ${M}_i$, and the specified confusion per pixel, $C$, such that:
\begin{equation}
\mathcal{C}=C^2\left[\displaystyle\sum\limits_{x,y} M_i\left(x,y\right)\right]^{\frac{1}{2}}.
\end{equation}
This value is then added in quadrature to the other error terms (in place of the blanks/randoms term).

In all cases, when the program performs any of the various normalisation `corrections' to fluxes (be they provided by the user or measured empirically in the form of the
minimum correction in Section \ref{sec: minimum aperture correction}, for example), the program ensures that fractional uncertainty is conserved.
%}}}
%}}}

\section{Testing using Simulated {\sc sdss} {\em r}-band imagery} \label{sec: rband sims}%{{{
%Details{{{
Before implementing the program on science images, we first test the code to ensure
it is performing as expected. Using synthetic data, we ensure that the code is returning
correct photometry by checking for recovery of known input fluxes. As the code
has many options that can be activated by the user,
testing of all possible permutations of the code's parameter flags was imperative.
Here we provide a brief sample of the tests performed on the program, and give examples of the outputs generated by the program
with the various options activated. In this way, we hope to demonstrate both the code's
functionality and versatility.
{\bbf Additional testing of the various parameters, that is not discussed here, has been
performed using the both simulated imaging and real imaging from GAMA. These tests focus mainly on
ensuring the correct functionality of each of the programs options, rather than testing the
scientific value of each of the settings. As this sort of testing is expected of any program,
we do not include discussion of these additional tests here. However, samples of these tests are included
in the `example' sections of the package documentation. }
%}}}

\subsection{Generating Simulated Images}\label{sec: simimagecreation} %{{{
To facilitate further testing of the program, we have incorporated
the ability to generate a (optical regime) simulated image into the main body of code. The function is designed
to generate a simulated image with galaxy characteristics based on an input catalogue (containing galaxy
locations and $2.5\times$Kron apertures), an input image (which dictates the dimension of the output
simulated image, and also dictates the noise characteristics of the output image), and
observation parameters used in calculating photon counts
(telescope collecting area, filter effective wavelength and width, exposure time, etc). Resolved galaxies generated by
the simulation function all exhibit perfect exponential profiles, and have simulated fluxes determined by the user's
choice of flux-weights. If no flux-weights are provided, galaxies will be scaled to have equal peak-flux.
Galaxies are generated via Monte Carlo integration of photon counts simulated for each galaxy's profile and
magnitude, up to a ceiling of $10^6$ photons per object. Beyond this point, galaxies are generated using analytic exponential
profiles. In this way, we naturally incorporate a realistic determination of the per-object shot noise into galaxies whose
flux can be influenced appreciably by this effect. Galaxies are convolved with the user defined PSF to emulate observation,
and are then added to the image, allowing for simple additive blending of objects.

Random noise is added to the image using $N_{\rm pix}$ random draws from a Gaussian with mean and standard
deviation equal to the modal flux and MAD RMS of the input image respectively.
In order to more realistically model true observations, the simulated image can have additional galaxies
added such that the galactic number counts follow a power-law model. We use a power-law
with functional form $\log_{10}N=0.38{\rm m_r}-4.37$, derived using low redshift object counts from the
Millennium Galaxy Catalogue \citep{Liske2003}. By `padding' the image
with low-brightness galaxies we emulate a more realistic sky, as these objects contribute non-negligibly to the
noise characteristics of the image. In low-resolution images, this means that we can correctly simulate
the existence of confusion noise in our simulation.

\begin{figure*}
\centering
\includegraphics[scale=0.40]{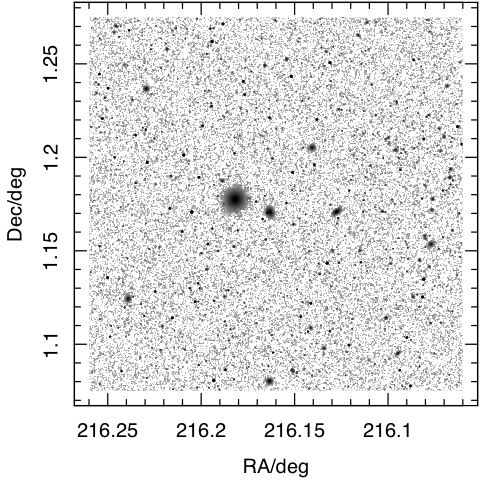}
\includegraphics[scale=0.40]{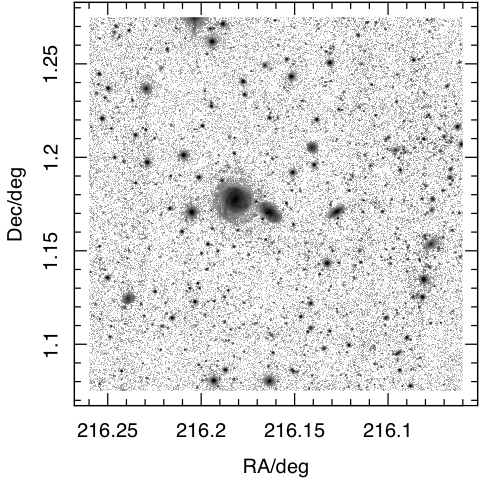}
\caption{An example of a simulated image (left) created from the GAMA galaxy catalogue and an input
\sdss\ {\em r}-band image, and the input image that it was based on (right). Galaxies are simulated with exponential
profiles, shot-noise, and physical backgrounds based on the input image. Images are asinh scaled, with white and black points
at 40\% and 90\% of the cumulative pixel density respectively.}
\label{fig: Sim Image}
\end{figure*}

Finally, the noise map can be convolved with a user-defined Gaussian, to simulate the process of `Gaussianisation' which
introduces correlations in the noise of the data image. This is required if the input image has been previously convolved with a Gaussian, as the
measured noise properties in the image (which are used to derive the noise generated in the simulation), will be different before and after
Gaussianisation. As such, after this optional convolution, the noise properties are again compared to the input image, and are corrected so that they match.
 The Figure
\ref{fig: Sim Image} shows an example of a simulated image created from the GAMA galaxy catalogue and an
input \sdss\ {\em r}-band image.

\begin{figure*}
\includegraphics[scale=0.30]{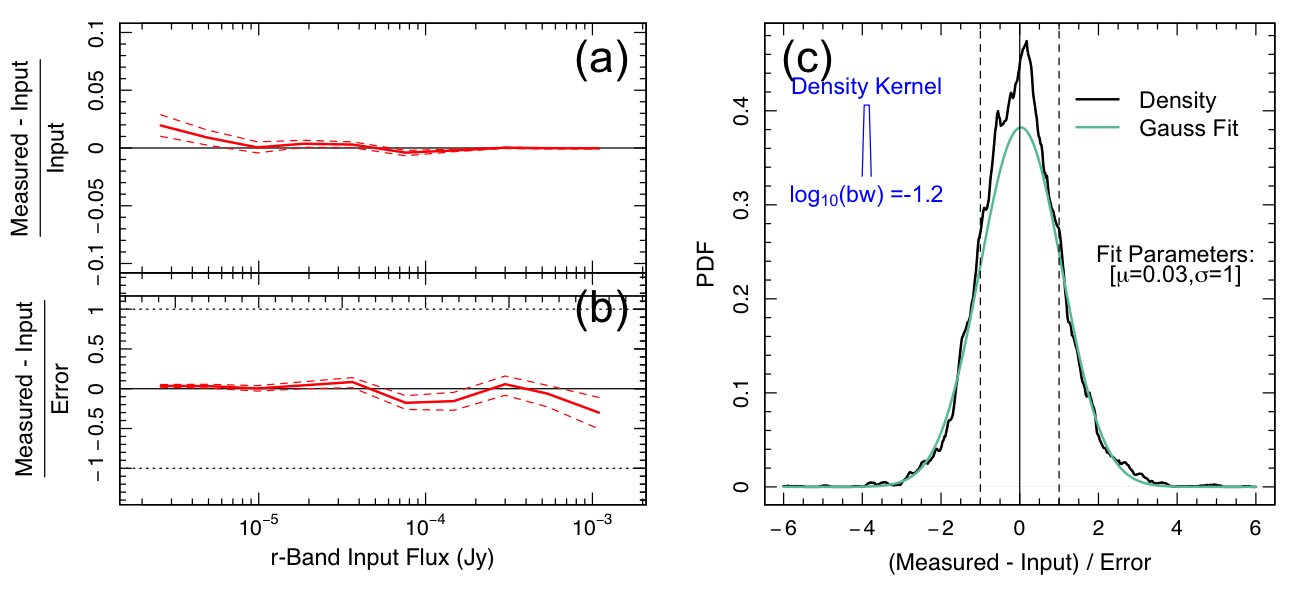}
\caption{{\bbf A comparison between the measured photometry (and uncertainties) to input photometry for our {\sc sdss} r-band simulation.
Panel (a) shows the running median (solid red) of the flux residual, with associated error bounds (dashed red), as a function of input flux.
Panel (b) shows the distribution of flux residuals divided by the measurement uncertainty; effectively a running median of `sigma deviation from truth'.
Panel (c) shows the kernel density of the sigma deviations.
Given appropriate uncertainties (that truly reflect the measurement error), this distribution should be a 0-mean gaussian with standard deviation of 1.
We fit a single component gaussian to the distribution, and find best fitting parameters $\mu=0.03$ and $\sigma=1.00$.
From this figure we can see that the measured photometry and uncertainties are both
a good representation of the input fluxes. } }\label{fig: Sim0Trumpet}
\end{figure*}

Using simulated galaxies, we are able to accurately compare the results of \lambdar\ to our known input flux.
Input fluxes are determined from the individual galaxy Monte Carlo Integrations, but prior to addition of
the sky-noise. Thus, we expect the output fluxes to demonstrate the standard `trumpet' behaviour, as demonstrated in
\cite{Driver2011}, and shown in Figure \ref{fig: Sim0Trumpet}. This behaviour arises because, for
fixed-distribution random sky-noise, a galaxy with lower apparent flux will experience greater perturbation.

%}}}

\subsection{Verification of Function Behaviour}\label{sec: testing functions}%{{{
%Details {{{
In addition to simple tests like those already described, we test the behaviour of some of the program's more
complicated and/or important functions, which are likely to be run by the typical user. We do this both to
test that the program is performing as expected in the general case, and to test the behaviour of the program
in exceptional cases.
%}}}
\subsubsection{Sky Estimation and Subtraction}\label{sec: testing skyest} %{{{
The sky estimate around every galaxy in the input catalogue is determined by fitting concentric annuli around each object and,
after optional iterative n-sigma clipping, fitting a running mean and median to each annular bin. This process is able to provide
robust sky estimates for each object, while also providing robust uncertainties and parameters that can be used in flagging poor/failed
estimates.

\begin{figure*}
\centering
\includegraphics[scale=0.20]{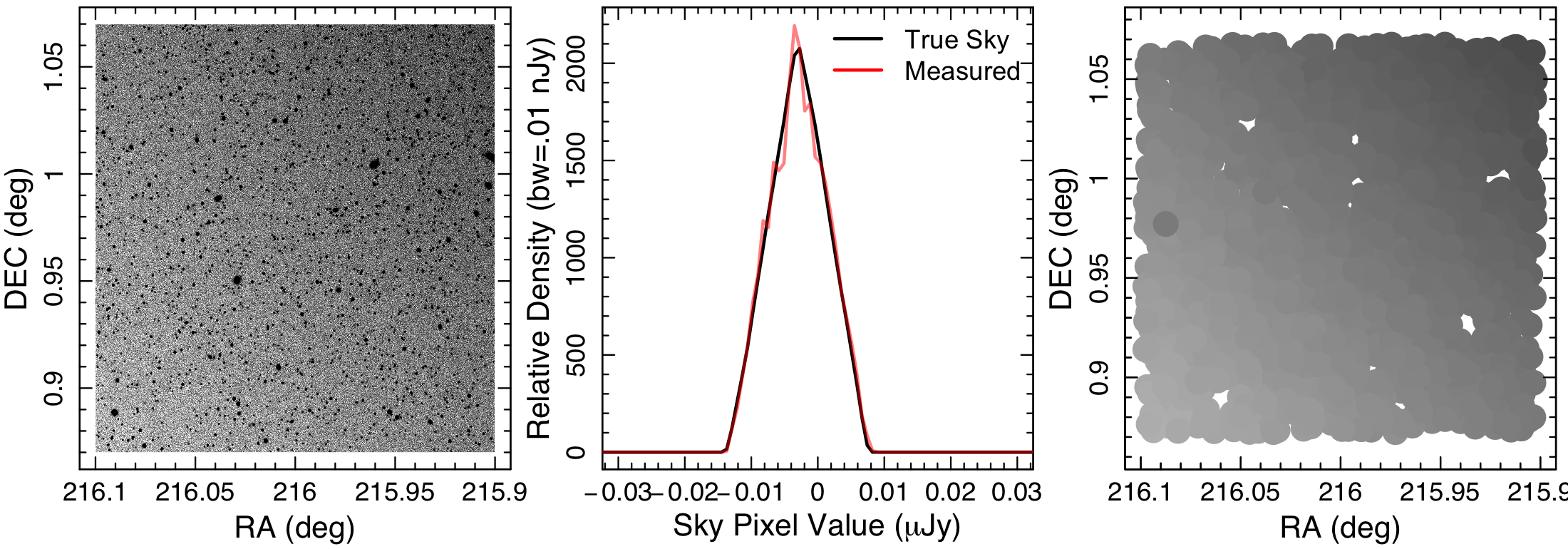}
\caption{Left: The simulated image with sky-gradient. Centre: Comparison between the input Sky Values and the measured sky
estimates from the program. Right; The on-sky distribution of estimates, with the same colour scaling as the simulated image (left).
}\label{fig: SkyEst_gradient}
\end{figure*}

%Behaviour with strong sky gradients %{{{
As our program determines sky estimates in concentric annuli around each object, we investigate first how the estimate behaves
in the regime where the sky value varies strongly on the same scale as the sky-estimate annuli. We simulate an astronomical image,
as described in Section \ref{sec: simimagecreation}, and apply a strongly varying sky of constant RMS. Figure \ref{fig: SkyEst_gradient}
shows the results of the program's sky estimation in this regime. From this image we can see that the sky estimate behaves well even
in this regime, returning estimates that follow the input sky gradient well.
%}}}

\begin{figure*}
\centering
\includegraphics[scale=0.20]{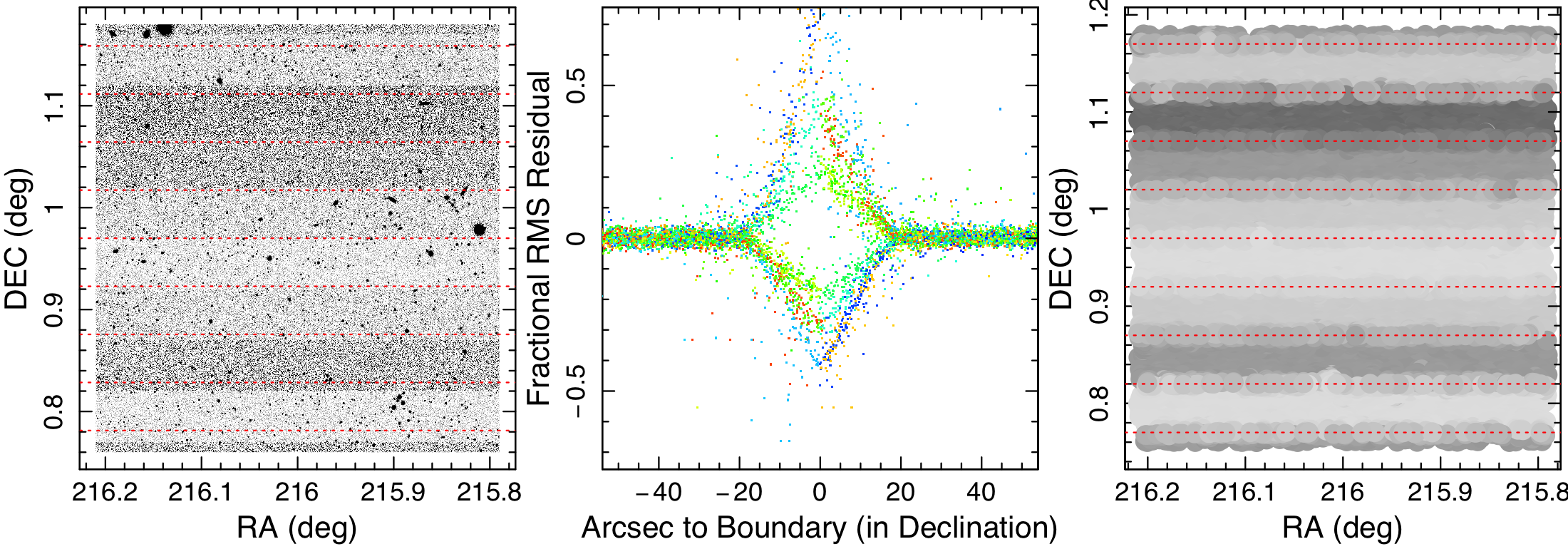}
\caption{Left: The simulated image with varying sky-RMS stripes. Centre: Comparison between the input Sky RMS Values and the measured sky
RMS estimates from the program, as a function of distance to the nearest RMS boundary. Points are coloured by Declination to demonstrate that
divergence from 0 occurs linearly as you approach a boundary, and that the degree of divergence correlates directly with the magnitude of the
discontinuity. Right: The on-sky distribution of estimates, where points closer to boundaries are plotted over others (to
show boundary effects). Colours in the right-hand panel have been scaled so that the measured
RMS value has the equivalent greyscale to the mean absolute sky value of that sky RMS the simulated image (left). Assuming skies are effectively
Gaussian, this allows direct visual comparison of colours in the right and left panels.
}\label{fig: SkyEst_RMSVars}
\end{figure*}

%Behaviour with non-uniform sky RMS %{{{
Astronomical images do not always have uniform/constantly varying sky. The most obvious example of this is in the
\sdss, where the drift-scan observations lead to sharp boundaries in sky behaviour (in both level and RMS). As such,
we test the code's ability to determine the sky RMS in the regime where we have sharp boundaries in the behaviour of the
sky. Figure \ref{fig: SkyEst_RMSVars} shows the results of the program's sky estimation in this regime. We see that the
returned RMS values are robust, except at the boundaries of the distributions. At boundaries we see that the returned RMS values tend
to lie within the range of the RMS values of the adjacent RMSs, and vary linearly with portion of aperture stamp in each segment.
%}}}

%}}}
\subsubsection{Object Deblends}\label{sec: object deblends}%{{{
To demonstrate the behaviour of the program's deblending routine, the program outputs information regarding deblend
fractions as a function of iteration. Additionally, the program optionally produces cutouts and CoGs for all/a sample of objects, so that
deblend behaviour may be examined by eye. An example of this has been shown already, in Figure \ref{fig: CoG}. While we do not show any
CoGs explicitly for our simulation here, we note that these are nonetheless output and are an available data product.
%}}}
%}}}

\subsection{Flux Measurements}\label{sec: optical sim fluxes} %{{{
We compare the fluxes returned by the program to those input to the simulation. Here we assume perfect source
detection, and thus use the known object locations and aperture parameters for our input catalogue. Figure
\ref{fig: Sim0Trumpet} shows this comparison, and shows good agreement between the input and returned fluxes.
As a result, we conclude that the program's flux measurement is being performed correctly.
%}}}
%}}}

\section{Testing using Simulated Deep FIR imagery from {\sc hermes}} \label{sec: fir sims}%{{{
%Details{{{
In addition to testing the program using optical simulations, we test the program
in the FIR regime using simulated Far-IR observations of the G10/COSMOS region. The mock imaging
utilises the semi-analytic models of \cite{Lacey2015}, synthetically observed
to mimic observation in the {\em Herschel} 250$\mu$m filter, using the same observation techniques and
integration times as were performed on observations of the G10/COSMOS field by the
{\em Herschel} Multi-tiered Extragalactic Survey ({\sc hermes}, \citealt{Oliver2012}). By
using the {\sc hermes} mock observations, we are able to test the program's behaviour in the
regime where detections are typically lower signal-to-noise and are more likely to be blended.

%}}}
\subsection{Flux Measurements}\label{sec: fir sim fluxes} %{{{
In order to accurately test our method of measuring FIR photometry in GAMA, we must select the objects for our testing catalogue in the same way that objects
are selected in GAMA. That is, we select targets that have either:
\begin{itemize}
\item an optical apparent magnitude equal to or brighter than $r=19.8$ mag in the SDSS r-band; or
\item a {\sc spire} $250\mu$m flux $\geq 4\sigma$ above the sky-RMS.
\end{itemize}
Specific details of how the GAMA photometric input catalogues are generated are described in Section \ref{sec: Catalogues}. Briefly, this
combined set of objects is required so that we can get measurements for all our targets of interest (\ie our optically selected sample), and
perform appropriate deblending of sources we can reliably identify as being contaminants not belonging to our target sample. Given our estimate of the
$1\sigma$ photometric uncertainty from the blanks routine (in Section \ref{sec: FIR testing functions}) of between $3.39$ and $4.47$ mJy,
we can estimate the $4\sigma$ limit of the image as being between AB magnitudes of $13.57$ and $13.27$. We therefore estimate the $4\sigma$ limit as being at an AB
magnitude of $13.5$, and select all objects in our simulation with input magnitudes brighter than this for our contaminant list definition.

\begin{figure*}
\includegraphics[scale=0.30]{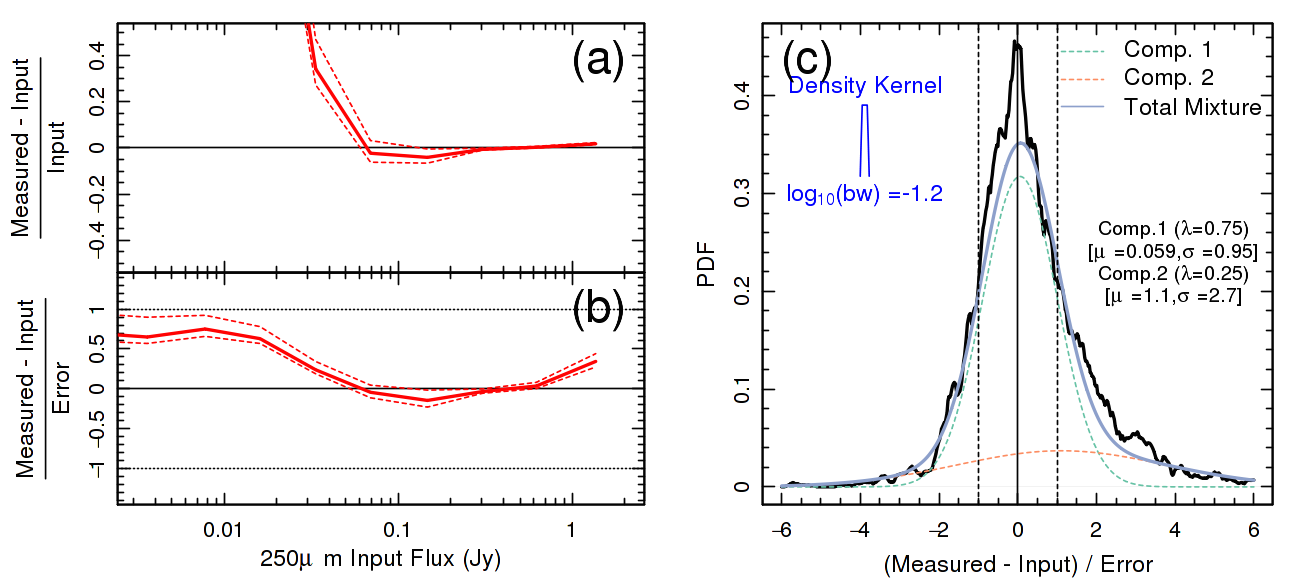}
\caption{A comparison between the measured photometry (and uncertainties) to input photometry for our {\sc hermes} simulation.
{\bbf Panels here are the same as in Figure \ref{fig: Sim0Trumpet}. Here, however, we fit a 2 component gaussian to the distribution
of sigma-deviations} to allow fitting of our expected (dominant) population of fluxes around a residual of 0, as well as
a population of pathological outliers (caused by contamination from sources not in our
contaminant list, both above and below the noise limit). We find that the dominant population is well approximated by a single gaussian component with
mean $0.06$ and sigma $0.95$. From this figure we can see that the measured photometry and uncertainties are both
a good representation of the input fluxes, despite flux boosting of faint sources.}\label{fig: HMock Trumpet}
\end{figure*}

Using this combined catalogue we measure photometry over 15 iterations with background subtraction and blanks estimation switched on.
We then compare the program's measured fluxes (and uncertainties) to the input fluxes. Figure \ref{fig: HMock Trumpet} shows the running median of the
fractional flux residual as a function of input flux (panel `a'), the running median of the sigma deviation from input as a function of input flux (panel `b'),
and the kernel density of sigma deviation from input (panel `c'). We fit a 2 component gaussian to the distribution to allow fitting of our
expected (dominant) population of fluxes around a residual of 0, as well as
a population of pathological outliers (caused by contamination from sources not in our
contaminant list, both above and below the noise limit). We find that the dominant population is well approximated by a single gaussian component with
mean $0.06$ and sigma $0.95$. As such, from these figures we can determine that fractional differences between input and measured photometry
are prominent (\ie $> 10\%$) only at the fainter end (fluxes $< 70mJy$) where contaminating flux boosting becomes significant (panel `a'). However, our uncertainties are appropriate
for the sources of error, as our median sigma deviation is constrained to within 0.5$\sigma$ of 0 for all fluxes brighter than $\sim 10mJy$ (panel `b').
Finally, our measurement errors are not
inappropriate over the whole sample, as our distribution of sigma deviation is well represented by a Gaussian with mean $\sim 0$ and sigma $\sim 1$ (panel `c'). The
secondary component in the sigma deviation distribution demonstrates the frequency of pathological failures, due to flux boosting of faint sources.

%}}}
\subsubsection{Impact of Contaminant Depth}\label{sec: contaminant lists} %{{{
\begin{figure*}
\centering
\includegraphics[scale=0.30]{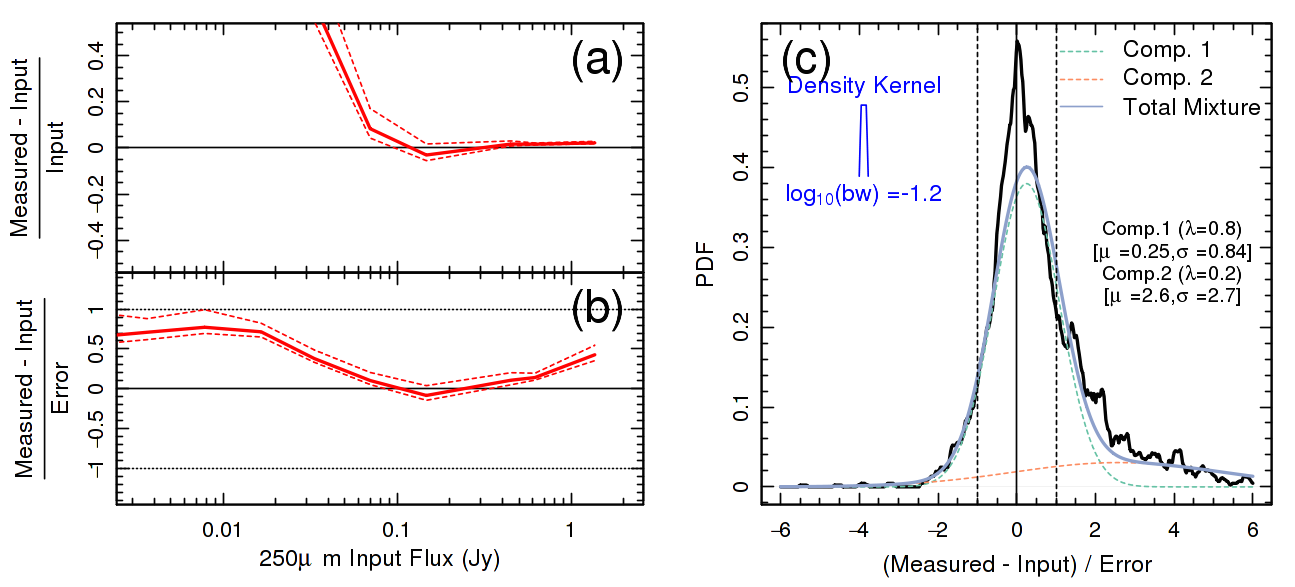}
\includegraphics[scale=0.30]{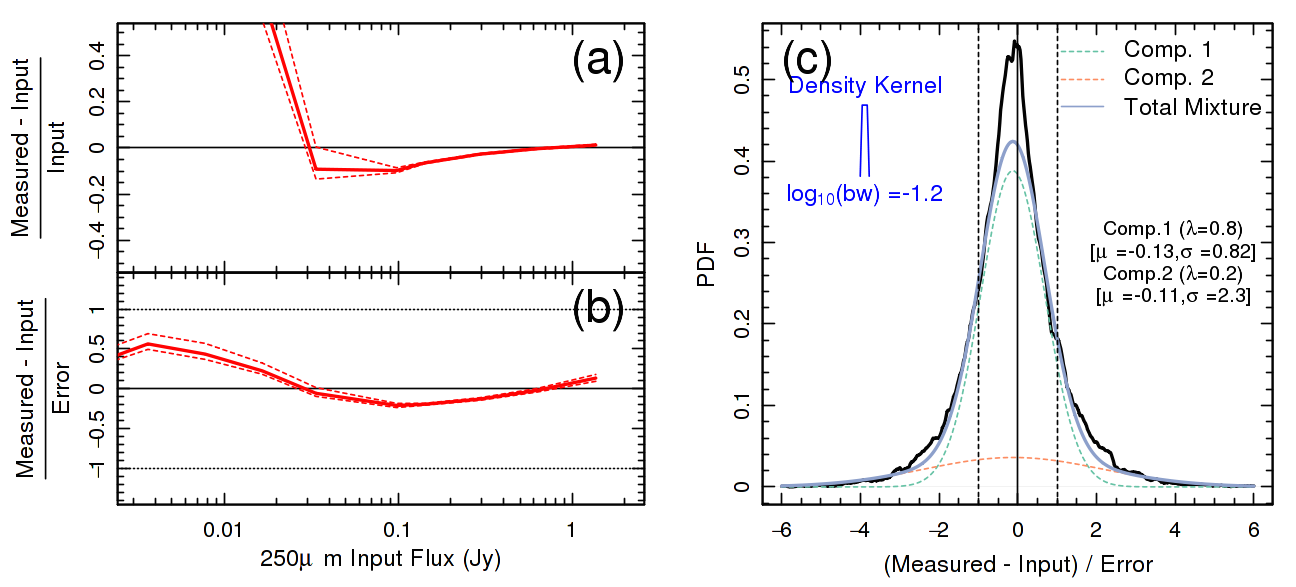}
\caption{Here we show the same as in Figure \ref{fig: HMock Trumpet}, but using contaminant lists cut at $6\sigma$ (top) and $2\sigma$ (bottom). From this
figure, we can see that the choice of contaminant list depth has important effects: a shallow list creates unwanted flux boosting of faint and bright sources,
while a deep list slows convergence and increases shredding of sources, {\bbf as seen by the systematic suppression of fluxes around $0.1$ Jy.}  }\label{fig: HMock Trumpet 2}
\end{figure*}

\begin{figure*}
\includegraphics[scale=0.30]{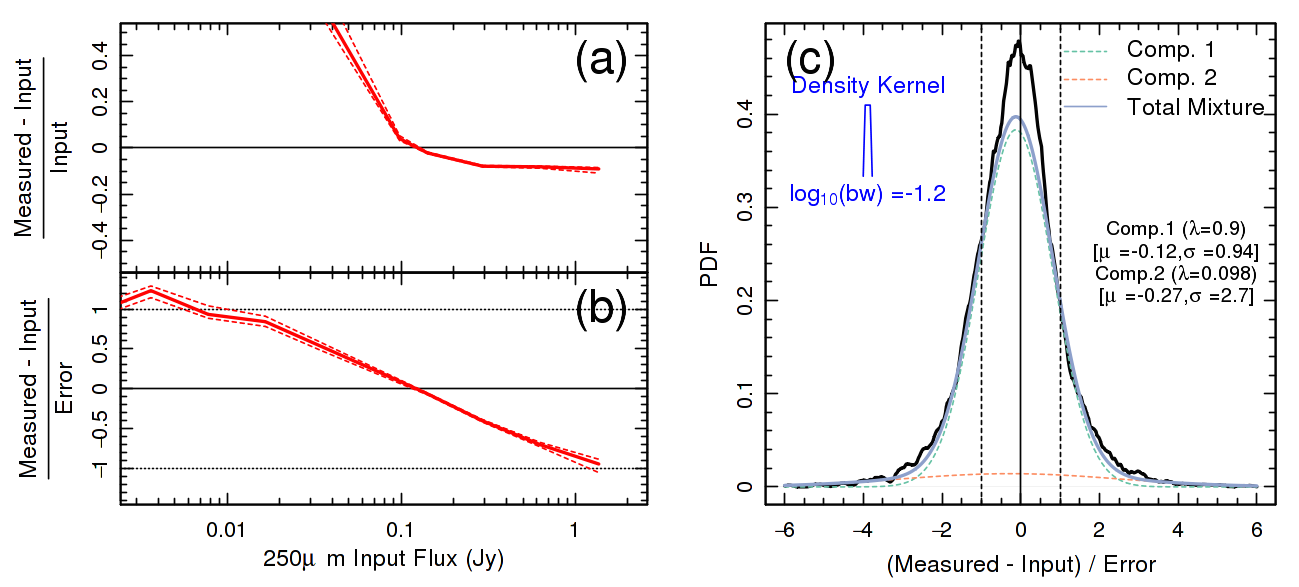}
\caption{The impact of iterating fluxes, accentuated by examining a case where we have a deep contaminant list, here cut at $2\sigma$. The figure is the same as the bottom section of
Figure \ref{fig: HMock Trumpet 2}, except here we show the fluxes as measured at the 0th iteration, where deblending is determined solely by on-sky position. The bias caused
is clearly evident in the systematic trend in sigma deviation when plotted against the simulation's input flux. }\label{fig: HMock convergence}
\end{figure*}

In Section \ref{sec: fir sim fluxes} we describe how we define and implement our `contaminant list' for this FIR simulation.
In this simulation, and in GAMA, we choose to define a contaminant list of objects that are strongly detected in each frequency range, using
conventional source extractors, but which can be reliably distinguished from our optically selected targets.
However, this raises the question of what is meant by `strongly detected', and how fluxes are affected by a change in this definition.
In the test above we implemented a $4\sigma$ cut on our contaminant list definition. Figure
\ref{fig: HMock Trumpet 2} demonstrates how flux measurements are impacted by using a contaminant list that is cut at $6\sigma$ and $2\sigma$ (Panel `a' and
`b' respectively), to demonstrate the impact of the choice of sigma cut.
From these figures, we can see that having a contaminant list that is too shallow means that there is a non-negligible increase in flux boosting of
faint and bright sources. When using a contaminant list that is much deeper, the fluxes are able to be more reliably deblended but take longer to converge. As a result,
in the same number of iterations, there is noticeable shredding of fluxes between sources; seen by the strong dip in the shape of the median distribution, and the negative
offset in the mean of the dominant population.
This effect will be more pronounced at lower iteration number. Figure \ref{fig: HMock convergence} shows the same distributions for the $2\sigma$-cut contaminant list
at iteration 0. As iteration 0 deblending is based only on sky-position, fluxes in any pixel are split equally between all sources with equal model coverage at that
pixel. The result is clear: faint sources start with too much flux, and bright sources start with too little flux. However, the deeper contaminant list has caused
the population of pathological failures to all but disappear.


Internally, the program distinguishes `science targets' from `contaminants' using an additional column specified in the input catalogue. The program
processes contaminants and targets identically, with the exception that contaminants that are not causally connected to a target are removed from
calculation. Similarly, photometry for contaminants is not included in the output catalogue. Causal connection is determined by whether the contaminant's
aperture array intersects with a science target's aperture array.
%}}}
\subsection{Verification of Randoms/Blanks Routine}\label{sec: FIR testing functions}%{{{
Using the {\sc hermes} mock imaging, we can also explore whether the program's internal randoms/blanks routine is able to recover
the expected noise and confusion properties of the image. To do this, we run the program with internal blanks routine
activated, and compare the RMS from this function with that measured when we run
blank apertures through the program using an externally derived blanks catalogue.
In the former case, the program returns a median blanks RMS of $3.87$ mJy, with quartile range $[3.39,4.47]$ mJy.
We then compare this value with that determined using the standard method of determining blank-apertures.
This is done by masking all sources in the catalogue, and generating
1000 random RA \& DEC positions in the field. We measure fluxes at each of these locations, and then fit the kernel density of these fluxes
(determined using a rectangular kernel of width $0.1$ mJy) with a Gaussian to determine the standard deviation. The blanks RMS measured in this way
is $3.45$ mJy.

%}}}

%}}}

\section{Updating GAMA Photometry: Comparing Measurements} \label{sec: comparison to PDR}%{{{
For the remainder of this paper, we detail the comparison between the photometry derived
from the GAMA PDR and the photometry derived by the \lambdar\ program. With the release of this new dataset, dubbed the GAMA
\lambdar\ Data Release (LDR), it is the hope of the authors that we will be able to subsequently
use this dataset for consistent panchromatic analysis of statistically relevant galaxy populations.

{\bbf The GAMA LDR contains $220,395$ sources, fewer than the $221,373$ galaxies presented in the GAMA PDR. The difference in
source counts is due to a comprehensive process of aperture definition whereby $1706$ sources were removed from the catalogue
by eye. Simultaneously, $728$ sources were created anew, that did not match a previously identified GAMA source. }

\subsection{Aperture Definition}\label{sec: aperture definition} %{{{
As the program does not perform an independent source detection, it is necessary to define an aperture
catalogue for use in this photometric analysis. {\bbf In the GAMA PDR, apertures in the optical and NIR were generated
using a single SExtractor run over the {\sc sdss} {\em r}-band imagery. Conversely, }
we use an aperture catalogue that is compiled through a
combination of SExtractor runs on {\sc sdss} {\em r}-band imaging, {\sc viking} Z-band imaging, and manual aperture creation.
This is done because it was apparent that simply running a single source extraction over the GAMA data was
not sufficient to create an aperture catalogue that was robust enough for our purposes.

\begin{figure*}
\centering
\includegraphics[scale=0.38]{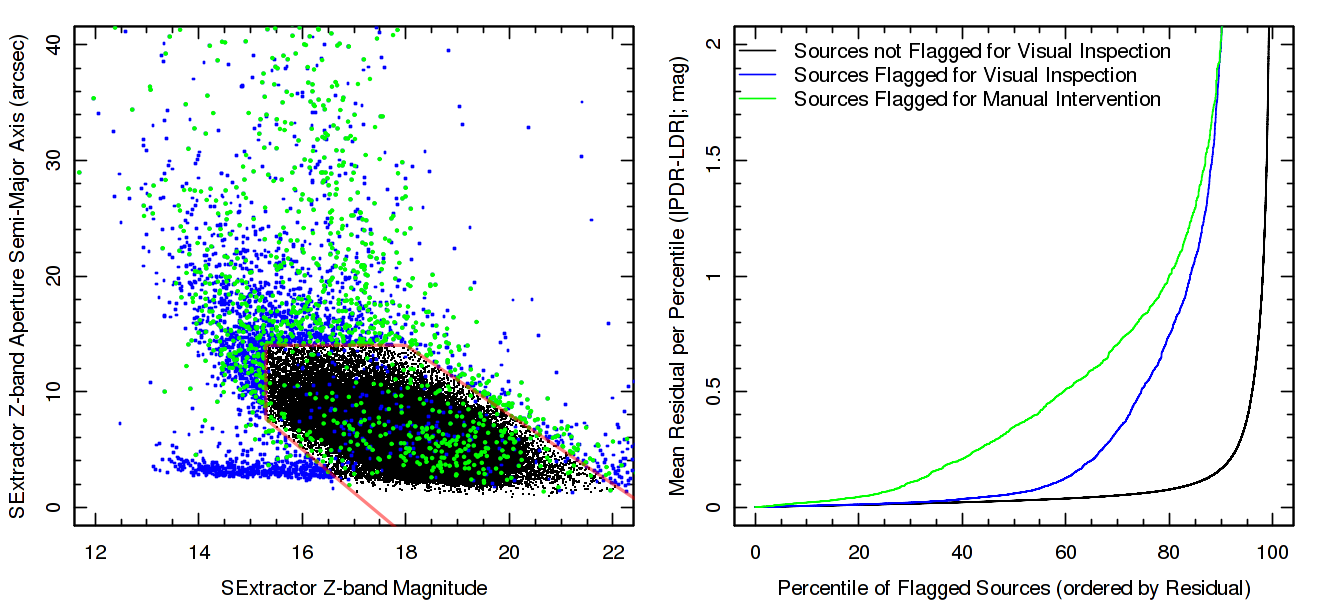}
\caption{{ \bbf The method of selecting objects with apertures requiring visual inspection. The left panel shows
the distribution of GAMA objects in SExtractor Auto magnitude
and semi-major axis length, as measured on in the {\sc viking} Z-band, for objects not flagged for visual inspection
(black), objects marked for visual inspection (blue), and objects identified (during visual inspection) as requiring manual
intervention (green). The selection criteria in surface brightness,
magnitude, and size used for determining objects requiring possible manual intervention
are shown as red lines; objects outside this boundary are all flagged for visual inspection.
Not shown are additional flagging criteria using on-sky position and Z-band coverage
(see Section \ref{sec: aperture definition}). The right panel shows the impact of our visual inspection and
intervention, quantified by the difference in PDR and LDR r-band magnitude. Colours are the same as in the left panel.
This figure shows that $\sim15\%$ of objects not flagged for visual inspection vary by $0.1$ mag or greater between
PDR and LDR, whereas $\sim40\%$ of objects flagged for visual inspection vary by $0.1$ mag or greater. For objects
that were identified as requiring manual intervention, more than $70\%$ of objects have magnitude differences of $0.1$
or larger.
} }\label{fig: aperture selections}
\end{figure*}

\begin{figure*}
\includegraphics[scale=0.30]{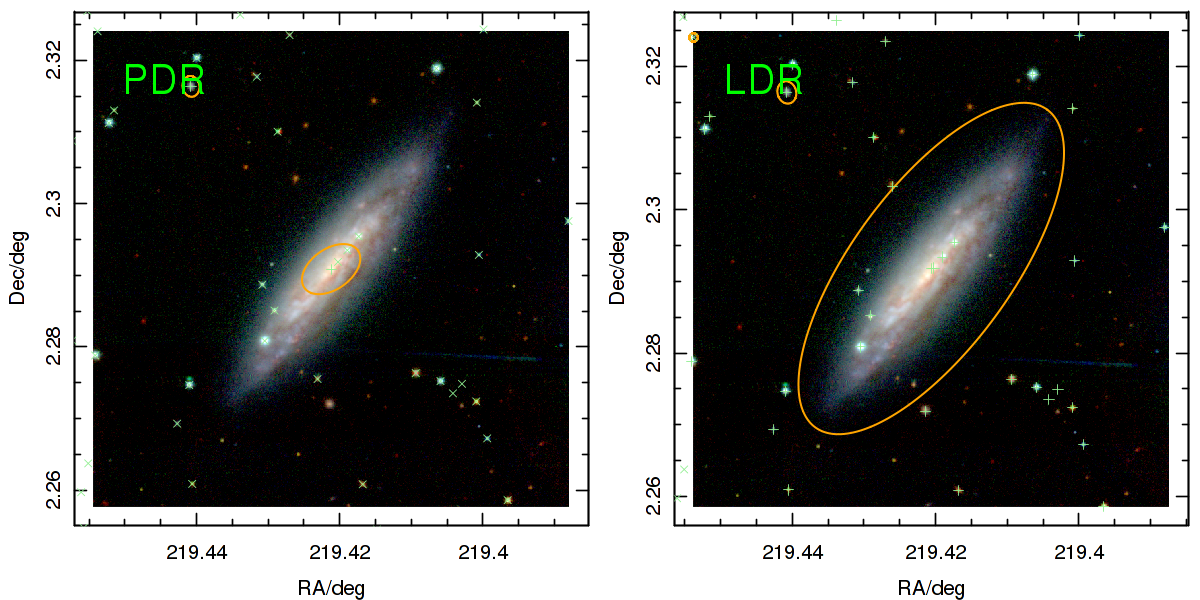}
\caption{Demonstration of an object whose aperture was flagged for manual intervention (left), and the aperture after correction (right).
}\label{fig: Aperture Fixes}
\end{figure*}

The aperture definition here follows the following prescription:
\begin{enumerate}
\item Run SExtractor over {\sc sdss} {\em r}-band, {\sc viking} Z-band `native', and {\sc viking} Z-band `convolved' images
\item Define criteria for determining possibly bad aperture definitions
\item Using a purpose-made visualisation tool, re-define problematic apertures.
\end{enumerate}
Here the `native' and `convolved' images refer to those at the native seeing and seeing convolved to $2^{\prime\prime}$ respectively
(see Section \ref{sec: PDR} for more information). {\bbf
For the determining which apertures required visual inspection, we used selection boundaries in size, magnitude, and average surface
brightness. Additionally, we include objects for visual inspection that have highly disparate sizes, magnitudes, and on-sky positions, when compared
to Sloan. The left-hand panel of Figure \ref{fig: aperture selections} shows an example of the measured sizes and magnitudes of GAMA objects after SExtraction
on the {\sc viking} Z-band native images. $4533$ objects were flagged for visual inspection (shown in blue), corresponding
to $\sim 2\%$ of all sources.

For visual inspection, `native' Z-band images were generated with apertures overlaid from each of the SExtractor runs
outlined above. During visual inspection each object is assigned one of these apertures, or (if no aperture is suitable)
it is marked for manual intervention. The manual intervention objects are fixed by hand using an online
aperture utility, allowing for
addition/removal of apertures, and modification of aperture parameters for objects already present in the catalogue. An
example of a manually fixed aperture is given in Figure \ref{fig: Aperture Fixes}. The object here was originally
flagged for visual inspection because of its anomalous surface-brightness, and was subsequently marked for
manual intervention. Of the $4533$ objects flagged for visual inspection, $702$ objects were flagged for manual
intervention.

The outcome of this process of flagging, visual inspection, and manual intervention, is shown in the right-hand panel of
Figure \ref{fig: aperture selections}. Here, we show the absolute change in {\sc sdss} {\em r}-band
magnitude between PDR and LDR, ranked smallest to largest, for all sources
not flagged for visual inspection (black), flagged for visual inspection (blue), and flagged for manual
intervention (green). From this figure, we can see that $\sim15\%$ of sources not flagged for visual inspection change by
more than $0.1$ magnitude between PDR and LDR. For sources flagged for visual inspection, we can see that the
fraction of sources that change by more than $0.1$ magnitude jumps to $\sim40\%$. Using the uninspected sample as a
baseline (effectively controlling for the difference in method between SExtractor and \lambdar), this indicates
that our visual inspection has had a substantial impact on the final flux estimates.
Further, for sample flagged for manual intervention the fraction increases to more than $70\%$.
}
%}}}

\subsection{Catalogues}\label{sec: Catalogues} %{{{
Having defined the apertures for the science targets, we then must determine what to define as
appropriate contaminant lists for analysis in each imaging band. We define 3 different contaminant lists, which reflects the 3 broad wavelength
regions probed by the GAMA multi-wavelength data: the UV-optical-NIR regime ({\sc fuv-}Ks), the MIR regime ({\sc w1-w2}), and the FIR regime
{\sc w3-500$\mu$m}. We choose these boundaries as they broadly mark the transitions between various contaminating sources; namely disk and halo stars,
additional dwarf stars, and high-redshift starburst galaxies respectively. As a result, the contaminant list required in the optical regime is
quite different to that required in the FIR, whereas the MIR and optical contaminants have a substantial overlap. Our contaminant lists are defined here.

In the optical, our contaminant list is defined using the GAMA Input Catalogue v06 (described in \cite{Liske2015}), and contains all stellar and galactic
objects that do not form part of the GAMA II galaxy sample. In the MIR, we use the sample of all objects that have been identified by the \wise\
team as not matching to a GAMA target as our contaminant list. In the FIR, we use the sample of all objects identified by the {\sc hatlas} team
as those not reliably matching a GAMA target (\ie with reliability parameter $<0.8$; see \citealt{Bourne2016}), as our contaminant list.
In each case, we perform a sky-match between the contaminants and
science targets, and exclude any contaminants that are within $1$ PSF FWHM in the detection band (\ie r-band in the optical, {\sc w1} in the MIR, and $250\mu$m in the FIR).
This is because targets within these limits are likely too close to be reliably detected as contaminating sources. We note that this is not technically the case in the FIR,
as spectral slope is a key indicator for the presence of a high-redshift contaminant; however in practice there are only three contaminants which fall within this limit in
the FIR contaminant list. As such we use the method across the board, for consistency.
%}}}

\subsection{Input parameters}\label{sec: input params}%{{{
For the determination of LDR photometry, we run the program with the settings presented in Table \ref{tab: lambdar params}.
{\bbf Here we justify our choices of each parameter, as this information will likely inform readers interested in applying the
program to other datasets. Parameters not stated in this table are left as default.

We implement a PSF convolution in all bands, including the optical and NIR (where apertures are defined). We do this because there
exist point-source objects in the prior catalogue is constructed with future convolution in mind, meaning that point-source objects are
given aperture-radii of 0.

We perform a local sky estimate in all bands except the {\em GALEX} FUV. The FUV imaging is Poissonian in nature, as the expected number
of sky photons per pixel is less than 1. The sky estimate routine in the program is not designed with Poissonian skies in mind, and it is not
clear that the program will behave sensibly in this regime. Fortunately, the FUV imaging has a probabilistic sky-estimate
incorporated into the imaging (see \cite{Andrae2014} for details).
As a result it is not necessary (or sensible) for us to perform our sky estimate on the FUV imaging.

We use PSF weighted photometry in all images, to improve extraction of fluxes at low signal-to-noise across the entire wavelength bandpass.
We also opt to use recursive descent aperture placement in all but the optical and NIR bands. This is because in the optical and NIR the
resolution is so high that apertures will always span a large number of pixels. In these bands we use quaternary aperture placement
(done by setting the number of aperture resampling iterations to 0).

Finally, we use pixel-flux weight weighting in all bands blueward of $12\mu$m, as the additional weighting can help the program more rapidly
converge to the best flux measurement. However, in the shallowest bands (\ie bands where less than 65\% of our target objects are detected at
$\ge 1\sigma$), pixel-flux weighting may act to produce more scatter in the measurements at low iteration numbers, though this is
predominantly conjecture. In any case, the choice of inclusion of pixel-flux weighting is largely inconsequential, as we choose to
iterate the flux determination, and use a large number of iterations (15).
}
%}}}

\subsection{Imaging Properties}\label{sec: imaging properties}%{{{
\begin{figure*}
\centering
\includegraphics[scale=0.365]{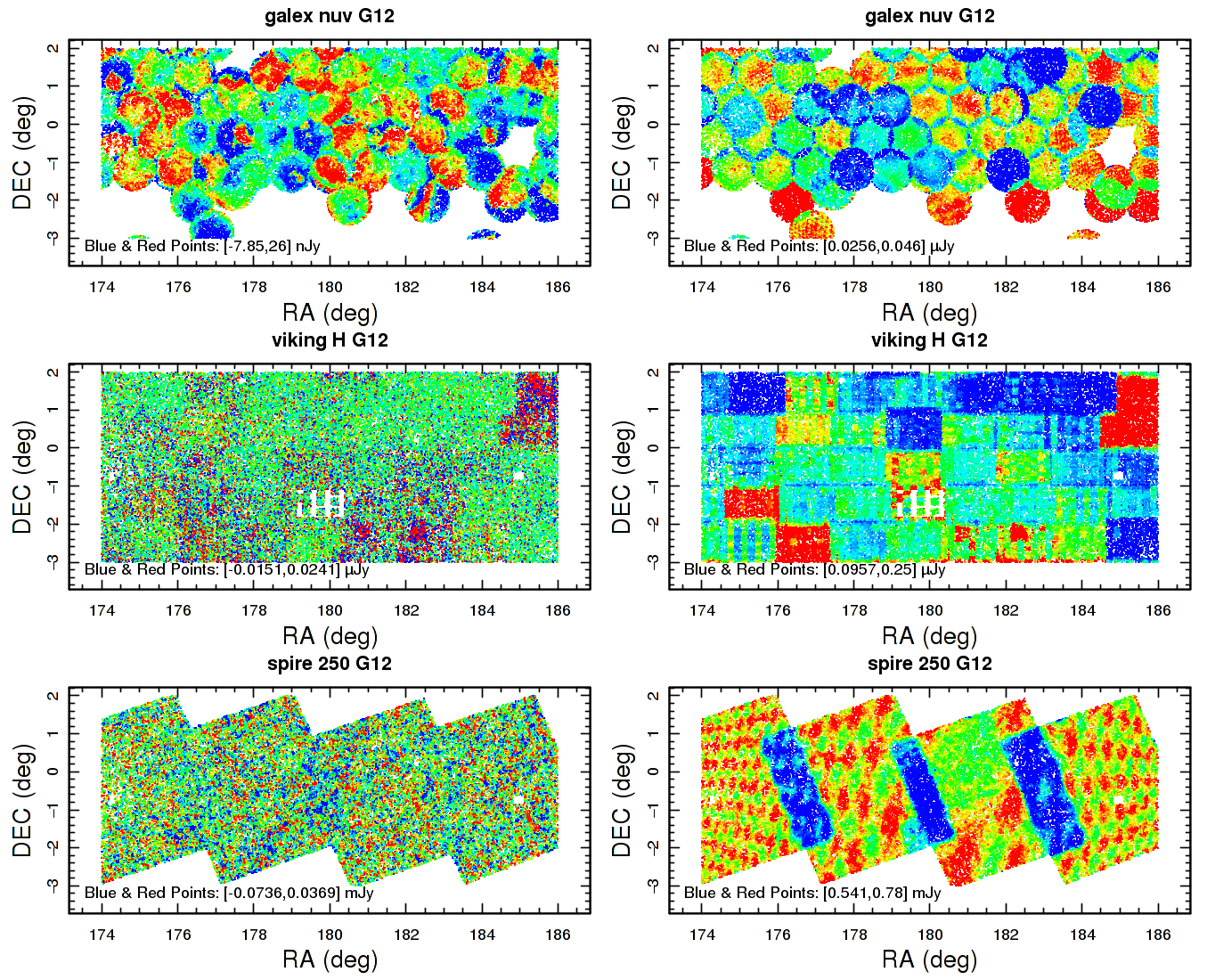}
\caption{A sample of the on-sky distributions of Local Sky Estimate (left) and Local Sky RMS (right) for 3 bands in the G12 field.
The top row is the {\sc \galex\ nuv}, the middle row is the {\sc viking h} band, and the bottom row is the {\sc spire} $250\mu$m band. In each image, the points are scaled with
blue and red points at the 10\% and 90\% values respectively. {\bbf The absolute values corresponding to these blue and red points are given in the bottom-left of each figure.} As these filters have all been sky-subtracted prior to running \lambdar, any variations
seen in the left column represents flux that will contribute adversely to the final measurement if not removed. Similarly, complex variation in the
sky RMS in the right column must be recognised for appropriate uncertainty estimation. We note that the pattern seen in the {\sc hatlas} mosaic sky-RMS (bottom right)
is a Moir\'e pattern induced when the mosaic was resampled onto the standard GAMA field centre (see \protect\citealt{Driver2016}), and is not present in the
original {\sc hatlas} imaging.}\label{fig: measured skyest}
\end{figure*}

Using the estimated values for the sky (both in value and RMS) from \lambdar, we can investigate the properties of the
imaging within the GAMA fields. Photometry in the GAMA PDR is measured, per-galaxy, on maps that have already had global backgrounds subtracted
\citep{Driver2016}, with the exception of the MIR where photometry has had local backgrounds estimated and subtracted at the time of
measurement \citep{Cluver2014}. As the imaging used by \lambdar\ is the same as
that used for measuring the PDR photometry (with the exception of the {\em Herschel} PACS bands; see Section \ref{sec: PDR}), any background
that we measure as requiring removal will therefore be present as an unrecognised
systematic in the PDR photometry. In Figure \ref{fig: measured skyest} we show a sample of the sky estimates measured by \lambdar, both in value
and in RMS, as a function of on-sky position. In these images we can see that there exists residual, systematic, variations in the sky level, as well as
complex structure in the sky RMS (which is important to consider when deriving flux uncertainties). As these images are used as-is for photometric
measurement in the PDR photometry, we conclude that there is likely to exist subtle systematic biases in the PDR measurements and uncertainties.
Conversely, as we characterise the imaging properties locally for every source in the LDR, we are able to remove any such biases.
%Sky Estimates
%}}}

\subsection{Flux Comparison}%{{{
\begin{figure}
\includegraphics[scale=0.30]{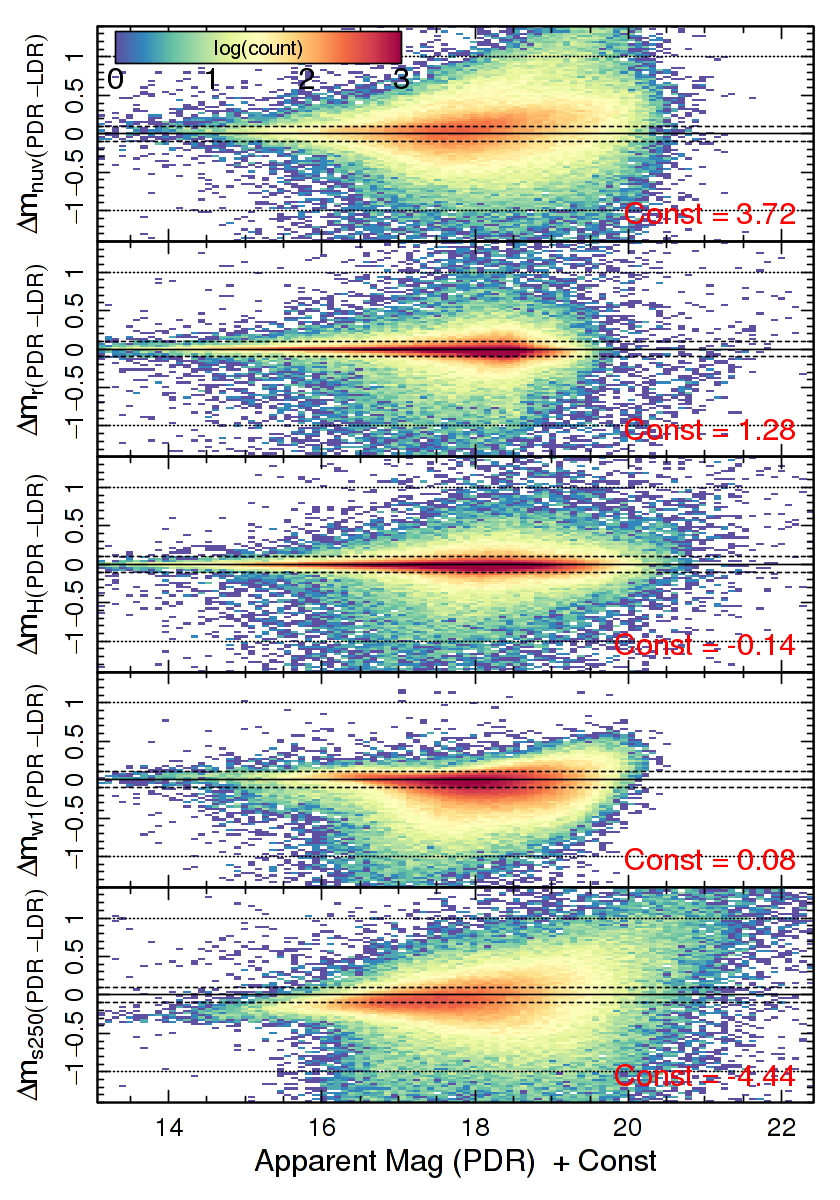}
\caption{`Trumpet' plots showing {\bbf difference in magnitude vs PDR magnitude for the \galex\ NUV, {\sc sdss} r, {\sc viking} H, \wise\ W1,
and {\sc spire} 250\um bands, demonstrating that the photometry is in broad agreement with the PDR photometry, but nonetheless
shows systemic differences caused by subtle differences in measurement methods. Each panel has a horizontal offset applied, which is
given in the lower right of each panel. Horizontal dashed and dotted lines are shown at $\pm$0.1 and 1 magnitude, respectively, for reference. } }\label{fig: trumpets}
\end{figure}

Figure \ref{fig: trumpets} shows trumpet plots for the \galex\ NUV, {\sc sdss} r, {\sc viking} K, \wise\ W1,
and {\sc pacs} $160\mu$m bands. A full compilation of trumpet plots in all bands can be found in Appendix \ref{sec: Full PDR Comparison}.
From these figures, we can see that the photometry from \lambdar\ agrees broadly with the photometry presented in the PDR,
however there are indeed variations in the distributions that cannot be explained on signal-to-noise grounds. For example, the
structure in the \wise\ trumpet is likely a combination of many effects, but is overarchingly due to
the differences in apertures (fixed size vs variable), and deblending (none vs some). {\bbf The difference in fixed and variable
aperture sizes can be seen in the systematic trend whereby LDR fluxes are fainter for faint PDR fluxes, and brighter for bright PDR
fluxes. This is because brighter objects are also typically larger, and a fixed size aperture will systematically miss flux. Similarly,
faint sources are typically small, and are much more sensitive to being contaminated by neighbouring sources.
As a result deblending creates an increased scatter downward (which is larger at the faint end of the distribution) as
fixed-size apertures with no deblending allow flux to be double-counted, whereas this is not possible in \lambdar's deblending. }

\begin{figure}
\includegraphics[scale=0.27]{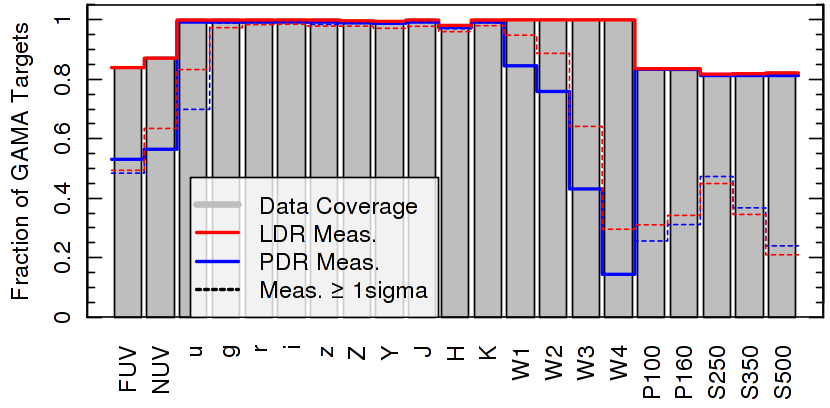}
\caption{The number of measurements made, per band, in each of the LDR (red) and PDR (blue) datasets. The fractional coverage of each band
is shown as the grey vertical bars. The benefit of performing matched aperture photometry is apparent; for each optically measured object (\ie 100\%
of GAMA targets), there exists a measurement in every band where there exists coverage. Conversely, the PDR dataset shows large numbers of missing
measurements where data exists in some bands (the UV and MIR particularly). {\bbf Further, we show the fraction of all sources that are detected at
$1\sigma$ or greater as dashed lines, for both the PDR and LDR. As the PDR photometry are signal-to-noise limited in the {\em GALEX} NUV and
{\em WISE}  MIR, the solid and dashed blue line in these bands overlap.} }\label{fig: coverage}
\end{figure}

In addition to showing the agreement between the measurements made in the PDR and LDR datasets, we also demonstrate the
utility of matched aperture photometry in terms of number of flux estimates. Figure \ref{fig: coverage} shows
the number of measurements made in the LDR and PDR datasets per band. In addition, the figure shows the fractional coverage
of each band. From this figure, the utility of matched aperture photometry is quite apparent; the LDR dataset has performed a measurement
for every object in every band for which there exists coverage, while the PDR has a substantial number of missing flux estimates in some bands
(particularly the UV and MIR), as no measurement was made. {\bbf Furthermore, we can see the gain in positively detected sources (\ie
signal-to-noise greater than 1) in the LDR dataset, over the often signal-to-noise limited PDR dataset. }

\begin{figure}
\includegraphics[scale=0.17]{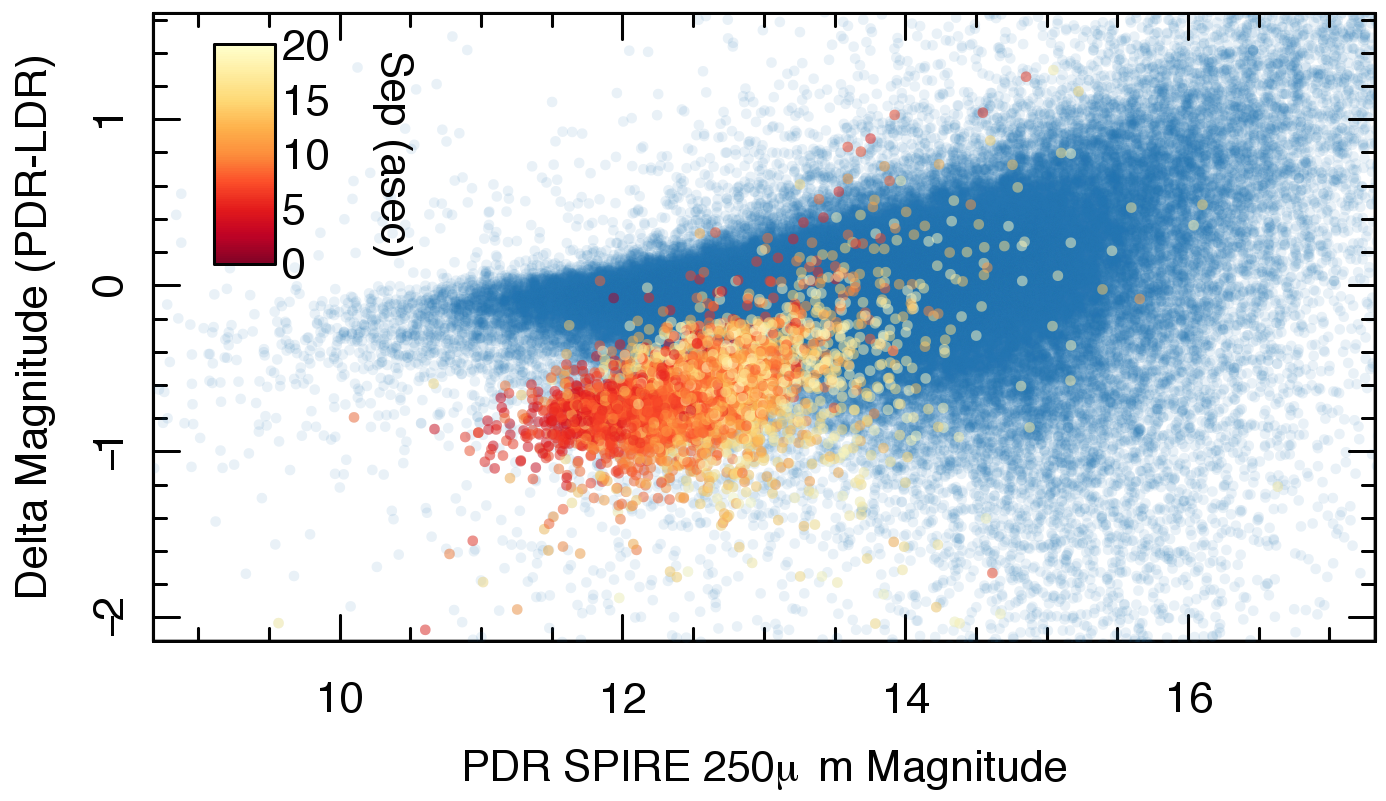}
\caption{Demonstration that the population of objects offset from 0 in the {\sc spire} 250$\mu$m trumpet is caused by deblending of nearby contaminants.
Here colouring is by separation between the object and its nearest-neighbour contaminant, in arcseconds. The blue points in the
background is the distribution of all objects not matched to a contaminant within $20^{\prime\prime}$, for reference.
}\label{fig: FIR cloud}
\end{figure}

%FIR Comparison %{{{
In the FIR trumpet plots, seen in Appendix \ref{sec: Full PDR Comparison}, there are two particular differences that are of note.
Firstly, there exists a population of objects with large magnitude offsets ($> 0.5$ mag)
at moderately bright PDR magnitudes (clearest in SPIRE $250\mu$m), which cannot be explained as variations due to backgrounds/noise.
Inspection of these objects shows that they are all objects that have been deblended from a high-redshift contaminating source (see the contaminant list definition
in Section \ref{sec: Catalogues}). This can be seen in Figure \ref{fig: FIR cloud}, where we show the residuals
between LDR and PDR fluxes, coloured by separation to the nearest high-z contaminant. From this figure, we can see that the
offset in the residuals changes systematically with distance to the nearest contaminant. The systematic trend with separation, and the fact that the cloud is dispersed
around a value of $f_{\rm LDR}/f_{\rm PDR} \sim 0.5$, suggests that these are objects for which \lambdar\ has deblended a contaminant that was not subtracted in
the PDR.
In the final trumpet plots in the appendix, this population is far more heavily dispersed than in Figure \ref{fig: FIR cloud}.

Secondly, the brightest FIR objects are systematically dimmer in the LDR dataset than in the PDR. This is because the \cite{Bourne2012} program implemented
our maximum normalisation factor by default, rather than the minimum factor {\bbf automatically employed in \lambdar}. As the brightest fluxes typically belong to the objects with the largest
apertures, the difference in normalisation factor becomes even more pronounced. Using our maximum factor, \lambdar\ recovers well the fluxes in the PDR at the
bright and faint ends; { \bbf however we opt to report our fluxes as calculated using the minimum correction as it makes fewer implicit assumptions about the
distribution of source flux (see Section \ref{sec: minimum aperture correction}). }

%}}}
%}}}

\subsection{Error Comparison}%{{{
\begin{figure}
\centering
\includegraphics[scale=0.23]{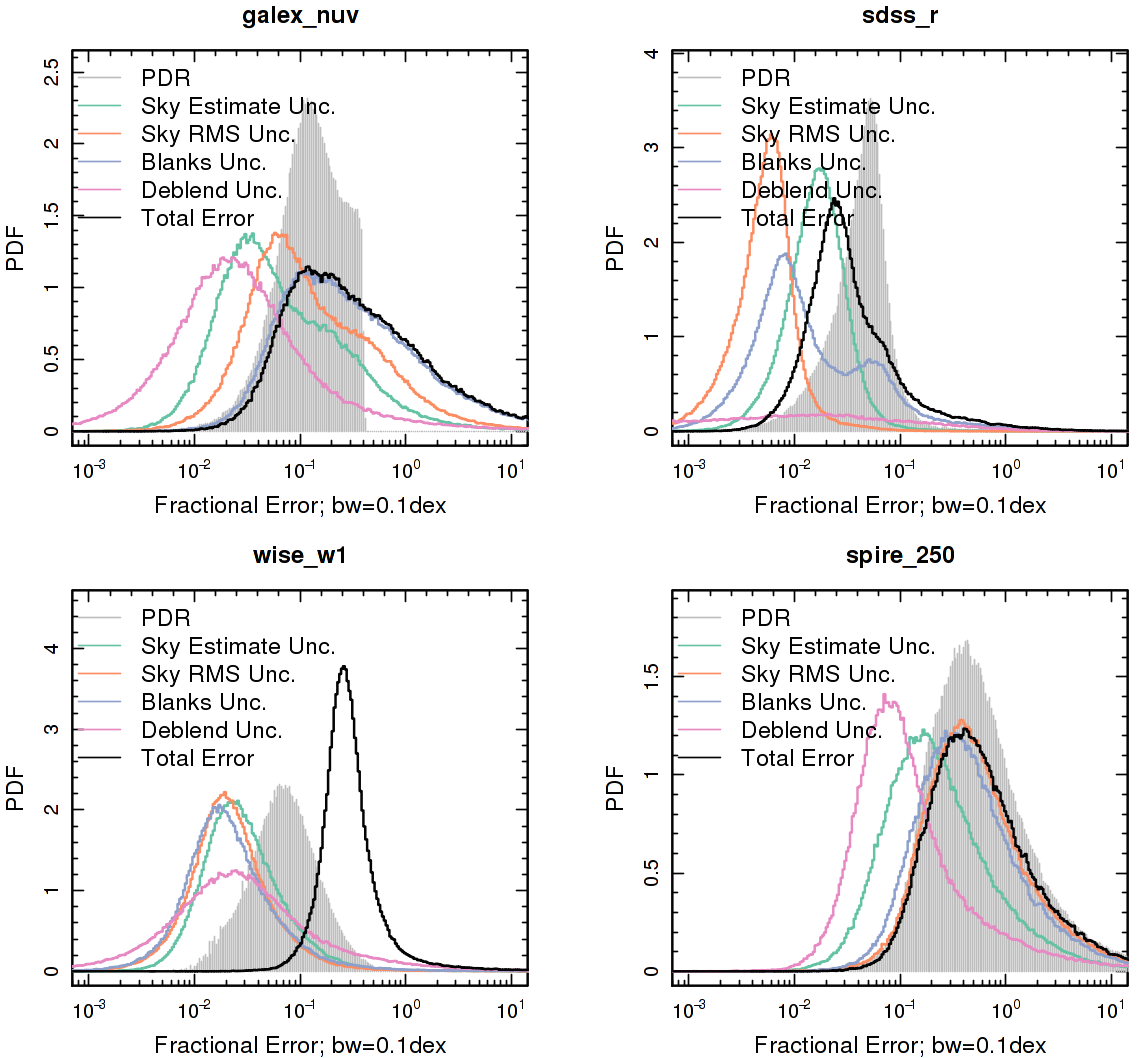}
\caption{Comparison of the uncertainties returned by the program, split into various components (coloured lines), compared to those in the GAMA PDR
(grey histogram). We note that typically our fractional uncertainties are consistent with those measured previously. This is not true in all bands, with the
SDSS z-band being a stand-out above with a factor of $\sim 2$ difference between the PDR and LDR final uncertainties. This is not necessarily unexpected, given that
the measurement methods used in the PDR and LDR datasets are different, the apertures are different,
the factors incorporated into the uncertainties are different, and the methods of determining those factors are different.
However, we can now be confident that the uncertainties are determined in a consistent manner, and are therefore  not going to create biases in
multiband SED analyses. }\label{fig: Error Components}
\end{figure}

We compare the uncertainties measured by \lambdar\ compared to those given in the GAMA PDR. Figure \ref{fig: Error Components}
shows a comparison between the uncertainties measured for a range of bands (split into individual components), compared to the uncertainties
present in the PDR release. A full compilation of error distributions can be found in Appendix \ref{sec: Full PDR Comparison}.
From these figures we can see that, while there exist differences in the error components in some bands, \lambdar\ is
typically returning uncertainties that are consistent with what was previously
determined. Furthermore, we can be confident that the uncertainties used in all the different bands are determined in a consistent manner, giving us confidence
that differences in uncertainties between PDR and LDR are real and likely due to differences in, for example, measurement methods.
This fact will ensure that we are not biased during SED fitting because of uncertainties in one or more bands being systemically under/over estimated
when compared to those in adjacent bands.
%}}}

\subsection{Colour Comparison}\label{sec: colours}%{{{
\begin{figure}
\includegraphics[scale=0.23]{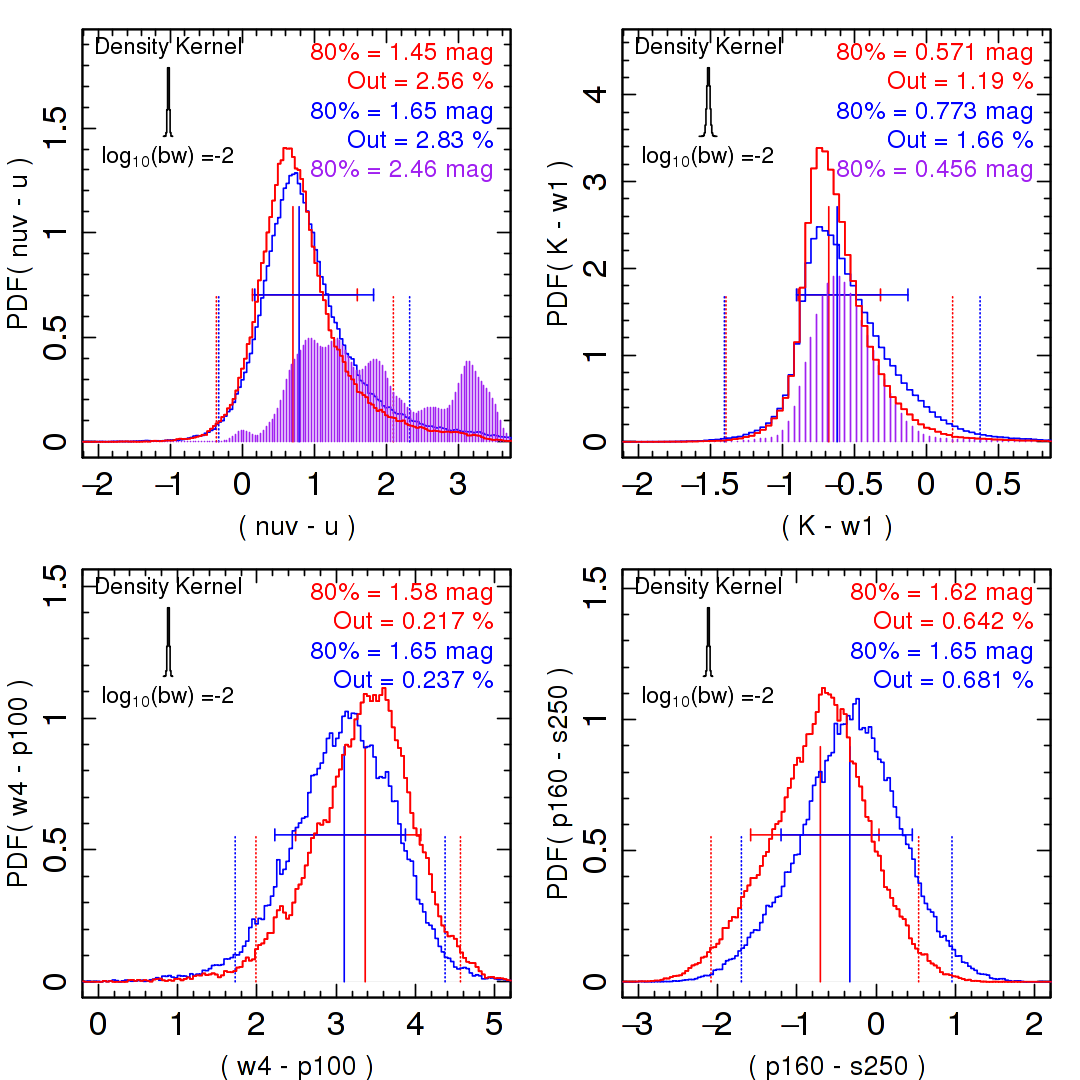}
\caption{Colour distributions comparing PDR (blue) and LDR (red) photometry for the \galex\ NUV - {\sc sdss} u, {\sc viking} K - \wise\ W1,
\wise\ W4 - {\sc pacs} 100, and {\sc pacs} 160 - {\sc spire} 250 bands. {\bbf These bands are selected for demonstration as they are colours
which cross facility boundaries, meaning that PDR photometry in these colours may be systematically inconsistent.}
In the first two panels, the purple histogram shows the colours of galaxies
that form the templates of \protect\cite{Brown2014}. {\bbf Inset text shows the effective colour width at 80\%, and the outlier rate, for
each distribution. The inset graph shows the density kernel used in calculating the PDFs. }
Samples here are matched so that only objects that are detected at (at least) $1\sigma$
in both bands, in both photometry samples, are present. This is done to remove complicated selection effects and objects with spurious colours. PDFs here
are generated using a kernel density estimator, with kernel as shown in the upper right of each panel.
}\label{fig: colours}
\end{figure}

Figure \ref{fig: colours} shows colour distributions for a sample of 5 colours in the 21-band PDR and LDR datasets. Again,
a full compilation of colours can be found in Appendix \ref{sec: Full PDR Comparison}.
From these distributions, we can determine two particular parameters of interest: the effective width of the distribution, and
the distribution outlier fraction.

\begin{figure}
\centering
\includegraphics[scale=0.28]{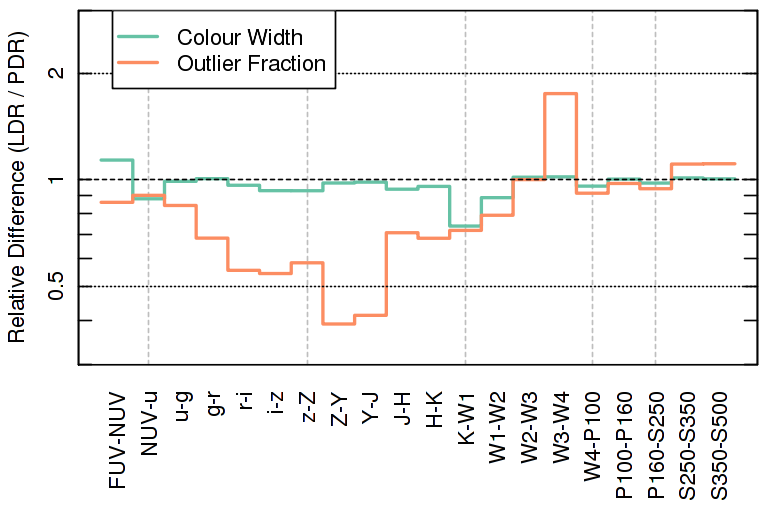}
\caption{Relative changes in the width (cyan) and outlier fraction (orange) of the colour distributions for adjacent bands in the PDR and LDR.
{\bbf Facility boundaries, where the colour shown uses fluxes from different instruments (and therefore typically different measurement methods),
are marked on the figure by grey vertical dotted lines.}
We can see that
the colour distributions in the LDR dataset are narrower than the PDR dataset, with the exceptions of the FUV-NUV, W2-W3, and W3-W4 colours. Similarly, we
see an improvement in the outlier fraction for all colours, with the exceptions of {\bbf the W3-W4 and internal {\sc spire} colours}. We note, in particular, that improvements in the
colour distribution widths are greatest (compared to adjacent colours) when crossing facility boundaries; \ie where measurement methods in the PDR change.
{\bbf Similar improvements in outlier fraction are seen at these boundaries also.} This indicates that the consistent measurement is having a large improvement in these bands, compared to the PDR. The dotted lines mark where the ratio is different by a
factor of 2, in denominator and numerator. Possible explanations for these differences seen in the FUV-NUV, W2-W3, and W3-W4, are described in the text. {\bbf Comparison of photometry in {\sc pacs} and {\sc spire} is presented in \protect\cite{Valiante2016}.}
}\label{fig: stacked panels}
\end{figure}

The effective width of the colour histograms is informative as, assuming there exists some
fundamental distribution of galactic colours, any measured distribution will trace the fundamental distribution convolved with a
Gaussian distribution (reflecting the combined measurement uncertainties for each galaxy). As a result, the distribution that is
measured to have a smaller effective width is therefore that with smaller measurement uncertainty. Due to the highly non-Gaussian shape
of the colour distributions, we use the width of the central 80\% of the distribution (\ie the number of magnitude separating the
10\% and 90\% limits) as our effective width. The 10\% and 90\% limits of the colour distributions are shown graphically in Figure
\ref{fig: colours} as a horizontal bar, coloured for each
distribution. Furthermore, Figure \ref{fig: stacked panels} shows the measured effective widths for
every adjacent colour in GAMA. The figure shows that the LDR colours are equivalent to (within the 0.01mag density bandwidth), or narrower than,
the PDR colours across the entire dataset, with the exceptions of the FUV-NUV, W2-W3, and
W3-W4 colours. We note that these three colours correlate with the bands containing the strongest sigma-cuts in the PDR data (see Figure \ref{fig: coverage}),
meaning that the distribution
of colours will likely be artificially narrow due to matching bias; objects with fluxes below the sigma-limit of the catalogue are incorrectly matched to contaminating
objects with fluxes above the sigma-limit. This effect is prominent when matching data that has been heavily sigma-cut to data of greater depth, and is exacerbated when
the sigma-cut data traces a fundamentally different range of populations to the deeper data, as is the case in the GAMA UV and MIR data. As the effect only works in
one direction (\ie low sigma source fluxes are replaced by high sigma contaminant fluxes, but never vice-versa), the result on the colour distribution is the removal
of noisy measurements, and replacement with strong detections. Given this effect, and assuming that the colour distribution of contaminating sources lies within the
limits of the distribution of target colours, the effect will cause a reduction in the effective width of the colour distribution when compared with the unbiased
distribution.
 We note that this effect can only occur when comparing photometry where one dataset has been modelled (as in the case of {\sc sdss model} magnitudes) rather than
measured directly, or where one data-set is subject to strong signal-to-noise selection. As this is not the case in the LDR photometric dataset, this cannot explain the
reduction in scatter that we see in each of our colour distributions.

The outlier fraction is similarly informative as it details the number of catastrophic outliers in the colour distribution. We define the
outlier fraction as the percentage of objects that are more than 0.5 dex beyond the 10\% and 90\% limits of the colour distribution. These points are
shown graphically in Figure \ref{fig: colours} as vertical dotted lines coloured for each distribution. Furthermore,
we show the outlier fraction measured for every adjacent colour
in GAMA in Figure \ref{fig: stacked panels}. The figure shows that the number of outliers is lower in each of the LDR colour distributions
compared to the PDR distributions, again with the exception {\bbf of the W3-W4 colour (for the same reason as above). We also see a modest increase
in outlier fraction in the internal {\sc spire} colours. As the methods for determining photometry in {\sc spire} are similar in the LDR and PDR
datasets, this increase can likely be attributed to differences in the choice of contaminant list. This is because the method is known to be similar,
and the colour distributions themselves are of equal width (indicating that the majority of sources are in agreement).
A detailed comparison of LDR photometry with individually tailored {\sc pacs} and {\sc spire} photometry
is provided in \cite{Valiante2016}.}  Most importantly, we note that
the outlier fractions at the boundaries between facilities (shown as vertical dotted lines) in the LDR are consistently lower than or equal to the
PDR, demonstrating that the consistent measurement of photometry is having a major impact in these colours.

\begin{figure}
\centering
\includegraphics[scale=0.33]{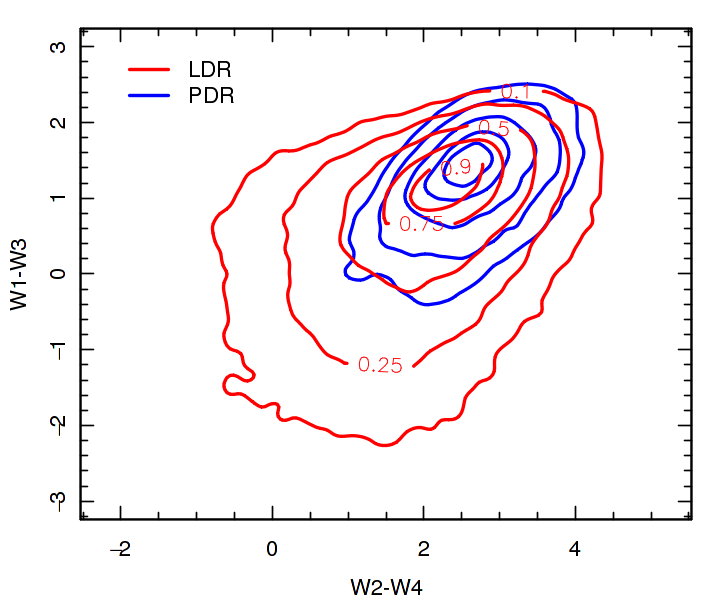}
\caption{Demonstration of the increased range of spectral colours that are able to be probed when performing matched aperture photometry, as bias against
weak emission is removed. Here we show how, in the Mid-IR, an increase in the catalogue depth allows us to push further into the colour-colour space, when
compared to the sigma-clipped measurements previously available. Contours are numbered by the fraction of sources outside the contour. The blue contours
show the distribution of PDR colours, while the red contours show the full distribution of LDR colours.
As the MIR data are useful for object classifications (see, \eg Figure 5 of
\protect\cite{Cluver2014}), we can determine that this expanded parameter-space is populated primarily by a range of morphological types, from
high-mass, low star formation rate ellipticals, to low-mass, moderately star-forming disk systems.}\label{fig: wise colours}
\end{figure}

Finally, it is particularly interesting to examine what new parameter space is opened for analysis when performing this sort of consistent matched aperture
photometry, compared to what was available previously in the PDR. Specifically, the far greater depth of measurement in the MIR allows us to explore the
highly important PAH emission, which traces (among other things) the hot emission from dust surrounding stellar nurseries.
By probing to lower fluxes, we open up a greater parameter space for investigation in this region. We demonstrate this in Figure \ref{fig: wise colours},
where the additional measurements made by \lambdar\ have pushed out down and left in the parameter space shown. We note that this increase in parameter
space cannot be explained by adding additional scatter to an additional sample of detections drawn from the same distribution as that in the PDR, as this
effect would cause a uniform broadening in all directions. In contrast, what we see is a distinct increase in parameter space in two directions, while the
other two colour-boundaries remain well defined.

%}}}

\subsection{The Final Product}%{{{
\begin{figure*}
\includegraphics[scale=0.4]{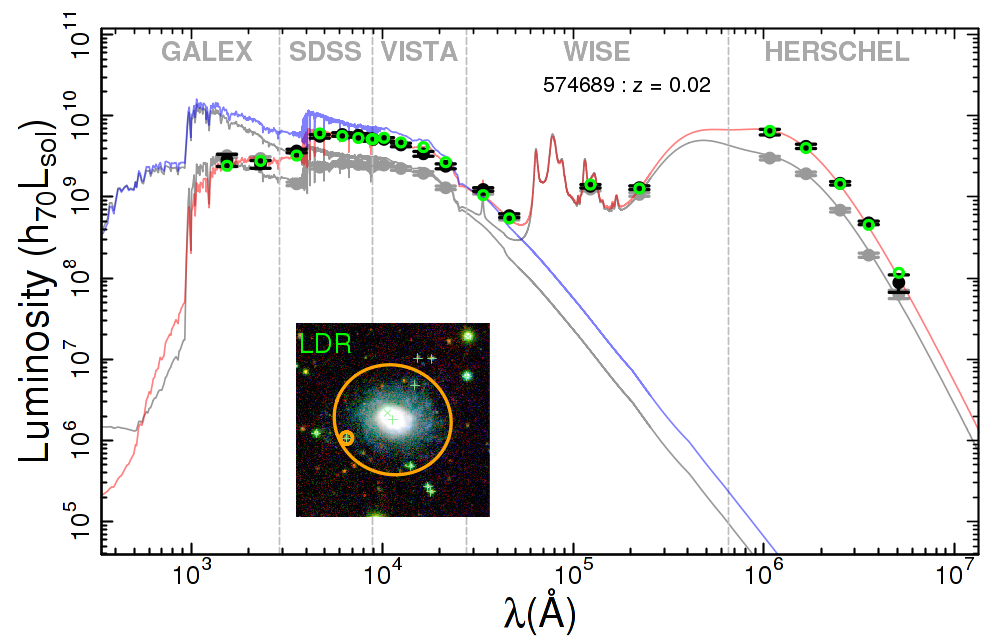}
\caption{{\bbf The panchromatic SEDs of GAMA object G574689. The grey SED is as determined
when using photometry from the PDR catalogue (\ie the same as Figure \ref{fig: bad aperture}), while the
coloured lines show the SED fit to photometry returned by \lambdar.
Note that after our procedure,
the aperture used for all bands is consistent (shown in the inset image) and the photometry is therefore also consistent.
Here the LDR photometry is in black, model
photometry is in green, unobscured SED is in blue, and the obscured SED is in red. As with Figure \ref{fig: bad aperture},
the inset is an RGB cutout using the {\sc viking} H - {\sc sdss} i - {\sc sdss} g bands.}
}\label{fig: fixedSED}
\end{figure*}

We began this paper with a demonstration of an object whose photometry was inconsistent across the full GAMA bandpass. It
would be remiss to not then demonstrate at the conclusion of the paper that this process had not, at the very least, been successful in
producing consistent panchromatic photometry in this case. As such, Figure \ref{fig: fixedSED} shows the photometry for this object, as
measured by \lambdar, with a fit performed by the energy-balance code {\sc magphys} \citep{daCunha2008,daCunha2011}.
We see that {\sc magphys} has been able to produce a better fit to the panchromatic SED, indicating that \lambdar\ has produced
photometry (and uncertainties) that are more consistent across the entire bandpass, and thus the SED fit is now a much more reliable representation
of the object's true panchromatic emission.

Comparing the fits from PDR and LDR, we find that the LDR SED is a better fit to the data $\chi^2_{\rm best} = 11.69 \rightarrow 2.11$. Taking each of the
output parameters (with indicated $1\sigma$ intervals) entirely at face-value, the LDR SED shows an
older system $t_{\rm form} (Gyr) = 9.997{\pm NA} \rightarrow 9.182^{+0.135}_{-0.000} $,
but whose mass-weighted and luminosity-weighted
ages are both younger ${\rm age}_m (Gyr) = 9.80{\pm NA} \rightarrow 8.79^{+0.29}_{-0.00}$ and ${\rm age}_r (Gyr) = 9.45{\pm NA}
\rightarrow 8.85^{+0.30}_{-0.00} $. The SED is
dustier $\log_{10} (M_D/M_\odot) = 6.81{\pm NA} \rightarrow 7.10^{+0.09}_{-0.03}$, but has maintained an equivalent stellar mass
$\log_{10} (M_*/M_\odot) = 9.507{\pm NA} \rightarrow 9.417^{+0.115}_{-0.005}$.
Bursts of star formation have been less recent $t_{\rm last burst} (Gyr) = 8.00{\pm NA} \rightarrow 8.473^{+0.88}_{-0.00} $, but the overall star formation
has been more sustained, as shown by a lower star formation timescale $\gamma (Gyr^{-1}) = 0.28{\pm NA} \rightarrow 0.14^{+0.00}_{-0.00}$ which determines
the overall star formation rate as a function of time ${\rm SFR}(t) \propto e^{-\gamma t}$ (neglecting bursts). Note that the $1\sigma$ uncertainties on the PDR SED
parameters are uniformly $\pm NA$; these have not been
forgotten, but rather are all not calculable. This is because the fit has been forced into an area of parameter space where there is limited modelling, meaning that the
PDF effectively becomes delta-function-like. In contrast, the LDR SED provides errors that are typically bound within an non-zero interval on one or both sides.

Full SED analysis of all galaxies in GAMA is left for an upcoming publication.

%}}}
%}}}

\section{Updating GAMA Photometry: Comparing Derived Properties}\label{sec: comparison to PDR II}%{{{
In addition to examining the change in the flux of measured sources, we also investigate how the new photometry
impacts the measurement of some particular properties of interest. Specifically, we examine how the photometry impacts the
measurement of stellar masses, star-formation rates, and the stellar-mass to star-formation-rate relation.

To begin, we estimate the stellar mass of every galaxy by fitting SEDs across the optical and NIR bands,
as described in \cite{Taylor2011}, for both the PDR and \lambdar\ photometry. The median residual between the stellar masses
for the two datasets is $-0.004\pm0.030$ dex, which is both consistent with no difference and
much lower than the median uncertainty of both datasets, which is $0.108$ and $0.111$ for the
PDR and LDR masses respectively. This is not surprising given the consistency in colours and fluxes across the optical and NIR bands, which are used
to estimate the stellar masses here.

Utilising the MIR and FIR data, we can estimate a star formation rate using physically motivated predictors. As a demonstration, we examine predictors using
both the {\sc \wise\ w4} and {\em Herschel} {\sc pacs} $100\mu$m band, for both the LDR and PDR dataset. We use luminosity-based SFR indicators for the W4 and $100\mu$m bands,
with the $100\mu$m indicator coming directly from \cite{Davies2016}, and the W4 indicator being derived in the same way.
Having determined stellar masses and SFR estimates for both the LDR and PDR datasets,
we can investigate how well the two datasets are able to recover the main sequence of
star-forming galaxies.

To measure the relation, we use the {\sc r} multi-dimensional Markov-Chain Monte-Carlo fitting package {\tt hyper.fit\footnote{\url{https://github.com/asgr/hyper.fit} }}
\citep{Robotham2015}. Without placing any selection
criteria on the data, other than the sigma-cut in W4 flux already implicit in the PDR dataset, we fit a linear relation of the form
${\rm log_{10}\left(SFR\right)} = \alpha {\rm log_{10} \left( M_\star\right)} + \beta$ to the distribution of stellar mass vs SFR,
using a Componentwise Hit-and-Run Metropolis (CHARM) MCMC optimisation. Using this method, we find a best-fit linear relationship for the PDR and LDR photometry. Parameters
of each of these fits are given in Table \ref{tab: linfits}. Included in these parameters is a value of $\sigma_{\rm orth}$ per fit, which is the intrinsic scatter
orthogonal to the best fit line.

Assuming that the relation is in some sense fundamental (\ie physically motivated), we are
able to argue (as we did in Section \ref{sec: colours}) that any reduction in the intrinsic scatter of the fit represents an improvement in the photometry
used in determining the fit components. We see a significant reduction in the intrinsic scatter of the fit for the W4 predictor when using the LDR photometry,
and see a consistent scatter when using the $100\mu$m predictor. However, we note that if we sigma-clip both datasets to $2\sigma$, thus decreasing the impact
of low significance and possibly spurious measurements, the intrinsic scatter about the fit for the PDR data increases significantly while
the LDR intrinsic scatter remains relatively consistent.
As such, we conclude that the {\sc w4} and $100\mu$m photometry in the LDR are both an improvement over the PDR. Nonetheless,
the LDR and PDR datasets return equivalent
relationships for each predictor. Note however, the substantial improvement in the number of measurements in the MIR
means that here we are able to increase our sample from $29,764$ estimates in the PDR to $127,524$ estimates in the LDR.

\begin{table}
\begin{tabular}{cccc}
\hline
Sample & $\alpha$ & $\beta$ & $\sigma_{\rm orth}$ \\
\hline
PDR $100\mu$m  & $0.62\pm0.05$    & $-5.69\pm0.56$ & $0.225\pm0.002$  \\
LDR $100\mu$m  & $0.61\pm0.04$    & $-5.56\pm0.36$ & $0.214\pm0.002$  \\
PDR {\sc w4}   & $0.75\pm0.06$    & $-6.98\pm0.68$ & $0.278\pm0.005$  \\
LDR {\sc w4}   & $0.75\pm0.08$    & $-7.06\pm0.85$ & $0.226\pm0.003$  \\
\hline
PDR($2\sigma$) $100\mu$m  & $0.62\pm0.04$    & $-5.53\pm0.43$ & $0.243\pm0.004$  \\
LDR($2\sigma$) $100\mu$m  & $0.60\pm0.02$    & $-5.37\pm0.23$ & $0.219\pm0.002$  \\
\hline
\end{tabular}
\caption{Fit parameters for the linear relationship between stellar mass and star formation rate, for both the PDR and LDR datasets, when deriving SFRs using
predictors based on $100\mu$m and {\sc w4} fluxes. The upper section of the table shows the fit to all available data, while the lower panel shows the fits when
fitting to data with measurements $\ge 2\sigma$. We can see that in each case, the LDR fits and PDR fits are equivalent, but the LDR fit has equivalent or reduced intrinsic
scatter. We therefore conclude that the LDR data are a more appropriate representation of the true underlying distribution. }\label{tab: linfits}
\end{table}
%}}}

\section{Updating GAMA Photometry: Data Release}\label{sec: Data Release}%{{{
In addition to releasing the program, we also release the various data-products that the program outputs for all galaxies in
the GAMA equatorial fields. The release is in the form of 24 machine-readable files (.csv), and is accessible via the GAMA Panchromatic
Swarp Imager ($\Psi$) website; \url{http://gama-psi.icrar.org}. The 24 files are:
\begin{itemize}
\item A summary file containing final photometry and uncertainties for all optically defined targets across all 21-bands of photometry;
\item Three input catalogues, containing the optical prior aperture information and contaminant lists, as described in Section \ref{sec: Catalogues};
\item 21 individual files containing details specific to the 21-bands in which photometry was measured.
\end{itemize}
The 21 files containing band-specific information each contain 50 columns, containing information about every objects' sky estimate, blanks measurement, deblend
solution, flux measurement, flux iteration, aperture normalisation, and any photometry warnings.

%}}}

\section{Conclusions}\label{sec: conclusions}%{{{
In this paper, we have presented a novel program for determining matched aperture photometry across images that are neither pixel- nor PSF-matched.
The program is sophisticated enough to reliably analyse imaging from the Far-UV to the Far-IR, and produces a substantial number of
data products to aid in photometric analysis, quality control, and error handling. We demonstrate that the program is able to return simulated
photometric values in both the high SNR, low confusion regime, as well as in the low SNR, high confusion regime.
We further demonstrate that the many available subroutines within the program, including (but not limited to) local sky estimation, blanks/randoms
correction, object deblending, and iterative flux measurement, behave well in all tested cases.

We run the program over 21-bands of photometry contained within the GAMA survey, and present comparisons between the photometry returned by the program to
those in the GAMA Panchromatic Data Release (PDR; \citealt{Driver2016}). We demonstrate that the photometry is both broadly consistent with what has come
previously, while still being an improvement over previous photometry, as determined by a decrease in the relative widths of colour distributions across facility boundaries,
an increase in the number of measurements, and greater consistency and reliability of uncertainties.

By fitting spectral energy distributions to the optical and near-IR photometry, we are able to measure stellar masses for all galaxies in our sample.
We compare stellar mass estimates derived from the GAMA PDR photometry to those derived from the \lambdar\ photometry, finding median residual between the
mass estimates of $-0.004\pm0.030$ dex.

Using the \lambdar\ program, we are able to increase the rate of measurements in low sensitivity images by forcing photometric measurements at optically motivated
positions. Using the program, for example, we make measurements in the \wise\ W4 band at the position of every GAMA target. The result is an increase in the number
of measurements, but also a systematic increase in the range of colours able to be probed in the \wise\ bands.

Using these stellar mass estimates and star formation rate indicators derived from {\em Herschel} {\sc pacs} $100\mu$m and {\sc \wise\ w4} luminosities, we measure a linear
fit to the star formation rate main sequence using the {\sc r} multi-dimensional Markov-Chain Monte-Carlo fitting package {\tt hyper.fit} \citep{Robotham2015}.
Comparing the relations we
derive using each predictor, for both the PDR and LDR datasets, we find good agreement. We note, however, that the relation derived using
the LDR dataset demonstrates a decrease in the intrinsic scatter about the star formation rate main sequence, indicating a reduction in random errors.

{\bbf From these tests, we conclude that the \lambdar\ photometry is indeed superior to that derived by table matching. }

Finally, we detail the data release to accompany this paper. Photometry measured using \lambdar\ has been made available through the GAMA
Panchromatic Swarp Imager ($\Psi$) website; \url{http://gama-psi.icrar.org/}, along with many relevant sub-products detailed here. These include sky estimates, deblend
fractions, normalisation factors, and more.
%}}}

\section{Acknowledgements} %{{{
{\bbf We thank the anonymous referee for a thorough reading of the paper and for their many
constructive comments. }
AHW and SKA are supported by the Australian Government's Department of Industry
Australian Postgraduate Award (APA). SB acknowledges funding support from the Australian Research Council
through a Future Fellowship (FT140101166).
LD and SJM acknowledge support from the ERC in the form of the Advanced Investigator Program, COSMICISM,
and the ERC Consolidator Grant CosmicDust.
NB acknowledges funding from the European Union Seventh Framework Programme (FP7/2007-2013) under grant agreement no. 312725.
GAMA is a joint European-Australasian project based around a spectroscopic campaign using the AAT. The
GAMA IC is based on data taken from the SDSS and the UKIRT Infrared Deep Sky Survey. Complementary imaging of the
GAMA regions is being obtained by a number of independent survey programmes including \galex\ MIS,
VST KiDS, VISTA VIKING, \wise, {\em Herschel}-ATLAS, GMRT and ASKAP providing UV to radio coverage. GAMA is funded
by the STFC (UK), the ARC (Australia), the AAO, and the participating institutions. The GAMA website is
\url{http://www.gama-survey.org/}. The {\em Herschel}-ATLAS is a project with Herschel, which is an ESA space observatory
with science instruments provided by European-led Principal Investigator consortia and with important participation from
NASA. The H-ATLAS website is \url{http://www.h-atlas.org/}. {\bbf Figures in this paper have been prepared using
the {\sc r} package {\tt magicaxis\footnote{\url{https://cran.r-project.org/package=magicaxis}}.}}
This research has made use of NASA's Astrophysics Data System.
%}}}
\bibliographystyle{mnras}
\bibliography{Wright_LAMBDAR_v2}

\begin{thebibliography}{}
\makeatletter
\relax
\def\mn@urlcharsother{\let\do\@makeother \do\$\do\&\do\#\do\^\do\_\do\%\do\~}
\def\mn@doi{\begingroup\mn@urlcharsother \@ifnextchar [ {\mn@doi@}
  {\mn@doi@[]}}
\def\mn@doi@[#1]#2{\def\@tempa{#1}\ifx\@tempa\@empty \href
  {http://dx.doi.org/#2} {doi:#2}\else \href {http://dx.doi.org/#2} {#1}\fi
  \endgroup}
\def\mn@eprint#1#2{\mn@eprint@#1:#2::\@nil}
\def\mn@eprint@arXiv#1{\href {http://arxiv.org/abs/#1} {{\tt arXiv:#1}}}
\def\mn@eprint@dblp#1{\href {http://dblp.uni-trier.de/rec/bibtex/#1.xml}
  {dblp:#1}}
\def\mn@eprint@#1:#2:#3:#4\@nil{\def\@tempa {#1}\def\@tempb {#2}\def\@tempc
  {#3}\ifx \@tempc \@empty \let \@tempc \@tempb \let \@tempb \@tempa \fi \ifx
  \@tempb \@empty \def\@tempb {arXiv}\fi \@ifundefined
  {mn@eprint@\@tempb}{\@tempb:\@tempc}{\expandafter \expandafter \csname
  mn@eprint@\@tempb\endcsname \expandafter{\@tempc}}}

\bibitem[\protect\citeauthoryear{{Abazajian} et~al.,}{{Abazajian}
  et~al.}{2009}]{Abazajian2009}
{Abazajian} K.~N.,  et~al., 2009, \mn@doi [\apjs]
  {10.1088/0067-0049/182/2/543}, \href
  {http://adsabs.harvard.edu/abs/2009ApJS..182..543A} {182, 543}

\bibitem[\protect\citeauthoryear{{Andrae}}{{Andrae}}{2014}]{Andrae2014}
{Andrae} E. e.~a.,  2014, \phdt, p. MPIfK

\bibitem[\protect\citeauthoryear{{Baldry} et~al.,}{{Baldry}
  et~al.}{2012}]{Baldry2012}
{Baldry} I.~K.,  et~al., 2012, \mn@doi [\mnras]
  {10.1111/j.1365-2966.2012.20340.x}, \href
  {http://adsabs.harvard.edu/abs/2012MNRAS.421..621B} {421, 621}

\bibitem[\protect\citeauthoryear{{Bertin} \& {Arnouts}}{{Bertin} \&
  {Arnouts}}{1996}]{Bertin1996}
{Bertin} E.,  {Arnouts} S.,  1996, \mn@doi [\aaps] {10.1051/aas:1996164}, \href
  {http://adsabs.harvard.edu/abs/1996A%26AS..117..393B} {117, 393}

\bibitem[\protect\citeauthoryear{{Bertin}, {Mellier}, {Radovich}, {Missonnier},
  {Didelon}  \& {Morin}}{{Bertin} et~al.}{2002}]{Bertin2002}
{Bertin} E.,  {Mellier} Y.,  {Radovich} M.,  {Missonnier} G.,  {Didelon} P.,
  {Morin} B.,  2002, in {Bohlender} D.~A.,  {Durand} D.,   {Handley} T.~H.,
  eds,  Astronomical Society of the Pacific Conference Series Vol. 281,
  Astronomical Data Analysis Software and Systems XI. p.~228

\bibitem[\protect\citeauthoryear{{Boquien} et~al.,}{{Boquien}
  et~al.}{2013}]{Boquien2013}
{Boquien} M.,  et~al., 2013, \mn@doi [\aap] {10.1051/0004-6361/201220768},
  \href {http://adsabs.harvard.edu/abs/2013A%26A...554A..14B} {554, A14}

\bibitem[\protect\citeauthoryear{{Bourne} et~al.,}{{Bourne}
  et~al.}{2012}]{Bourne2012}
{Bourne} N.,  et~al., 2012, \mn@doi [\mnras]
  {10.1111/j.1365-2966.2012.20528.x}, \href
  {http://adsabs.harvard.edu/abs/2012MNRAS.421.3027B} {421, 3027}

\bibitem[\protect\citeauthoryear{{Bourne et al.}}{{Bourne et
  al.}}{2016}]{Bourne2016}
{Bourne et al.} 2016, accepted

\bibitem[\protect\citeauthoryear{{Brown} et~al.,}{{Brown}
  et~al.}{2014}]{Brown2014}
{Brown} M.~J.~I.,  et~al., 2014, \mn@doi [\apjs] {10.1088/0067-0049/212/2/18},
  \href {http://adsabs.harvard.edu/abs/2014ApJS..212...18B} {212, 18}

\bibitem[\protect\citeauthoryear{{Bundy}, {Hogg}, {Higgs}, {Nichol}, {Yasuda},
  {Masters}, {Lang}  \& {Wake}}{{Bundy} et~al.}{2012}]{Bundy2012}
{Bundy} K.,  {Hogg} D.~W.,  {Higgs} T.~D.,  {Nichol} R.~C.,  {Yasuda} N.,
  {Masters} K.~L.,  {Lang} D.,   {Wake} D.~A.,  2012, \mn@doi [\aj]
  {10.1088/0004-6256/144/6/188}, \href
  {http://adsabs.harvard.edu/abs/2012AJ....144..188B} {144, 188}

\bibitem[\protect\citeauthoryear{{Cameron}}{{Cameron}}{2011}]{Cameron2011}
{Cameron} E.,  2011, \mn@doi [\pasa] {10.1071/AS10046}, \href
  {http://adsabs.harvard.edu/abs/2011PASA...28..128C} {28, 128}

\bibitem[\protect\citeauthoryear{{Camps} \& {Baes}}{{Camps} \&
  {Baes}}{2015}]{Camps2015}
{Camps} P.,  {Baes} M.,  2015, \mn@doi [Astronomy and Computing]
  {10.1016/j.ascom.2014.10.004}, \href
  {http://adsabs.harvard.edu/abs/2015A%26C.....9...20C} {9, 20}

\bibitem[\protect\citeauthoryear{{Capak} et~al.,}{{Capak}
  et~al.}{2007}]{Capak2007}
{Capak} P.,  et~al., 2007, \mn@doi [\apjs] {10.1086/519081}, \href
  {http://adsabs.harvard.edu/abs/2007ApJS..172...99C} {172, 99}

\bibitem[\protect\citeauthoryear{{Cluver} et~al.,}{{Cluver}
  et~al.}{2014}]{Cluver2014}
{Cluver} M.~E.,  et~al., 2014, \mn@doi [\apj] {10.1088/0004-637X/782/2/90},
  \href {http://adsabs.harvard.edu/abs/2014ApJ...782...90C} {782, 90}

\bibitem[\protect\citeauthoryear{{Conroy}}{{Conroy}}{2013}]{Conroy2013}
{Conroy} C.,  2013, \mn@doi [\araa] {10.1146/annurev-astro-082812-141017},
  \href {http://adsabs.harvard.edu/abs/2013ARA%26A..51..393C} {51, 393}

\bibitem[\protect\citeauthoryear{{Da Cunha} \& {Charlot}}{{Da Cunha} \&
  {Charlot}}{2011}]{daCunha2011}
{Da Cunha} E.,  {Charlot} S.,  2011, {MAGPHYS: Multi-wavelength Analysis of
  Galaxy Physical Properties}, Astrophysics Source Code Library (\mn@eprint
  {ascl} {1106.010})

\bibitem[\protect\citeauthoryear{{Da Cunha}, {Charlot}  \& {Elbaz}}{{Da Cunha}
  et~al.}{2008}]{daCunha2008}
{Da Cunha} E.,  {Charlot} S.,   {Elbaz} D.,  2008, \mn@doi [\mnras]
  {10.1111/j.1365-2966.2008.13535.x}, \href
  {http://adsabs.harvard.edu/abs/2008MNRAS.388.1595D} {388, 1595}

\bibitem[\protect\citeauthoryear{{Davies} et~al.,}{{Davies}
  et~al.}{2015}]{Davies2015}
{Davies} L.~J.~M.,  et~al., 2015, \mn@doi [\mnras] {10.1093/mnras/stv1241},
  \href {http://adsabs.harvard.edu/abs/2015MNRAS.452..616D} {452, 616}

\bibitem[\protect\citeauthoryear{{Davies} et~al.,}{{Davies}
  et~al.}{2016}]{Davies2016}
{Davies} L.~J.~M.,  et~al., 2016, \mn@doi [\mnras] {10.1093/mnras/stv2573},
  \href {http://adsabs.harvard.edu/abs/2016MNRAS.455.4013D} {455, 4013}

\bibitem[\protect\citeauthoryear{{De Santis}, {Grazian}, {Fontana}  \&
  {Santini}}{{De Santis} et~al.}{2007}]{deSantis2007}
{De Santis} C.,  {Grazian} A.,  {Fontana} A.,   {Santini} P.,  2007, \mn@doi
  [New.~Astron.] {10.1016/j.newast.2006.10.004}, \href
  {http://adsabs.harvard.edu/abs/2007NewA...12..271D} {12, 271}

\bibitem[\protect\citeauthoryear{{De Vaucouleurs}}{{De
  Vaucouleurs}}{1948}]{deVaucouleurs1948}
{De Vaucouleurs} G.,  1948, Annales d'Astrophysique, \href
  {http://adsabs.harvard.edu/abs/1948AnAp...11..247D} {11, 247}

\bibitem[\protect\citeauthoryear{{Driver} et~al.,}{{Driver}
  et~al.}{2011}]{Driver2011}
{Driver} S.~P.,  et~al., 2011, \mn@doi [\mnras]
  {10.1111/j.1365-2966.2010.18188.x}, \href
  {http://adsabs.harvard.edu/abs/2011MNRAS.413..971D} {413, 971}

\bibitem[\protect\citeauthoryear{{Driver} et~al.,}{{Driver}
  et~al.}{2016}]{Driver2016}
{Driver} S.~P.,  et~al., 2016, \mn@doi [\mnras] {10.1093/mnras/stv2505}, \href
  {http://adsabs.harvard.edu/abs/2016MNRAS.455.3911D} {455, 3911}

\bibitem[\protect\citeauthoryear{{Dunne} et~al.,}{{Dunne}
  et~al.}{2011}]{Dunne2011}
{Dunne} L.,  et~al., 2011, \mn@doi [\mnras] {10.1111/j.1365-2966.2011.19363.x},
  \href {http://adsabs.harvard.edu/abs/2011MNRAS.417.1510D} {417, 1510}

\bibitem[\protect\citeauthoryear{{Eales} et~al.,}{{Eales}
  et~al.}{2010}]{Eales2010}
{Eales} S.,  et~al., 2010, \mn@doi [\pasp] {10.1086/653086}, \href
  {http://adsabs.harvard.edu/abs/2010PASP..122..499E} {122, 499}

\bibitem[\protect\citeauthoryear{{Elbaz} et~al.,}{{Elbaz}
  et~al.}{2011}]{Elbaz2011}
{Elbaz} D.,  et~al., 2011, \mn@doi [\aap] {10.1051/0004-6361/201117239}, \href
  {http://adsabs.harvard.edu/abs/2011A%26A...533A.119E} {533, A119}

\bibitem[\protect\citeauthoryear{{Erwin}}{{Erwin}}{2014}]{Erwin2014}
{Erwin} P.,  2014, {Imfit: A Fast, Flexible Program for Astronomical Image
  Fitting}, Astrophysics Source Code Library (\mn@eprint {ascl} {1408.001})

\bibitem[\protect\citeauthoryear{{Freeman}}{{Freeman}}{1970}]{Freeman1970}
{Freeman} K.~C.,  1970, \mn@doi [\apj] {10.1086/150474}, \href
  {http://adsabs.harvard.edu/abs/1970ApJ...160..811F} {160, 811}

\bibitem[\protect\citeauthoryear{{Graham}, {Driver}, {Petrosian}, {Conselice},
  {Bershady}, {Crawford}  \& {Goto}}{{Graham} et~al.}{2005}]{Graham2005b}
{Graham} A.~W.,  {Driver} S.~P.,  {Petrosian} V.,  {Conselice} C.~J.,
  {Bershady} M.~A.,  {Crawford} S.~M.,   {Goto} T.,  2005, \mn@doi [\aj]
  {10.1086/444475}, \href {http://adsabs.harvard.edu/abs/2005AJ....130.1535G}
  {130, 1535}

\bibitem[\protect\citeauthoryear{{Griffin} et~al.,}{{Griffin}
  et~al.}{2010}]{Griffin2010}
{Griffin} M.~J.,  et~al., 2010, \mn@doi [\aap] {10.1051/0004-6361/201014519},
  \href {http://adsabs.harvard.edu/abs/2010A%26A...518L...3G} {518, L3}

\bibitem[\protect\citeauthoryear{{Grogin} et~al.,}{{Grogin}
  et~al.}{2011}]{Grogin2011}
{Grogin} N.~A.,  et~al., 2011, \mn@doi [\apjs] {10.1088/0067-0049/197/2/35},
  \href {http://adsabs.harvard.edu/abs/2011ApJS..197...35G} {197, 35}

\bibitem[\protect\citeauthoryear{{Hildebrandt} et~al.,}{{Hildebrandt}
  et~al.}{2012}]{Hildebrandt2012}
{Hildebrandt} H.,  et~al., 2012, \mn@doi [\mnras]
  {10.1111/j.1365-2966.2012.20468.x}, \href
  {http://adsabs.harvard.edu/abs/2012MNRAS.421.2355H} {421, 2355}

\bibitem[\protect\citeauthoryear{{Hill}, {Driver}, {Cameron}, {Cross}, {Liske}
  \& {Robotham}}{{Hill} et~al.}{2010}]{Hill2010}
{Hill} D.~T.,  {Driver} S.~P.,  {Cameron} E.,  {Cross} N.,  {Liske} J.,
  {Robotham} A.,  2010, \mn@doi [\mnras] {10.1111/j.1365-2966.2010.16374.x},
  \href {http://adsabs.harvard.edu/abs/2010MNRAS.404.1215H} {404, 1215}

\bibitem[\protect\citeauthoryear{{Hill} et~al.,}{{Hill}
  et~al.}{2011}]{Hill2011}
{Hill} D.~T.,  et~al., 2011, \mn@doi [\mnras]
  {10.1111/j.1365-2966.2010.17950.x}, \href
  {http://adsabs.harvard.edu/abs/2011MNRAS.412..765H} {412, 765}

\bibitem[\protect\citeauthoryear{{Jarrett}, {Chester}, {Cutri}, {Schneider},
  {Skrutskie}  \& {Huchra}}{{Jarrett} et~al.}{2000}]{Jarrett2000}
{Jarrett} T.~H.,  {Chester} T.,  {Cutri} R.,  {Schneider} S.,  {Skrutskie} M.,
   {Huchra} J.~P.,  2000, \mn@doi [\aj] {10.1086/301330}, \href
  {http://adsabs.harvard.edu/abs/2000AJ....119.2498J} {119, 2498}

\bibitem[\protect\citeauthoryear{{Jarrett} et~al.,}{{Jarrett}
  et~al.}{2012}]{Jarrett2012}
{Jarrett} T.~H.,  et~al., 2012, \mn@doi [\aj] {10.1088/0004-6256/144/2/68},
  \href {http://adsabs.harvard.edu/abs/2012AJ....144...68J} {144, 68}

\bibitem[\protect\citeauthoryear{{Jarrett} et~al.,}{{Jarrett}
  et~al.}{2013}]{Jarrett2013}
{Jarrett} T.~H.,  et~al., 2013, \mn@doi [\aj] {10.1088/0004-6256/145/1/6},
  \href {http://adsabs.harvard.edu/abs/2013AJ....145....6J} {145, 6}

\bibitem[\protect\citeauthoryear{{Kelvin} et~al.,}{{Kelvin}
  et~al.}{2012}]{Kelvin2012}
{Kelvin} L.~S.,  et~al., 2012, \mn@doi [\mnras]
  {10.1111/j.1365-2966.2012.20355.x}, \href
  {http://adsabs.harvard.edu/abs/2012MNRAS.421.1007K} {421, 1007}

\bibitem[\protect\citeauthoryear{{Kelvin} et~al.,}{{Kelvin}
  et~al.}{2014}]{Kelvin2014}
{Kelvin} L.~S.,  et~al., 2014, \mn@doi [\mnras] {10.1093/mnras/stt2391}, \href
  {http://adsabs.harvard.edu/abs/2014MNRAS.439.1245K} {439, 1245}

\bibitem[\protect\citeauthoryear{{Kron}}{{Kron}}{1980}]{Kron1980}
{Kron} R.~G.,  1980, \mn@doi [\apjs] {10.1086/190669}, \href
  {http://adsabs.harvard.edu/abs/1980ApJS...43..305K} {43, 305}

\bibitem[\protect\citeauthoryear{{Kuijken}}{{Kuijken}}{2008}]{Kuijken2008}
{Kuijken} K.,  2008, \mn@doi [\aap] {10.1051/0004-6361:20066601}, \href
  {http://adsabs.harvard.edu/abs/2008A%26A...482.1053K} {482, 1053}

\bibitem[\protect\citeauthoryear{{Lacey} et~al.,}{{Lacey}
  et~al.}{2015}]{Lacey2015}
{Lacey} C.~G.,  et~al., 2015, preprint, \href
  {http://adsabs.harvard.edu/abs/2015arXiv150908473L} {} (\mn@eprint {arXiv}
  {1509.08473})

\bibitem[\protect\citeauthoryear{{Laidler} et~al.,}{{Laidler}
  et~al.}{2007}]{Laidler2007}
{Laidler} V.~G.,  et~al., 2007, \mn@doi [\pasp] {10.1086/523898}, \href
  {http://adsabs.harvard.edu/abs/2007PASP..119.1325L} {119, 1325}

\bibitem[\protect\citeauthoryear{{Liske}, {Lemon}, {Driver}, {Cross}  \&
  {Couch}}{{Liske} et~al.}{2003}]{Liske2003}
{Liske} J.,  {Lemon} D.~J.,  {Driver} S.~P.,  {Cross} N.~J.~G.,   {Couch}
  W.~J.,  2003, \mn@doi [\mnras] {10.1046/j.1365-8711.2003.06826.x}, \href
  {http://adsabs.harvard.edu/abs/2003MNRAS.344..307L} {344, 307}

\bibitem[\protect\citeauthoryear{{Liske} et~al.,}{{Liske}
  et~al.}{2015}]{Liske2015}
{Liske} J.,  et~al., 2015, \mn@doi [\mnras] {10.1093/mnras/stv1436}, \href
  {http://adsabs.harvard.edu/abs/2015MNRAS.452.2087L} {452, 2087}

\bibitem[\protect\citeauthoryear{{Madau} \& {Dickinson}}{{Madau} \&
  {Dickinson}}{2014}]{Madau2014}
{Madau} P.,  {Dickinson} M.,  2014, \mn@doi [\araa]
  {10.1146/annurev-astro-081811-125615}, \href
  {http://adsabs.harvard.edu/abs/2014ARA%26A..52..415M} {52, 415}

\bibitem[\protect\citeauthoryear{{Mancone}, {Gonzalez}, {Moustakas}  \&
  {Price}}{{Mancone} et~al.}{2013}]{Mancone2013}
{Mancone} C.~L.,  {Gonzalez} A.~H.,  {Moustakas} L.~A.,   {Price} A.,  2013,
  \mn@doi [\pasp] {10.1086/674431}, \href
  {http://adsabs.harvard.edu/abs/2013PASP..125.1514M} {125, 1514}

\bibitem[\protect\citeauthoryear{Martin, Papastergis, Giovanelli, Haynes,
  Springob  \& Stierwalt}{Martin et~al.}{2010}]{Martin2010}
Martin A.~M.,  Papastergis E.,  Giovanelli R.,  Haynes M.~P.,  Springob C.~M.,
   Stierwalt S.,  2010, The Astrophysical Journal, 723, 1359

\bibitem[\protect\citeauthoryear{{Merlin} et~al.,}{{Merlin}
  et~al.}{2015}]{Merlin2015}
{Merlin} E.,  et~al., 2015, \mn@doi [\aap] {10.1051/0004-6361/201526471}, \href
  {http://adsabs.harvard.edu/abs/2015A%26A...582A..15M} {582, A15}

\bibitem[\protect\citeauthoryear{{Morrissey} et~al.,}{{Morrissey}
  et~al.}{2007}]{Morrissey2007}
{Morrissey} P.,  et~al., 2007, \mn@doi [\apjs] {10.1086/520512}, \href
  {http://adsabs.harvard.edu/abs/2007ApJS..173..682M} {173, 682}

\bibitem[\protect\citeauthoryear{{Oliver} et~al.,}{{Oliver}
  et~al.}{2012}]{Oliver2012}
{Oliver} S.~J.,  et~al., 2012, \mn@doi [\mnras]
  {10.1111/j.1365-2966.2012.20912.x}, \href
  {http://adsabs.harvard.edu/abs/2012MNRAS.424.1614O} {424, 1614}

\bibitem[\protect\citeauthoryear{{Papastergis}, {Cattaneo}, {Huang},
  {Giovanelli}  \& {Haynes}}{{Papastergis} et~al.}{2012}]{Papastergis2012}
{Papastergis} E.,  {Cattaneo} A.,  {Huang} S.,  {Giovanelli} R.,   {Haynes}
  M.~P.,  2012, \mn@doi [\apj] {10.1088/0004-637X/759/2/138}, \href
  {http://adsabs.harvard.edu/abs/2012ApJ...759..138P} {759, 138}

\bibitem[\protect\citeauthoryear{{Patterson}}{{Patterson}}{1940}]{Patterson1940}
{Patterson} F.~S.,  1940, Harvard College Observatory Bulletin, \href
  {http://adsabs.harvard.edu/abs/1940BHarO.914....9P} {914, 9}

\bibitem[\protect\citeauthoryear{{Petrosian}}{{Petrosian}}{1976}]{Petrosian1976}
{Petrosian} V.,  1976, \mn@doi [\apjl] {10.1086/182253}, \href
  {http://adsabs.harvard.edu/abs/1976ApJ...209L...1P} {209, L1}

\bibitem[\protect\citeauthoryear{{Pilbratt} et~al.,}{{Pilbratt}
  et~al.}{2010}]{Pilbratt2010}
{Pilbratt} G.~L.,  et~al., 2010, \mn@doi [\aap] {10.1051/0004-6361/201014759},
  \href {http://adsabs.harvard.edu/abs/2010A%26A...518L...1P} {518, L1}

\bibitem[\protect\citeauthoryear{{Poglitsch} et~al.,}{{Poglitsch}
  et~al.}{2010}]{Poglitsch2010}
{Poglitsch} A.,  et~al., 2010, \mn@doi [\aap] {10.1051/0004-6361/201014535},
  \href {http://adsabs.harvard.edu/abs/2010A%26A...518L...2P} {518, L2}

\bibitem[\protect\citeauthoryear{{Popescu}, {Tuffs}, {Dopita}, {Fischera},
  {Kylafis}  \& {Madore}}{{Popescu} et~al.}{2011}]{Popescu2011}
{Popescu} C.~C.,  {Tuffs} R.~J.,  {Dopita} M.~A.,  {Fischera} J.,  {Kylafis}
  N.~D.,   {Madore} B.~F.,  2011, \mn@doi [\aap] {10.1051/0004-6361/201015217},
  \href {http://adsabs.harvard.edu/abs/2011A%26A...527A.109P} {527, A109}

\bibitem[\protect\citeauthoryear{{R Core Team}}{{R Core Team}}{2015}]{R}
{R Core Team} 2015, R: A Language and Environment for Statistical Computing.
R Foundation for Statistical Computing, Vienna, Austria, \url
  {https://www.R-project.org/}

\bibitem[\protect\citeauthoryear{{Robotham} \& {Obreschkow}}{{Robotham} \&
  {Obreschkow}}{2015}]{Robotham2015}
{Robotham} A.~S.~G.,  {Obreschkow} D.,  2015, \mn@doi [\pasa]
  {10.1017/pasa.2015.33}, \href
  {http://adsabs.harvard.edu/abs/2015PASA...32...33R} {32, e033}

\bibitem[\protect\citeauthoryear{{Scoville} et~al.,}{{Scoville}
  et~al.}{2007}]{Scoville2007}
{Scoville} N.,  et~al., 2007, \mn@doi [\apjs] {10.1086/516585}, \href
  {http://adsabs.harvard.edu/abs/2007ApJS..172....1S} {172, 1}

\bibitem[\protect\citeauthoryear{{S{\'e}rsic}}{{S{\'e}rsic}}{1963}]{Sersic1963}
{S{\'e}rsic} J.~L.,  1963, Boletin de la Asociacion Argentina de Astronomia La
  Plata Argentina, \href {http://adsabs.harvard.edu/abs/1963BAAA....6...41S}
  {6, 41}

\bibitem[\protect\citeauthoryear{{Strauss} et~al.,}{{Strauss}
  et~al.}{2002}]{Strauss2002}
{Strauss} M.~A.,  et~al., 2002, \mn@doi [\aj] {10.1086/342343}, \href
  {http://adsabs.harvard.edu/abs/2002AJ....124.1810S} {124, 1810}

\bibitem[\protect\citeauthoryear{{Sutherland} et~al.,}{{Sutherland}
  et~al.}{2015}]{Sutherland2015}
{Sutherland} W.,  et~al., 2015, \mn@doi [\aap] {10.1051/0004-6361/201424973},
  \href {http://adsabs.harvard.edu/abs/2015A%26A...575A..25S} {575, A25}

\bibitem[\protect\citeauthoryear{{Taylor} et~al.,}{{Taylor}
  et~al.}{2011}]{Taylor2011}
{Taylor} E.~N.,  et~al., 2011, \mn@doi [\mnras]
  {10.1111/j.1365-2966.2011.19536.x}, \href
  {http://adsabs.harvard.edu/abs/2011MNRAS.418.1587T} {418, 1587}

\bibitem[\protect\citeauthoryear{{Vaccari} \& {HELP Consortium}}{{Vaccari} \&
  {HELP Consortium}}{2015}]{Vaccari2015}
{Vaccari} M.,  {HELP Consortium} T.,  2015, preprint, \href
  {http://adsabs.harvard.edu/abs/2015arXiv150806444V} {} (\mn@eprint {arXiv}
  {1508.06444})

\bibitem[\protect\citeauthoryear{{Valiante et. al.}}{{Valiante et.
  al.}}{2016}]{Valiante2016}
{Valiante et. al.} 2016, accepted

\bibitem[\protect\citeauthoryear{{Vika}, {Bamford}, {H{\"a}u{\ss}ler}  \&
  {Rojas}}{{Vika} et~al.}{2013}]{Vika2013}
{Vika} M.,  {Bamford} S.,  {H{\"a}u{\ss}ler} B.,   {Rojas} A.,  2013, Memorie
  della Societa Astronomica Italiana Supplementi, \href
  {http://adsabs.harvard.edu/abs/2013MSAIS..25...41V} {25, 41}

\bibitem[\protect\citeauthoryear{{Walcher}, {Groves}, {Budav{\'a}ri}  \&
  {Dale}}{{Walcher} et~al.}{2011}]{Walcher2011}
{Walcher} J.,  {Groves} B.,  {Budav{\'a}ri} T.,   {Dale} D.,  2011, \mn@doi
  [\apss] {10.1007/s10509-010-0458-z}, \href
  {http://adsabs.harvard.edu/abs/2011Ap%26SS.331....1W} {331, 1}

\bibitem[\protect\citeauthoryear{{Wright} et~al.,}{{Wright}
  et~al.}{2010}]{Wright2010}
{Wright} E.~L.,  et~al., 2010, \mn@doi [\aj] {10.1088/0004-6256/140/6/1868},
  \href {http://adsabs.harvard.edu/abs/2010AJ....140.1868W} {140, 1868}

\bibitem[\protect\citeauthoryear{{York} et~al.,}{{York}
  et~al.}{2000}]{York2000}
{York} D.~G.,  et~al., 2000, \mn@doi [\aj] {10.1086/301513}, \href
  {http://adsabs.harvard.edu/abs/2000AJ....120.1579Y} {120, 1579}

\makeatother
\end{thebibliography}
\appendix%{{{
\section{Sky Estimates and Nebuliser}\label{sec: Nebuliser}
\begin{figure*}
\includegraphics[scale=0.50]{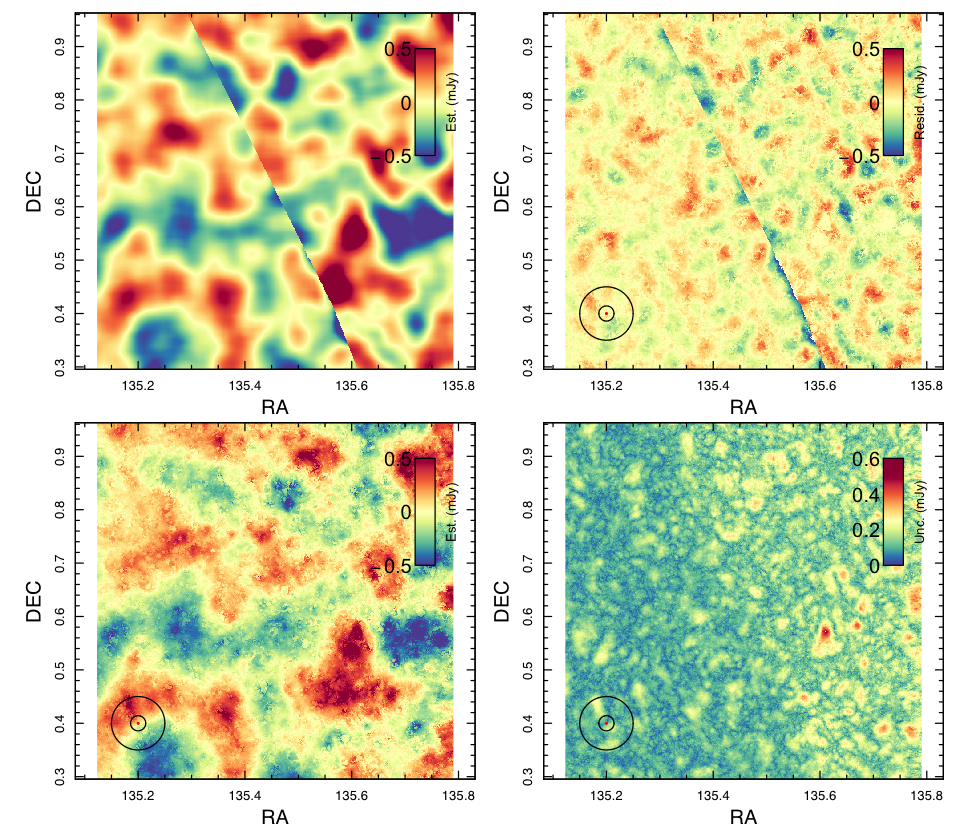}
\caption{Comparison between the small background variations in a sub-region of the G09 PACS $100\mu$m mosaic, as measured by nebuliser (top-left)
with those measured by the \lambdar\ sky-estimate routine
(bottom-left). The residual between the two estimates is shown in the top-right panel, and the uncertainty on the \lambdar\ measurement is shown in the
bottom-right. The diagonal discontinuity in the nebuliser distribution lies along the boundary of two frames. Because of the frame-overlap, there is single-depth
data to the right of the join, and double-depth data to the left of the join. This is reflected in the \lambdar\ uncertainty panel, where uncertainties are systematically
higher. In each of the \lambdar\ panels, a graphic demonstration of the annuli used in measuring the sky is shown (in black) in the bottom left.
The PSF FWHM is also shown (in red), for reference. }\label{fig: Nebuliser}
\end{figure*}

Comparison between low-level variations as measured by \lambdar's sky estimation routine and as measured by the nebuliser routine presented in
\cite{Valiante2016}. Figure \ref{fig: Nebuliser} shows the two estimates as a function of RA/DEC. Here the \lambdar\ measurements were made by
running the sky estimate routine centred on every pixel in the image, {\em without} masking of known targets, using a single 3-sigma clip, and with
annuli between 17 and 60 pixels ($\sim1^\prime$ to $3^\prime$) in radius. These annuli are shown graphically (in black) in the bottom right of
each \lambdar\ panel, along with the PSF FWHM (in red).
From these figures we conclude that the sky-estimate routine is a sufficiently capable tool of removing subtle variations in the background in the
absence of a removal by nebuliser.

\section{Flux Iteration and Deblending}\label{sec: Iterative Deblending}
The outcome of iterative deblending and flux determination for a range of simulated flux ratios. Note that, as (in these tests) we always begin (Iteration 0)
from a state where the model flux ratio is unity, we will always converge from the same direction: bright objects will begin seemingly dimmer, and dim objects
will begin seemingly brighter. The result of this is that our final fluxes will, in cases of highly non-unity flux ratios, tend to underestimate the brighter object's
flux and overestimate the dimmer object's flux. Importantly, these tests showcase that deblending of unresolved sources with flux ratio $< 0.02$ typically
does not converge to a correct solution within 10 iterations.

\begin{figure*}
\includegraphics[scale=0.28]{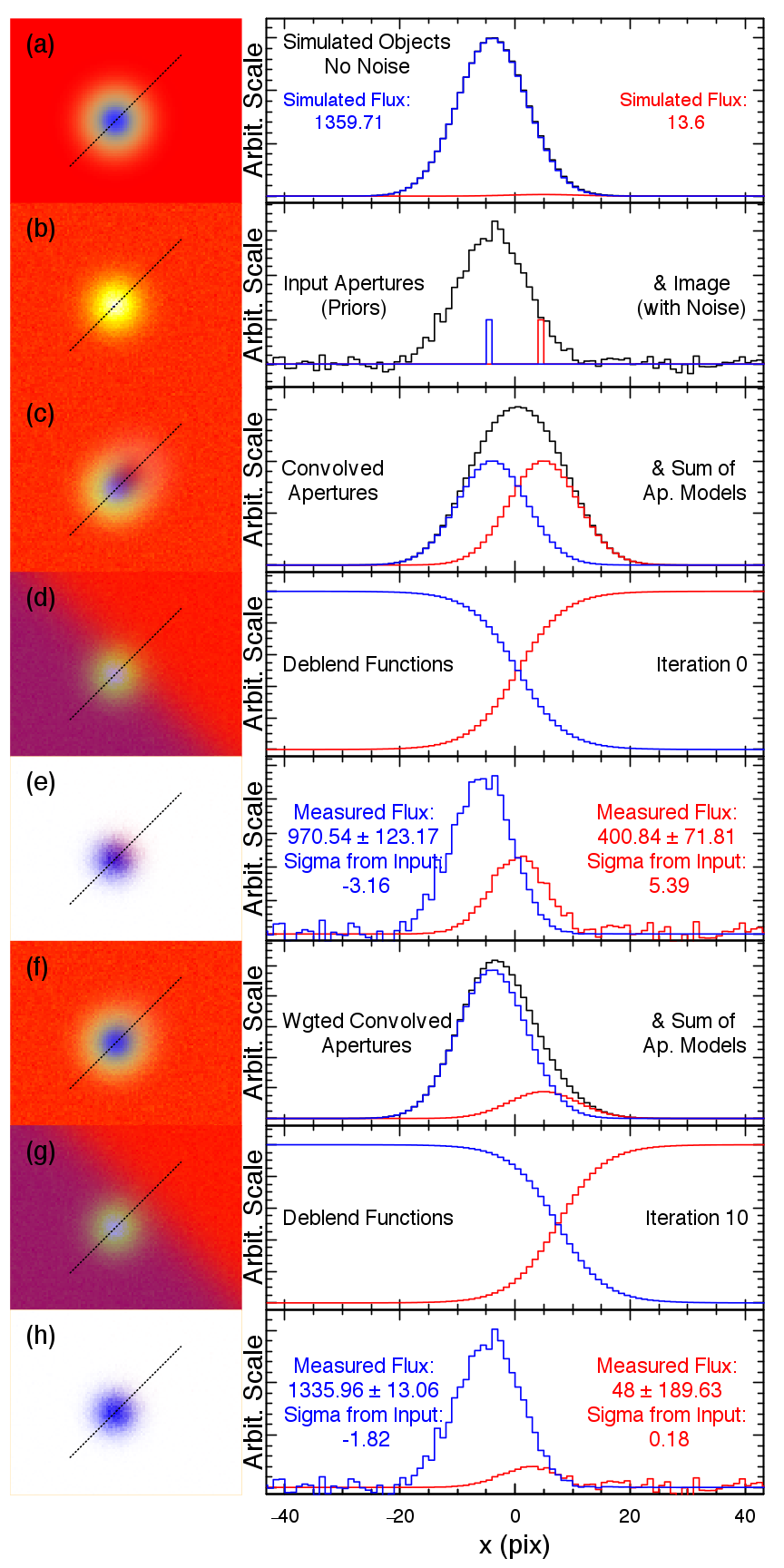}\hspace{2pt}\includegraphics[scale=0.28]{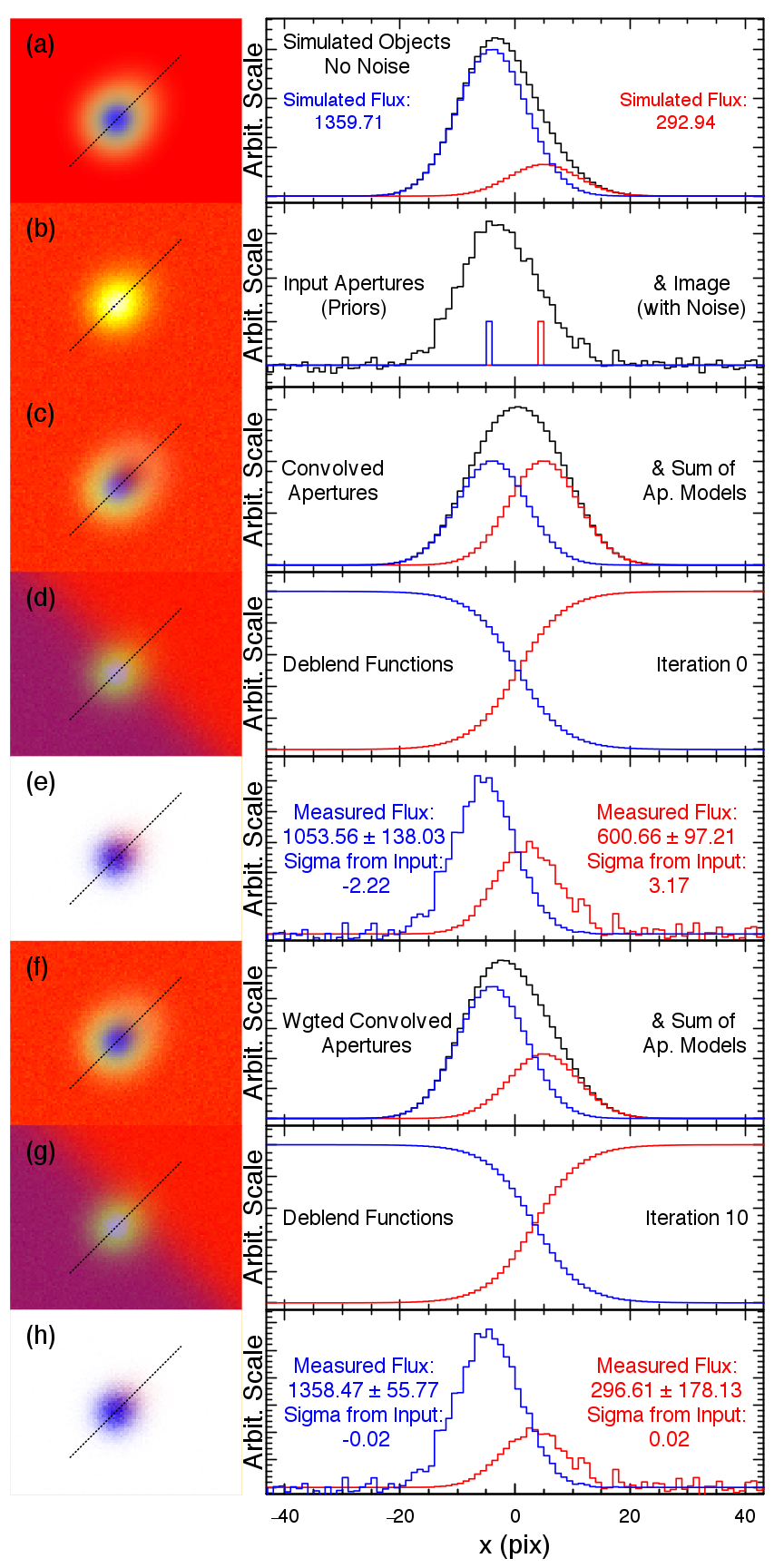}
\caption{Iterative Deblending of complex sources in \lambdar. The rows of each figure show:
The two simulated objects (red \& blue respectively) and the total flux profile (black), before addition of noise (row `a'); The simulated total flux profile
after addition of noise (black), and the apertures provided to the \lambdar\ program (red \& blue; row `b'); The convolved apertures and their sum with no weighting applied
(red, blue, and black respectively; row `c'); The iteration 0 deblend functions for the red \& blue sources, determined using the distributions from the panel above (row `d'); The simulated
image multiplied by the iteration 0 deblend functions (this is the so-called `deblended images'), for the red and blue sources respectively. Also shown are the measured
fluxes and uncertainties at iteration 0 (row `e'); The iteration 10 weighted convolved apertures for the red \& blue sources, and their sum (row `f'); The iteration 10 deblend function (row `g'); The
iteration 10 deblended images, and the measured iteration 10 fluxes with uncertainties (row `h'). In these two simulations, two point sources have been simulated with very
different fluxes. In the left panel, the source shown in blue has simulated flux 100x brighter than the companion source, shown in red.
In the right panel, the blue source has flux $\sim$5x brighter than the companion source. In the first case, the program is unable to converge (within uncertainties) in 10 iterations; this highlights the problem of providing catalogues
that are too deep for the imaging under analysis.
In the second case however, despite the two sources being very
different in their respective brightnesses, the program converges to within uncertainties within the
10 iterations shown here.
}\label{fig: appendixA}
\end{figure*}
\begin{figure*}
\includegraphics[scale=0.28]{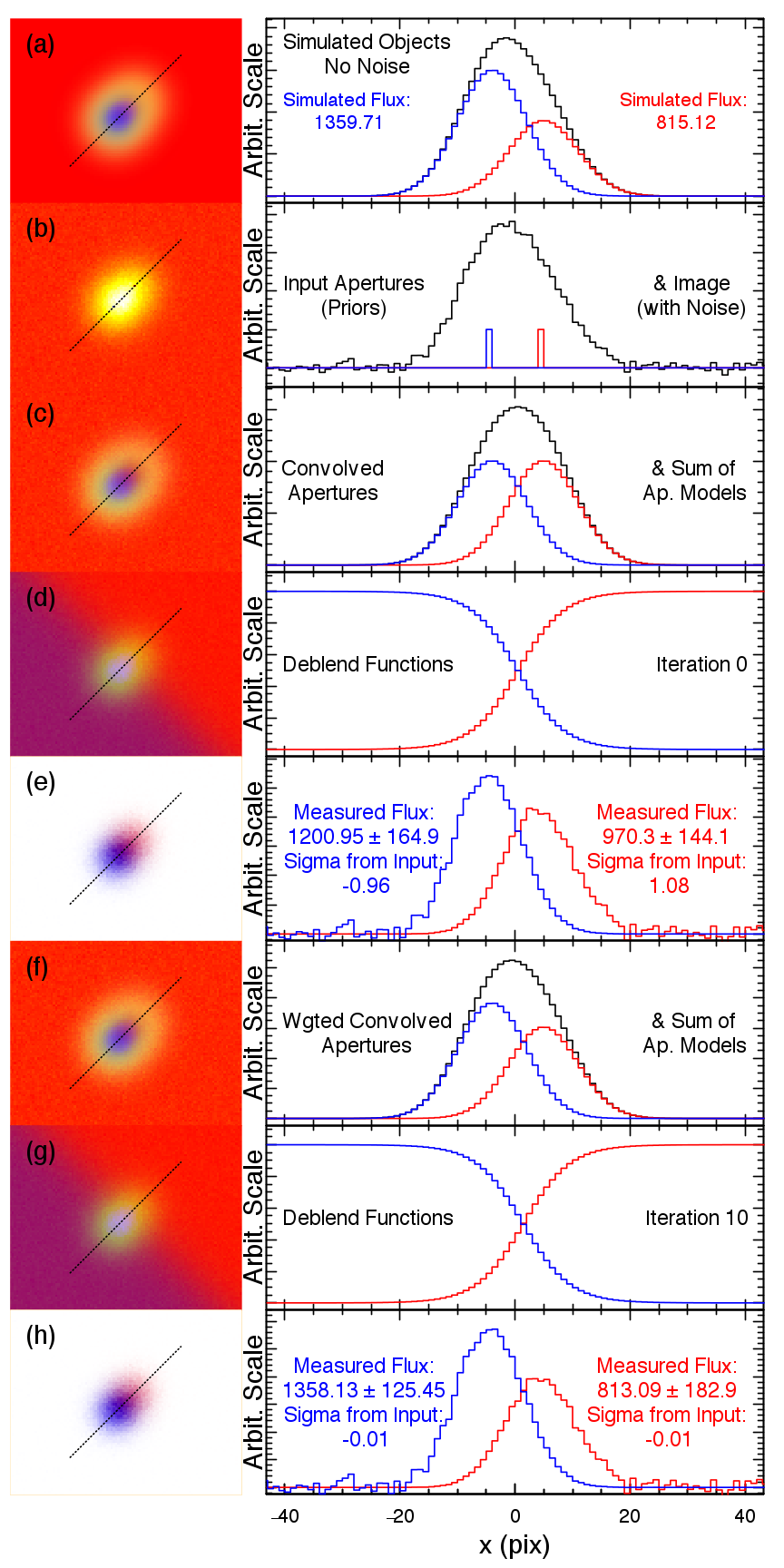}\hspace{2pt}\includegraphics[scale=0.28]{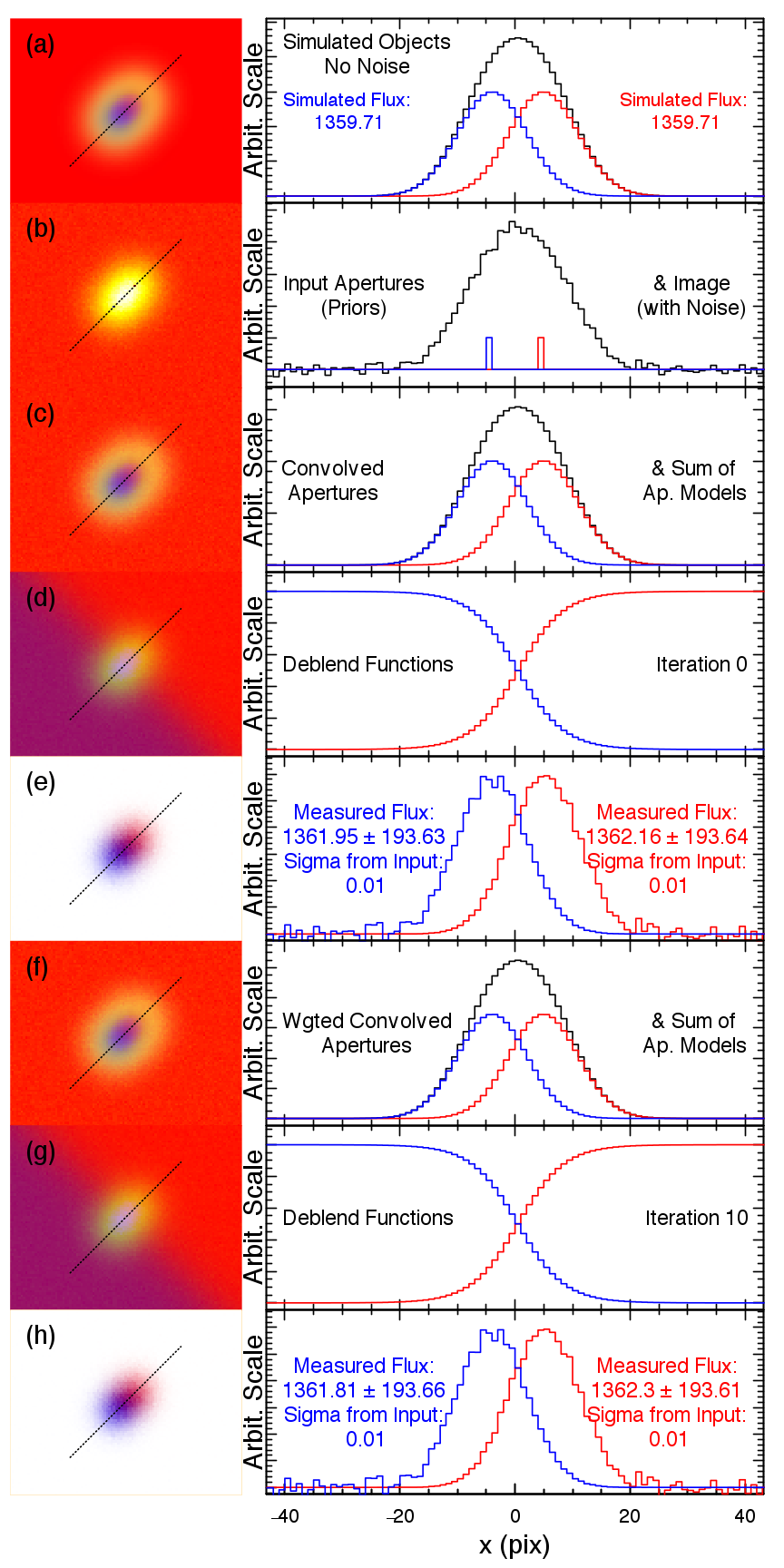}
\caption{Figure \ref{fig: appendixA} continued. In these two simulations the blue source has flux that is $\sim$1.5x brighter (left) and equal to the companion (right).
Again, the program converges to the solution within uncertainties within the 10 iterations.}
\end{figure*}

\section{Comparisons to GAMA PDR}\label{sec: Full PDR Comparison}
Here we show the full comparisons between the GAMA PDR photometry and that derived from the program. Note that in all of these figures, photometry has been calculated
with the parameters detailed in Table \ref{tab: lambdar params}, running \lambdar\ version 0.14. Included here are trumpet plots (Figure \ref{fig: trumpets}), colour
distributions (Figure \ref{fig: colours}), and error components (Figure \ref{fig: Error Components}), for all 21 photometric bands.

\begin{figure*}
\includegraphics[scale=0.35]{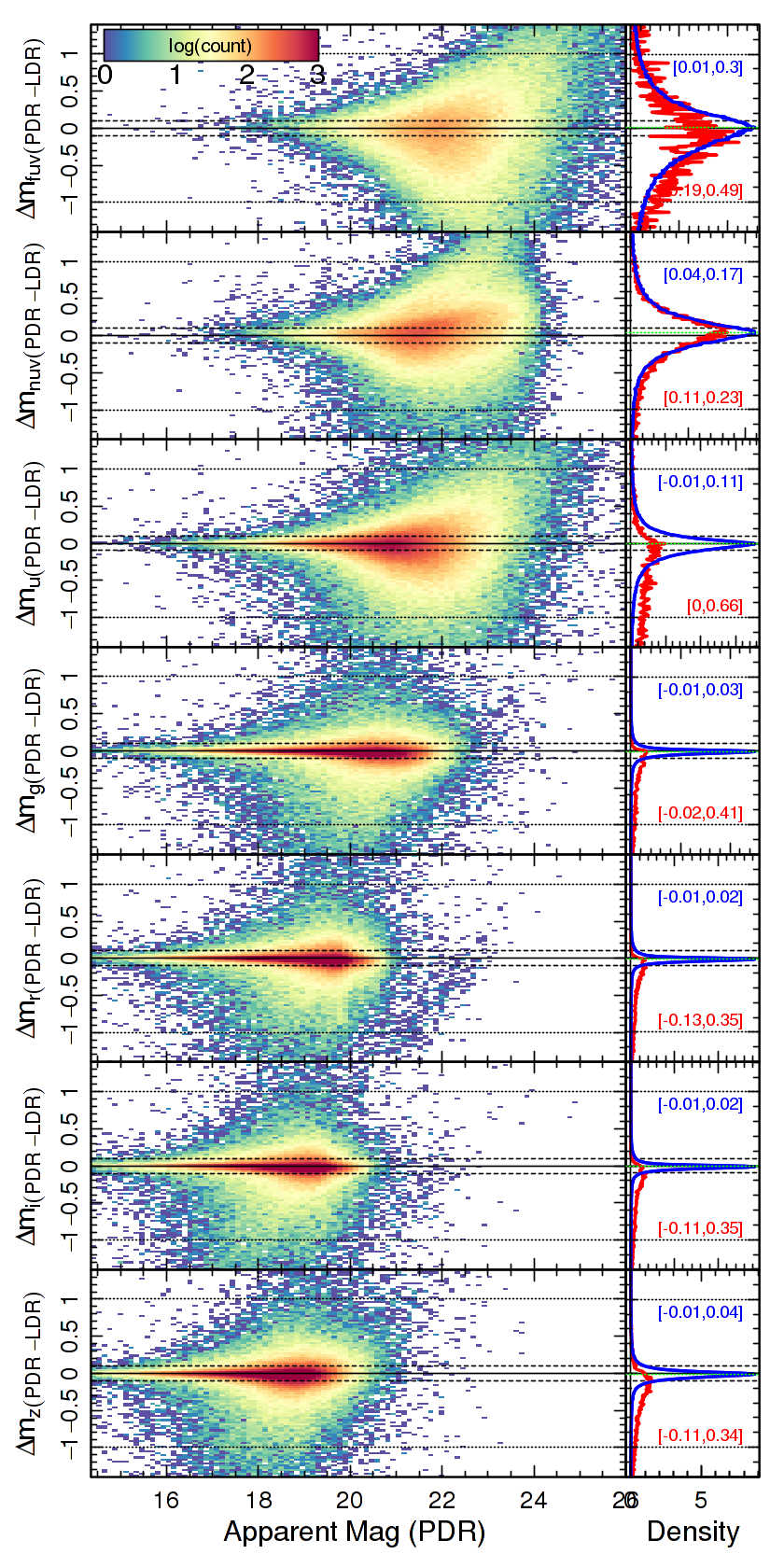}
\caption{Trumpet plots for the full 21-band photometry in GAMA. The left column is the same as Figure \ref{fig: trumpets}. The right column shows the projected distribution of residuals, split into resolved sources (blue), and
point sources (red). The mode and RMS of each distribution is also shown for each of the projected density distributions,
in blue and red. Horizontal solid, dashed, and dotted lines show the 0, 0.1, and 1 magnitude residual points, for reference.
}\label{fig: appendixB}
\end{figure*}
\begin{figure*}
\includegraphics[scale=0.35]{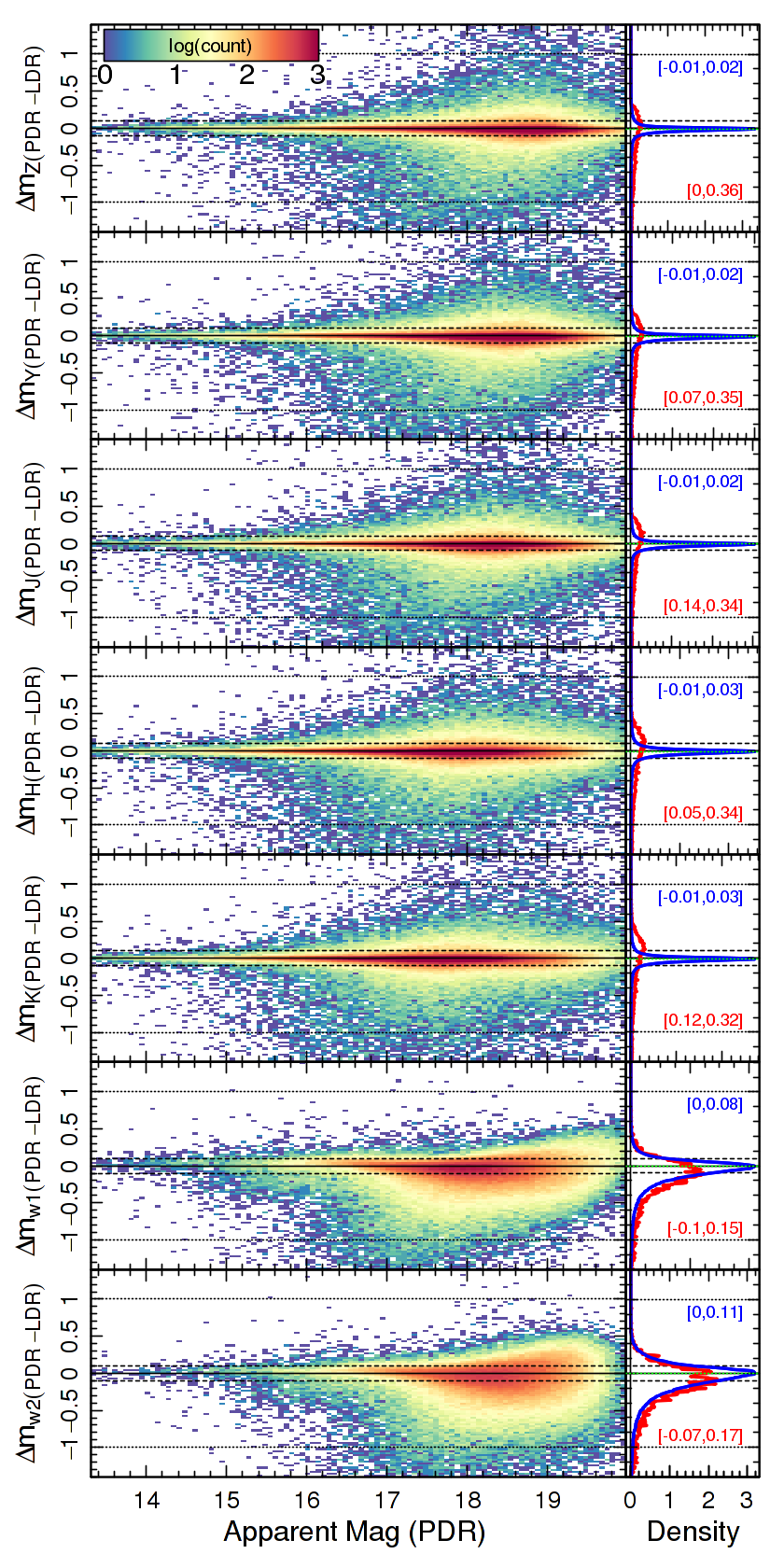}
\caption{Figure \ref{fig: appendixB} continued.}
\end{figure*}
\begin{figure*}
\includegraphics[scale=0.35]{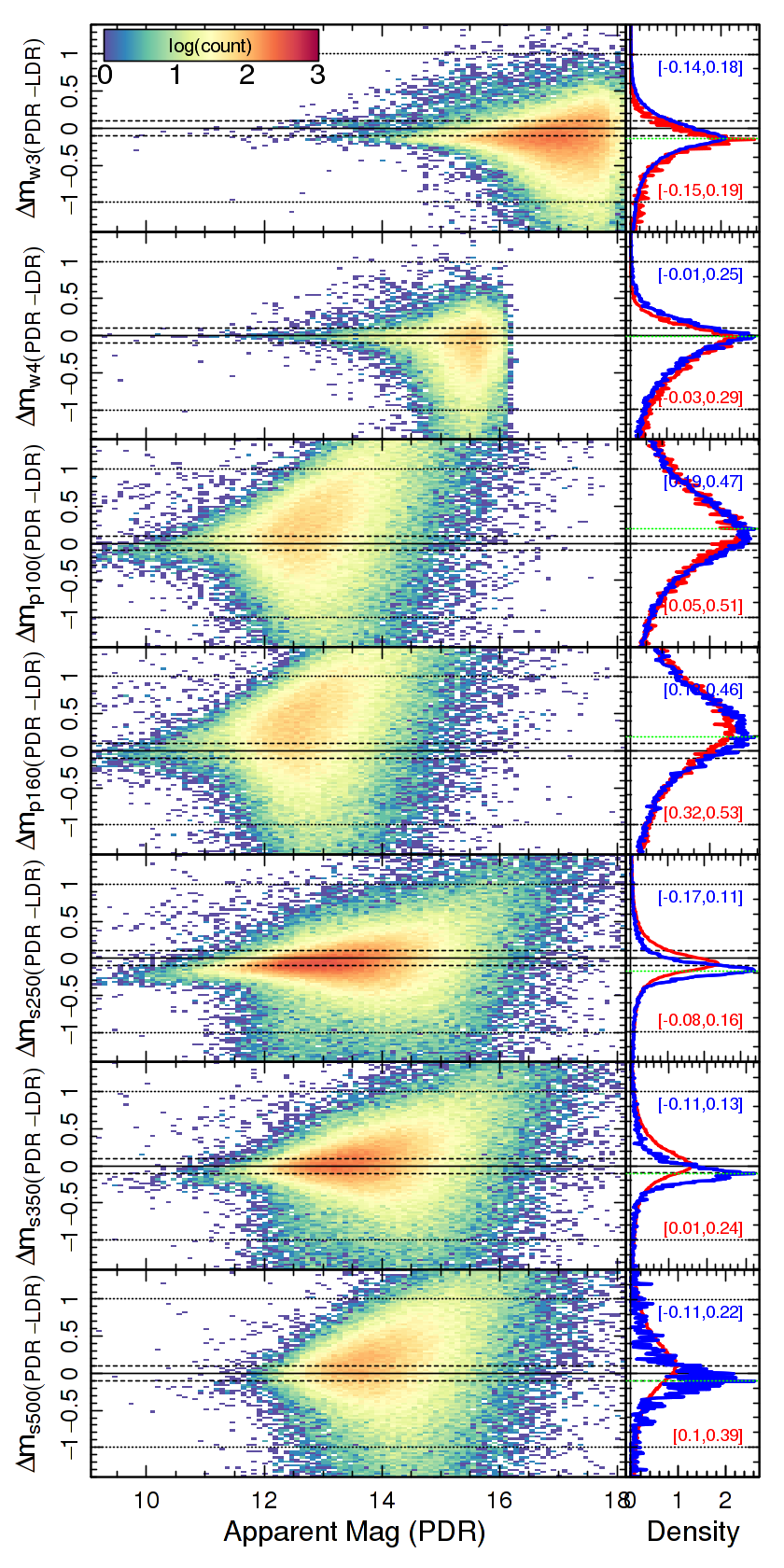}
\caption{Figure \ref{fig: appendixB} continued.}
\end{figure*}

\begin{figure*}
\includegraphics[scale=0.18]{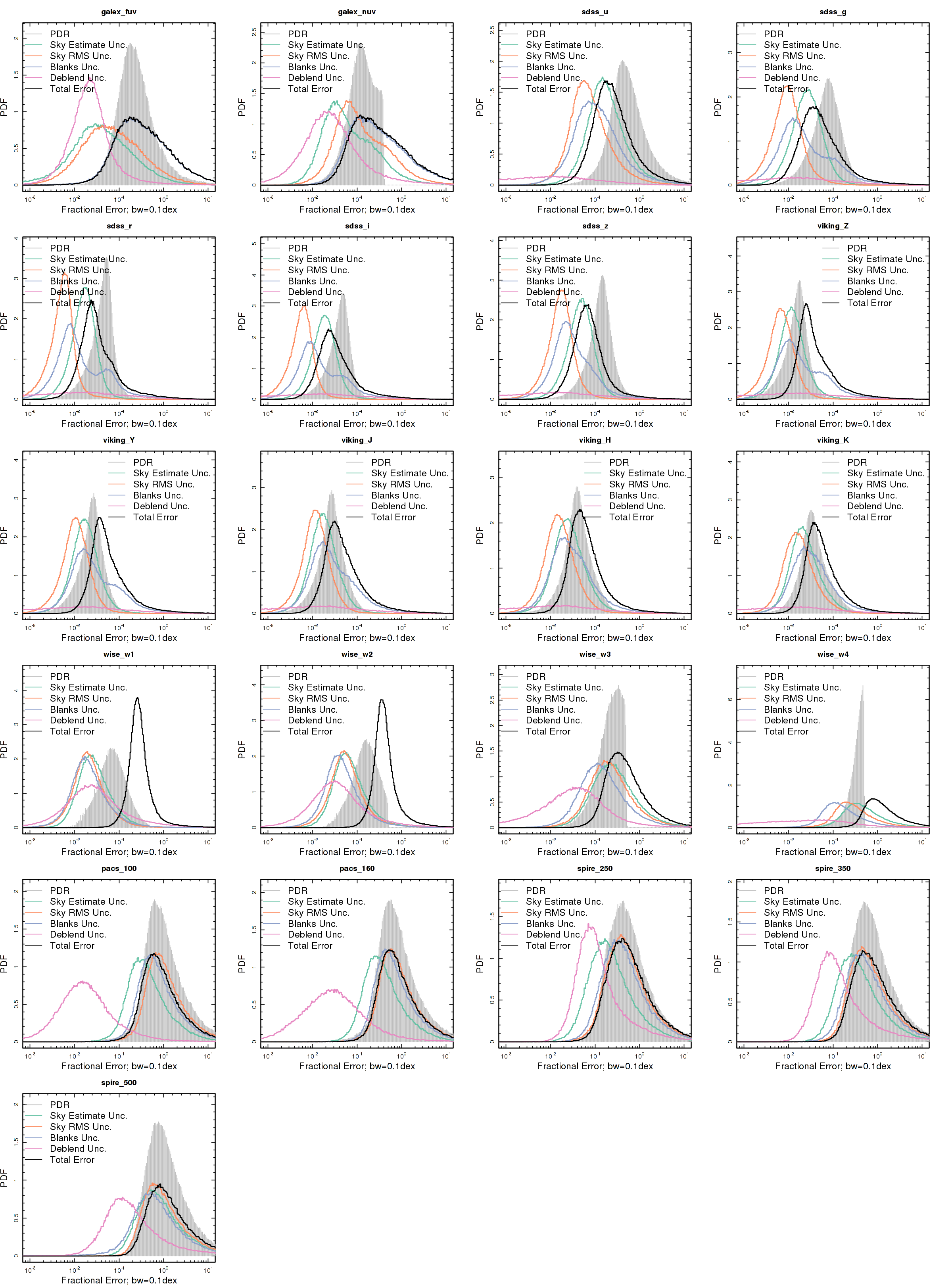}
\caption{Error Distributions for the full 21-band photometry in GAMA. Details of the panels are given in the caption of Figure \ref{fig: Error Components}.}\label{fig: appendixB2}
\end{figure*}

\begin{figure*}
\includegraphics[scale=0.25]{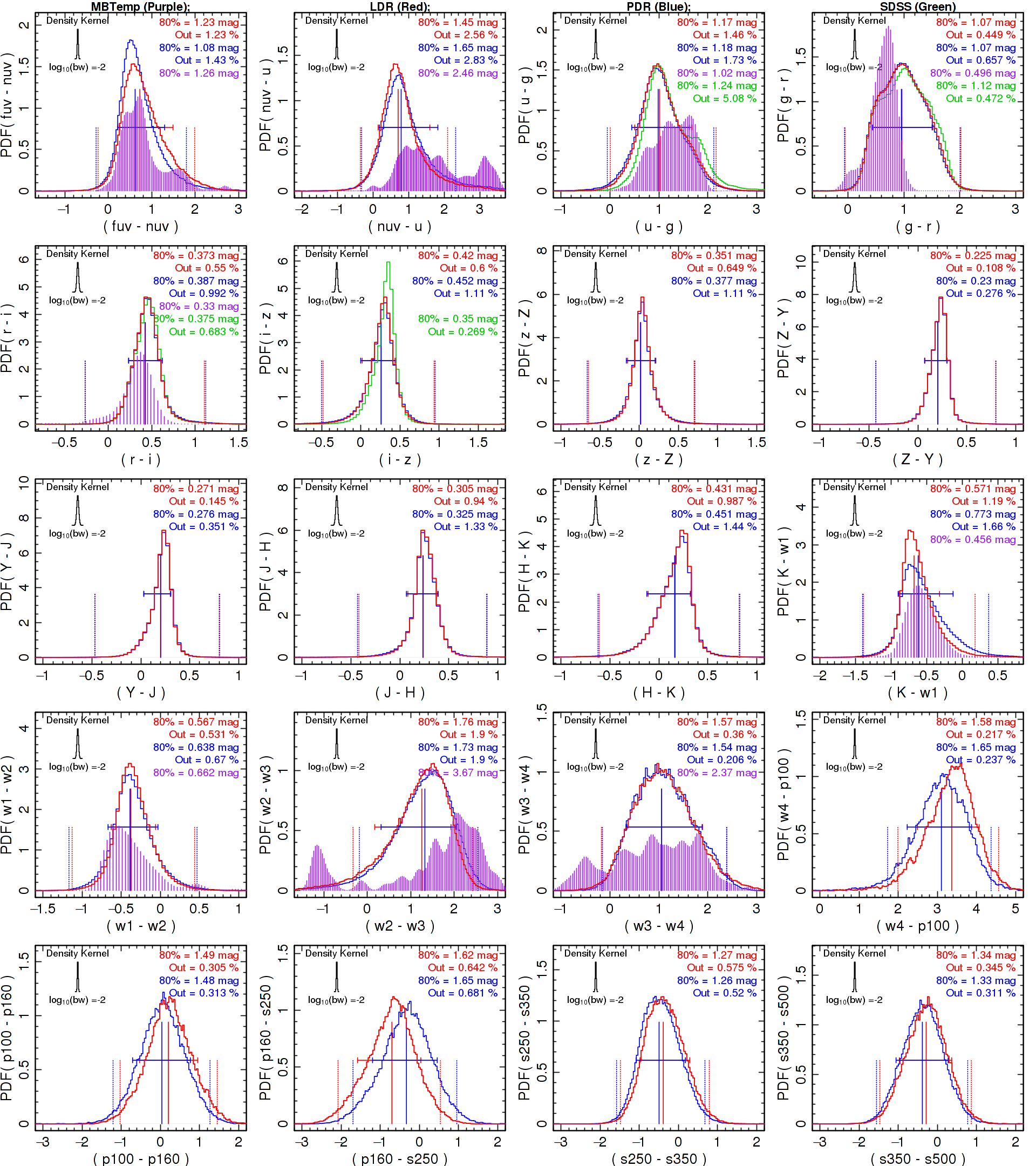}
\caption{Colour Distributions for the full 21-band photometry in GAMA. Details of the panels are given in the caption of Figure \ref{fig: colours}.}\label{fig: appendixB3}
\end{figure*}
%}}}

\end{document}